\documentclass[11pt]{article}

\usepackage[a4paper,includeheadfoot,top=1cm,bottom=2cm,left=2cm,right=2cm]{geometry}
\usepackage{here}
\usepackage{graphicx}
\usepackage[colorlinks=true,urlcolor=blue]{hyperref}
\usepackage{authblk}
\usepackage{caption}
\usepackage{xcolor}
\usepackage{soul}
\usepackage{amsmath}
\usepackage{float}
\usepackage[section]{placeins}
\usepackage{multicol}
\usepackage[
backend=biber,
style=numeric,
sorting=none
]{biblatex}
\graphicspath{{Figs/}}

\title{%Shafranov shift effects in burning plasmas
Energetic particles, Shafranov shift and finite $\beta$ effects on TAE, KBM and ITG instabilities in global electromagnetic gyrokinetic simulations

}

\author[1]{B.~Rofman}
\author[1]{G.~Di~Giannatale}
\author[2]{A.~Mishchenko}
\author[1]{E.~Lanti}
\author[3]{A.~Bottino}
\author[3]{T.~Hayward-Schneider}
\author[6]{J.N.~Sama}
\author[5]{A.~Biancalani}
\author[4]{B.F.~McMillan}
\author[1]{S.~Brunner}
\author[1]{L.~Villard}

\affil[1]{Ecole Polytechnique F\'ed\'erale de Lausanne, Swiss Plasma Center, CH-1015 Lausanne, Switzerland}
\affil[2] {Max-Planck-Institut f\"ur Plasmaphysik, Greifswald, Germany}
\affil[3]{Max-Planck-Institut f\"ur Plasmaphysik, Garching, Germany}
\affil[4]{University of Warwick, Department of Physics, UK}
\affil[5]{De Vinci Higher Education, De Vinci Research Center, 92916 Paris, France}
\affil[6]{Universit\'e de Lorraine, CNRS, IJL, Nancy, France}
\date{}             

\begin{document}

\maketitle

\begin{abstract}

Burning plasma is computationally challenging to simulate due to the multi-scale interactions between energetic particles (EPs), Alfven eigenmodes, and microinstabilities that drive turbulence. Many studies circumvent this difficulty by focusing on a single instability and making the corresponding simplifying assumptions. However, these approximations do not necessarily preserve the global instability spectrum, leading to conflicting results and inconsistencies. In this work, we identify the minimal set of assumptions needed to model a burning plasma self-consistently with the modes and species in the system.  Using the global gyrokinetic code ORB5, to systematically evaluate the impact of commonly adopted assumptions on the plasma response. We find it is essential to include Shafranov shift and finite $\beta$ effects from all magnetically confined species. Otherwise, unphysical electromagnetic modes like internal kinks and kinetic ballooning modes (KBMs) appear to dominate the instability spectrum. The EP contribution to the Shafranov shift is particularly important, stabilizing both the toroidal ion temperature gradient (ITG) and toroidal Alfven eigenmode (TAE) at the longer wavelengths (low toroidal mode numbers). For TAEs, these effects significantly reduce the linear growth rate, which saturates instead of proportionally increasing with EP fraction. In the nonlinear regime ITG-driven heat and particle fluxes are unaffected by the Shafranov shift in self-consistent magnetic equilibria. While the the nonlinear saturation level of the TAE remains unchanged across all cases, unlike for the ITG case Shafranov shift does reduce the TAE-driven EP fluxes.

\end{abstract}

%======================================================================
\section{Introduction} 
%======================================================================

Burning plasmas exhibit a broad instability spectrum driven by steep thermal gradients and energetic particles (EPs), posing a major challenge for predictive simulations. To reduce computational cost, many gyrokinetic studies use simplifying approximations such as electrostatic plasma response, adiabatic electrons, \cite{dimits2000comparisons} local "flux-tube" \cite{pueschel2008gyrokinetic}, circular concentric (ad hoc), or $\beta_{MHD}=0$ magnetic geometries  \cite{biancalani2016linear}. These assumptions are very useful in isolating and studying specific modes. However, they often do not reflect the physical conditions of the entire system \cite{mazzi2023study}. Consequently, it remains unclear which approximations preserve the full dispersion relation.  

A key requirement for predictive burning plasma simulations is to capture the multi-scale electromagnetic effects of EPs in systems with strong thermal gradients. The total effect of EPs on plasma stability is non-trivial to predict due to competing (stabilizing and destabilizing) mechanisms. On the one hand they excite a wide range of Alfv\'en eigenmodes (AEs) which in turn, act to expel the resonating EP from the plasma \cite{heidbrink1994behaviour}. On the other hand, there are experimental and numerical evidence of turbulence mitigation and improved confinement in the presence of EPs \cite{sama2024ion, na2025fast}. For self-consistency, when EP pressure is comparable to the thermal plasma pressure, $\beta_{\rm EP}\sim\beta_{\rm th}$, it must be accounted for in the magnetic equilibrium force balance, i.e. the Grad-Shafranov equation \cite{grad1958hydromagnetic, shafranov1958magnetohydrodynamical,shafranov1966plasma}. This results in a substantial increase in Shafranov shift which typically stabilizes the plasma \cite{coppi1979ideal, ramos1991theory, aleynikova2018kinetic}. 

\begin{figure} [h!]
\begin{center}
\includegraphics[width=0.495\textwidth]{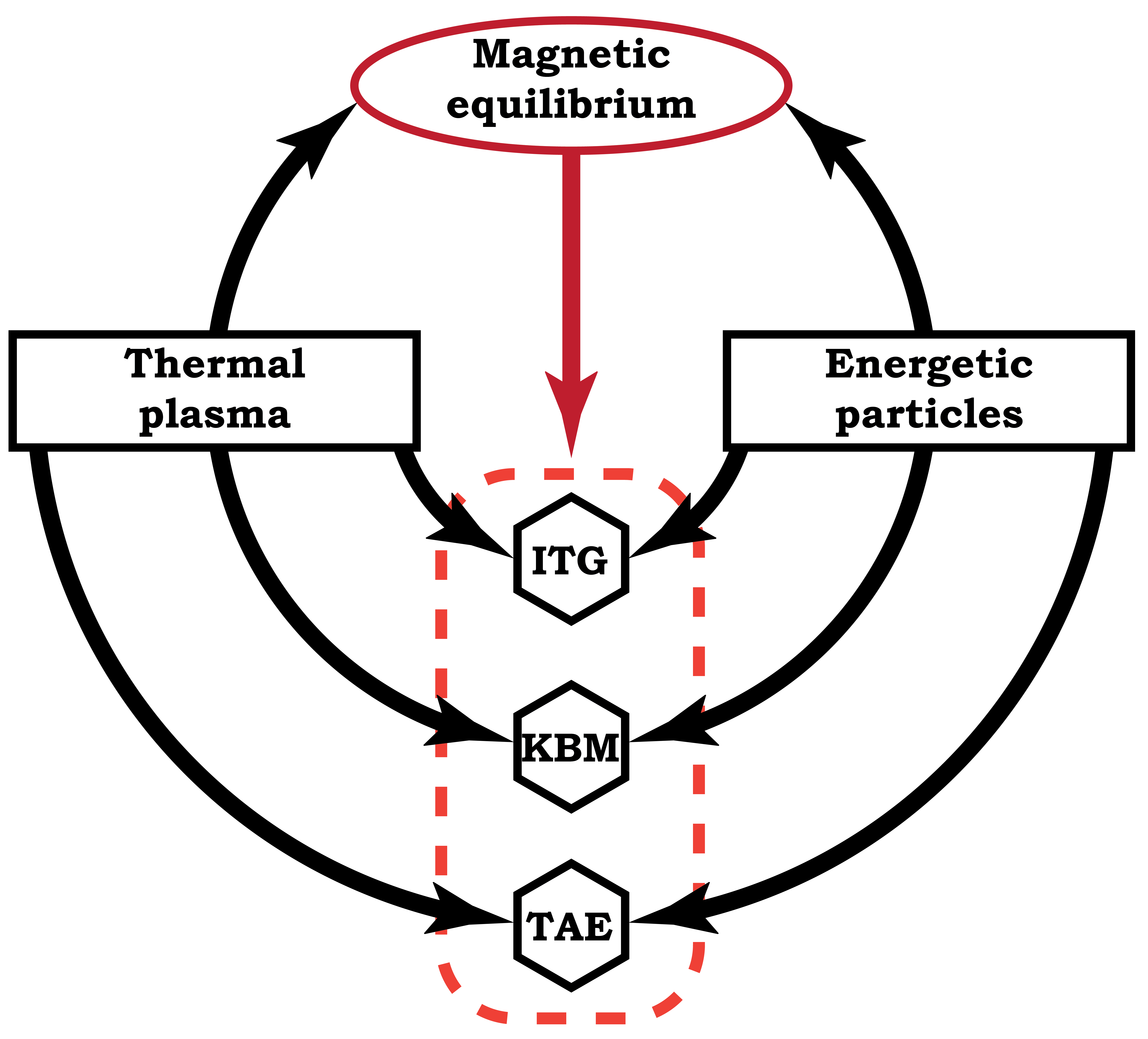}
\caption{\label{FIG:schematic} \it
Schematic illustration of the burning plasma components in a burning plasma system considered in this work.}
\end{center}
\end{figure}

In this work we do not focus on a single instability, but rather examine a wide range of the spectrum, (up to $k_\theta \rho^* \approx 0.6$) were we expect to find several principle instabilities naturally arising from the sources of free energy in the system. At the shorter wavelengths we consider the toroidal ion temperature gradient (ITG) \cite{coppi1967instabilities, guzdar1983ion, kim1993electromagnetic} microinstability.
Driven by temperature fluctuations on the ion Larmur radius scale, these are mainly electrostatic modes in nature. At longer wavelengths we can encounter the pressure driven, yet electromagnetic, kinetic ballooning mode (KBM) \cite{tang1980kinetic}. Finally, at the longest wavelengths we expect to find electromagnetic  toroidal Alfv\'en eigenmodes (TAEs) \cite{cheng1986low, chen2016physics} which are resonantly excited by the energetic particles. While these modes would be excited at different parts of the spectrum, they still must coexist in the same self-organizing system, simultaneously interacting with the thermal plasma, the energetic particles and the magnetic equilibrium (Fig. \ref{FIG:schematic}. As a result, assumptions made in the investigation of a single instability should still maintain the global spectrum. The goal of this work is to map the minimal set of assumptions needed to self-consistently simulate unstable modes in a burning plasma system. 

The rest of the work is organized as follows: using the global gyrokinetic code ORB5 \cite{lanti2020orb5} we first establish the instability spectrum of the thermal plasma by starting from electrostatic cases with adiabatic electrons in a $\beta_{MHD}=0$ magnetic equilibrium, and progressively introducing kinetic electrons, electromagnetic effects, and realistic magnetic geometries. Energetic particles are then introduced to study their kinetic effects on ITGs and KBMs, before accounting for the self-consistent equilibrium modification. Next, we study the impact on the TAE spectrum of increasing the EP fraction in a self-consistent system. Finally, nonlinear simulations of selected modes connect between magnetic equilibrium consistency, the linear spectrum and the resulting turbulent transport.

\FloatBarrier
%======================================================================
\section{Simulation setup and parameters}
%======================================================================

In this work we use ORB5 \cite{lanti2020orb5}, a global, multi-species, Lagrangian PIC code, which solves the nonlinear, electromagnetic, gyrokinetic Vlasov–Maxwell system. The fields are represented by the potentials $(\phi,A_\parallel)$ using B-splines on a 3D finite element grid in Fourier space. Throughout this work, except for the internal kink simulations, we use a correction term to the drift velocity $v_d$, which accounts for the parallel magnetic field perturbations $\delta B_{\parallel}$ to first order \cite{antlitz2026comparison}. 

To reduce the Monte Carlo sampling noise ORB5 implements a field-aligned Fourier filter in a straight-field-line coordinate representation $(s,\theta^*,\varphi)$. This allows us to keep only the $m \in [nq - \Delta m \ ,\ nq + \Delta m]$ modes. where, $\Delta m = 5$, $q$ is the safety factor, $n$ and $m$ are the toroidal and poloidal mode numbers. In an axisymmetric geometry the toroidal modes are linearly decoupled, which allows us to simulate a restricted set of toroidal mode numbers, $n$, while including a large range of poloidal mode numbers $m$. 

\begin{figure}[htbp]
    \centering

    % Left large figure
    \begin{minipage}[c]{0.53\textwidth}
        \centering
        \includegraphics[width=\linewidth]{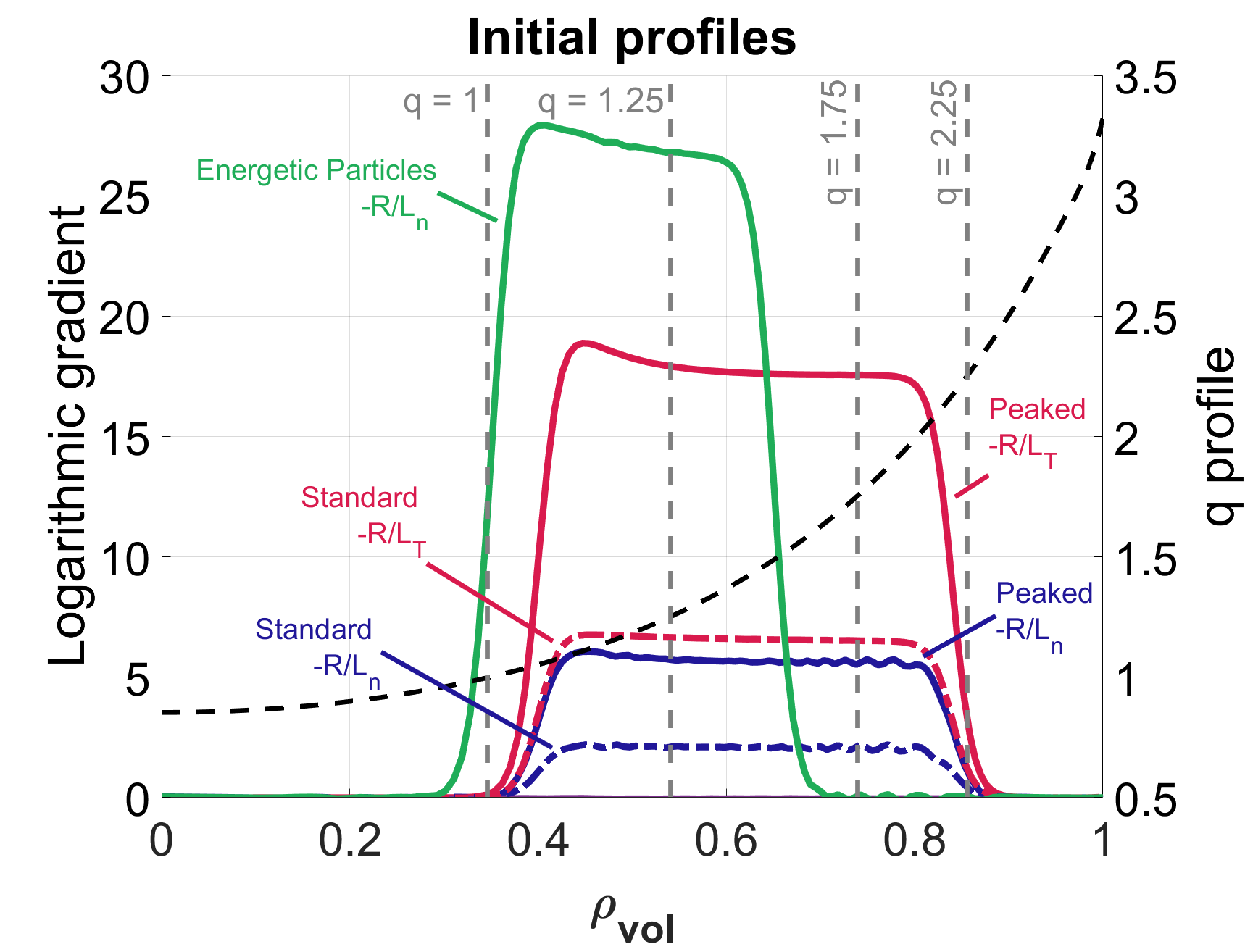}
       % \caption*{(a) Large figure}
    \end{minipage}
    \hfill
    % Right 2x2 matrix of small figures
    \begin{minipage}[c]{0.46\textwidth}
        \centering

        \includegraphics[width=0.49\linewidth]{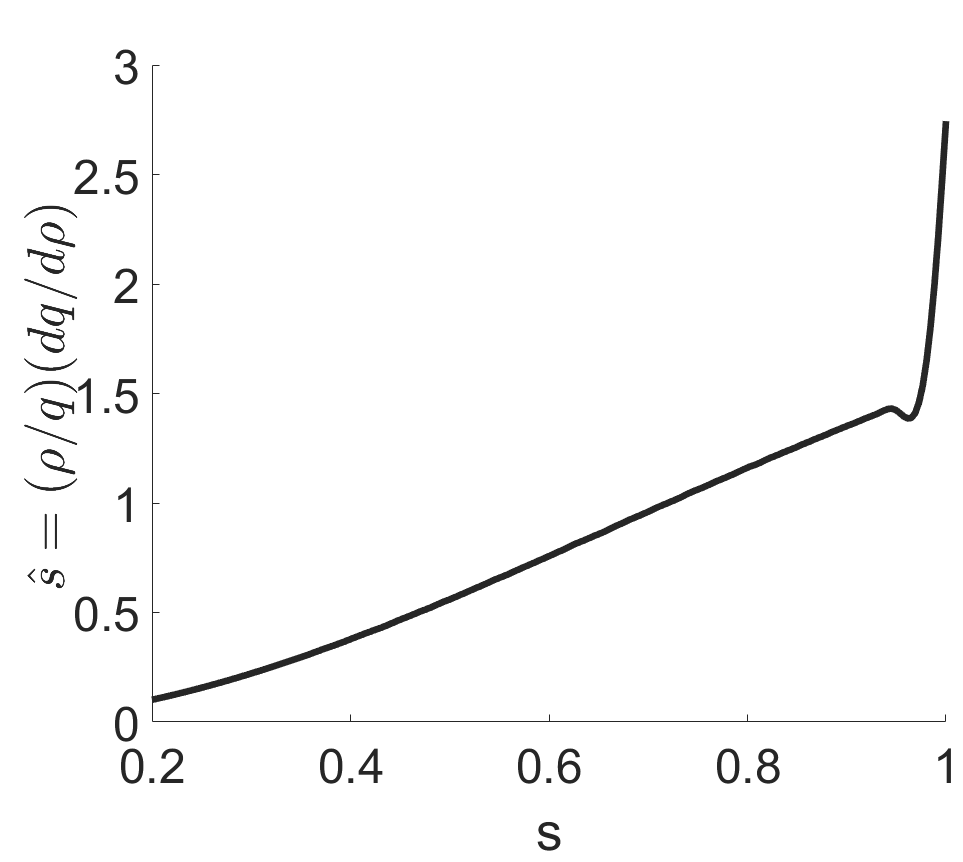}
        \hfill
        \includegraphics[width=0.49\linewidth]{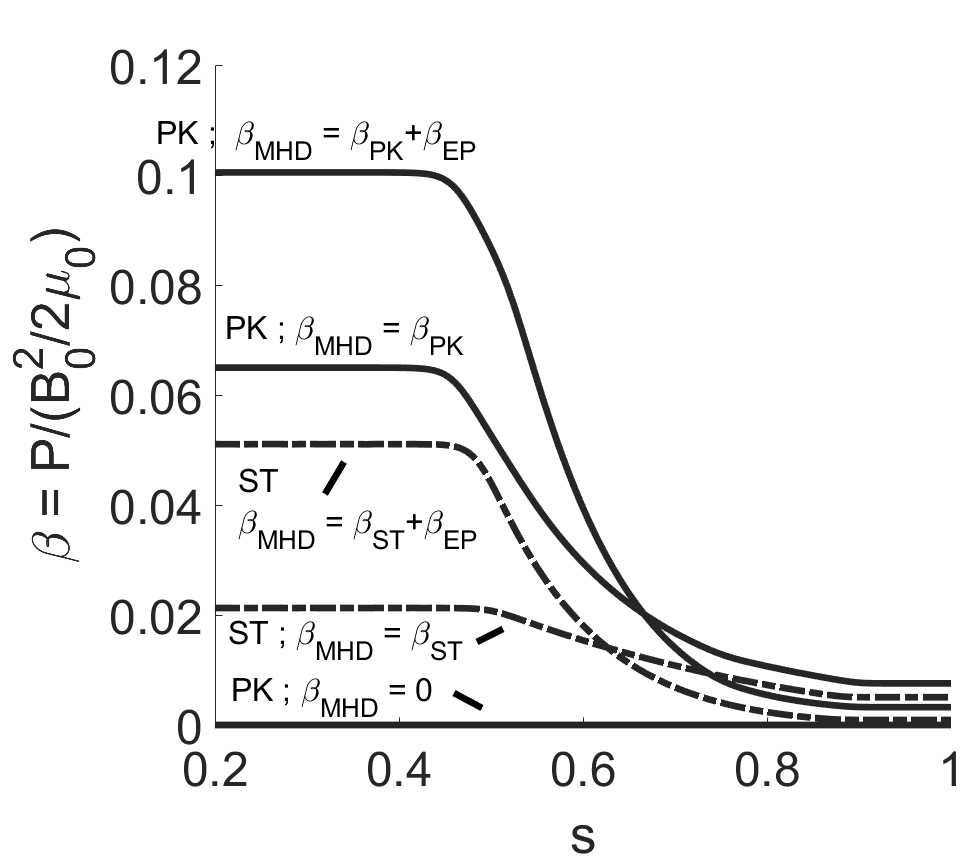}

        %\vspace{0.5em}

        \includegraphics[width=0.49\linewidth]{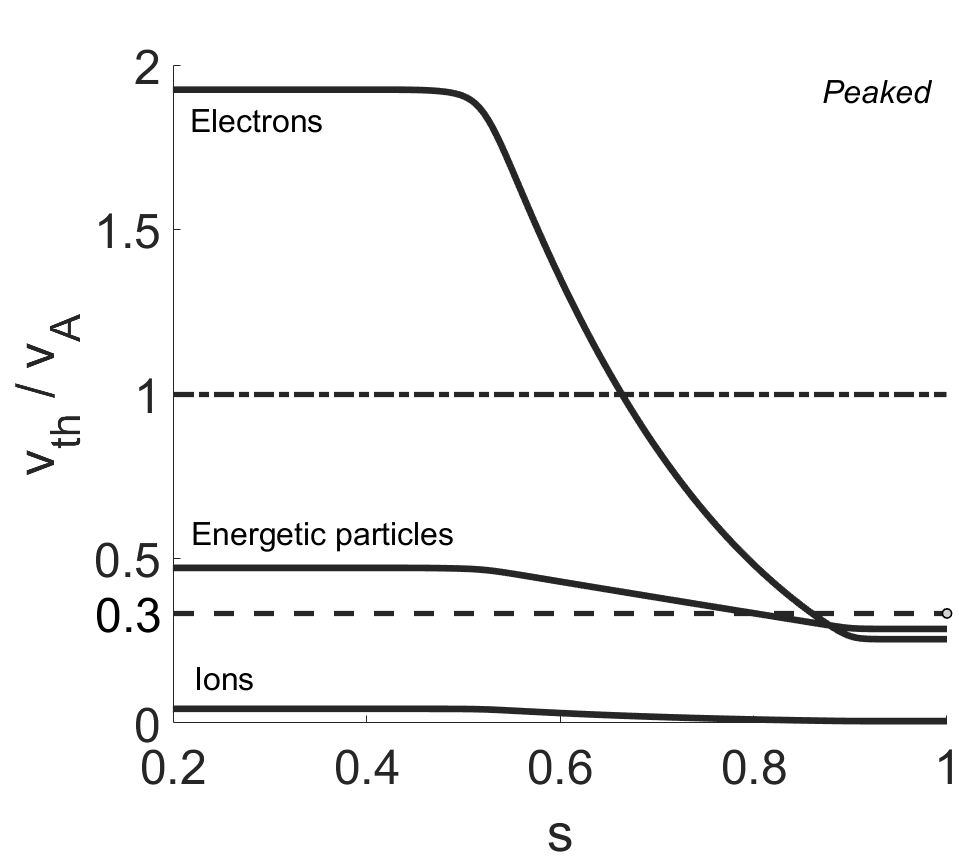}
        \hfill
        \includegraphics[width=0.49\linewidth]{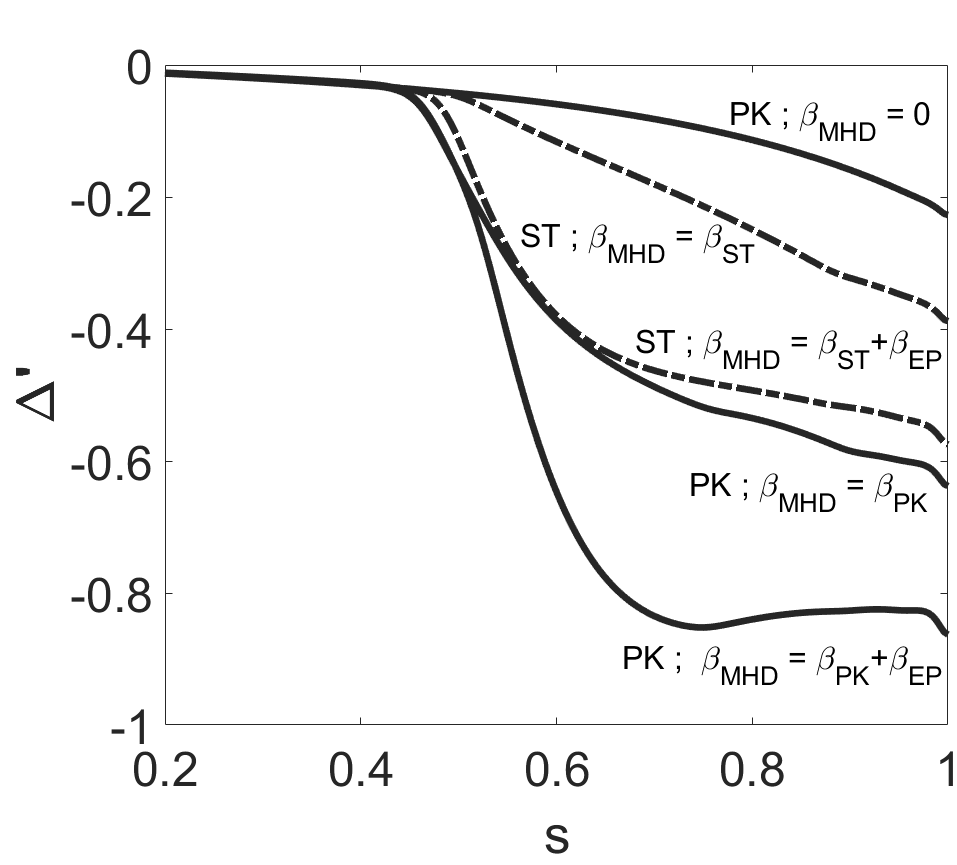}

       % \caption*{(b) Four small figures}
    \end{minipage}
    \caption{\label{FIG:Combined_profiles_particles} \it On the left: initial temperature and density (flux-surface-averaged) logarithmic gradients, for ions, electrons, and EPs. The profiles of the bulk species (ions and electrons) change between the standard (ST) and peaked (PK) cases, while the EP profiles remain the same. The q profile is indicated by the black dashed line. The vertical gray dashed lines mark the mode rational surfaces $(nq = m + 1/2)$ of the $n = 2$ TAE. On the right: global shear, $\beta$, $v_{th}/v_A$, and  $\Delta^\prime$. Where $\beta_{MHD}$ is used for the MHD reconstruction, and $\beta_x$ is the contribution from x. 
    %$\Delta^\prime$ and the $\beta$ profiles of the MHD equilibrium for the main 5 cases in this work.: $\beta = 0$, not consistent ST and PK equilibria based on pressure arising solely from the bulk profiles, and consistent ST and PK equilibria which account for $1\%$ EP pressure as well.
    }
\end{figure}

In the nonlinear simulations we use a modified Krook operator both as noise control and as a source with a damping/drive of about $5\%$ of the linear growth rate. As noise control the Krook operator is designed to conserve flux-surface-averaged moments such as density, energy, parallel flows and residual zonal flows. In gradient-driven like simulations, the Krook operator acts as a source/sink on the particles to maintain the original profile, e.g. the temperature gradient in ITG. Being only $5\%$ of the linear growth rate, it allows for limited profile relaxation while avoiding decaying turbulence simulations where after the initial saturation the profiles relax and there is no more drive \cite{mcmillan2008long}. Such methods are sufficient to investigate the early nonlinear phase, while flux driven simulations with more complex noise control operator are needed to reliably investigate longer time scales \cite{di2025collisional}.

A recently introduced 'pullback' scheme with a mixed - variables representation \cite{mishchenko2019pullback} which deals well with the so called "cancellation problem", allows for electromagnetic linear and nonlinear simulations with gyro-kinetic species to be performed at a reasonable computational effort. This allowed us to successfully capture a ITG-KBM transition \cite{cole2021tokamak}, investigate the interactions between EPs, Alfv\'enic modes and microinstabilities \cite{biancalani2021gyrokinetic, hayward2022multi, sama2024ion, ivanov2026dynamics}, and carry out extensive numerical linear and nonlinear validation campaigns \cite{vlad2021linear, vlad2025state}.  

We perform our simulations with a $\Delta t = 0.5 [\Omega^{-1}_{c,i}]$ for numerical stability, and use $40 M$ numerical markers per species, and per toroidal mode number $n$. Working on the finite elements grid $n_s\times n_{\theta^{*}}\times n_{\varphi}$, we use a grid of $128 \times 128 \times 64$ for the TAE cases and a $512\times (1024,512) \times256$ grids for the ITG cases. 

The plasma parameters are inspired by the Cyclone Base Case (CBC) \cite{dimits2000comparisons}: major radius $R_0 = 1.7\ m$, minor radius $a = 0.61\ m$, and magnetic field on axis $B_0 = 1.9\ T$. We assume collisionless, hot plasma with $T_i = T_e$ and we keep the machine size constant with $1/\rho^*(s=0.5) = 180$, were $\rho^* = \rho_s/a$, $\rho_s = c_s/\Omega_{c,i}$ is the ion sound Larmor radius, with $c_s = \sqrt{T_e/m_i}$ being the ion sound speed. As a result, the peaked profiles will have a larger local $\rho^\ast$ in the core - and a lower $\rho^\ast$ near the edge when compared to the standard profiles. We treat the ions and EPs gyro-kinetically, with 4-16 points in the Larmor ring, and the electrons drift-kinetically because of their relatively small orbit width. Due to computational limitations, we use slightly heavy electrons with a mass ratio of $m_e/m_i = 1/1000$ %(vs. the physical ratio of $m_e/m_i = 1/3670$ for deuterium).

We use CHEASE \cite{lutjens1996chease}, a fixed boundary Grad - Shafranov solver to generate a set of circular finite $\beta$ ideal-MHD equilibria, all with the same q profile. Each equilibrium corespondents to a different pressure profile and thus Shafranov shift. Where, Shafranov shift is the plasma equilibrium response to the internal confined pressure which can be expressed as $\Delta(s) = R_{mag}-R_{geom}(s)$, where $R_{mag}$ is the location of the magnetic axis, $R_{geom}(s) = (R_{max}(s)+R_{min}(s))/2$, and $s=\sqrt{\psi/\psi_{a}}$ is a flux surface label.
We refer to the normalized pressure used to recreate the MHD equilibrium as $\beta_{MHD}$. This $\beta_{MHD}$ can be a function of the thermal plasma pressure, EP pressure, both, or neither. In summary, we consider different MHD equilibria with $\beta_{MHD} = [0,\beta_{ST},\beta_{PK},\beta_{ST} + \beta_{EP}, \beta_{ST} + \beta_{EP}]$, respectively. Where ST and PK stand for standard and peaked profiles as shown in Figure \ref{FIG:Combined_profiles_particles}. In summery the profiles are found Table \ref{tab:gradients}

\begin{table}[!h]
\begin{center}
\caption{Temperature and density logarithmic gradients for the 3 cases in studies in this work.}
\label{tab:gradients}
\begin{tabular}{ | c | c | c | c | } 
 \hline
 Profiles & $ -R/L_T $ & $ -R/L_n $ & $\eta_{i,e} = \frac{R/L_T}{R/L_n}$ \\ [0.5ex] 
 \hline
 \hline
 Low temperature (LT) & $6.23$ & $6.23$ & $1$ \\ 
 \hline
 Standard (ST) & $6.72$ & $2.195$ & $3.06$ \\ 
 \hline
 Peaked (PK) & $19.25$ & $6.23$ & $3.09$ \\ 
 [1ex] 
 \hline
 %\hline

\end{tabular}

\end{center}
\end{table}

The Alfv\'en Eigenmodes are excited by energetic particles with either $1\%$ or $3\%$ dilution from the bulk ions. We take the EP mass to be equal to ion mass $m_{EP} = m_i$, and the effect on bulk density profiles is considered negligible. The EPs have a constant initial temperature that is 120 times hotter than the bulk ions $\tau_{EP} = T_{EP}/T_i \stackrel{s=0.5}{=} 120$.

Figure \ref{FIG:Combined_profiles_particles} presents the species temperature and density logarithmic gradients in standard and peaked cases. With the resulting $\Delta'$, and the $\beta_{MHD}$ profiles shown on the left. We chose the EP logarithmic density gradient with several goals in mind. First, to induce the desired instability while avoiding other instabilities that might exist in the system, e.g. an internal kink due to the $q < 1$ profile at the core $(s < 0.35)$. Second, to induce the microinstabilities and the AEs at a similar radial location, and third, to obtain a system where both the low-$n$ AE and the mid-$n$ ITG have comparable growth rates.  

%======================================================================
\section{Results and Discussion}
%======================================================================
\section{Linear physics}
%======================================================================
In the following section we will use linear simulations to map the dispersion relation of our system. Typically, ITG instabilities are studied in the electrostatic regime, assuming adiabatic electron response, and a circular concentric magnetic geometry. However, this set of assumptions is incompatible with a burning plasma which includes EPs and AEs that are electromagnetic and kinetic in nature. Therefore we start from the electrostatic and adiabatic case and build up the complexity of the system by adding electromagnetic effects, and EPs.

%======================================================================
\subsection{Part I - Thermal plasma}
%======================================================================
First we focus on the thermal bulk plasma without energetic particles, where gradients in temperature and pressure drive ITG and KBM instabilities.

\subsubsection{Electron models in electrostatic cases}
\begin{figure} [h!]
\begin{center}
\includegraphics[width=\textwidth]{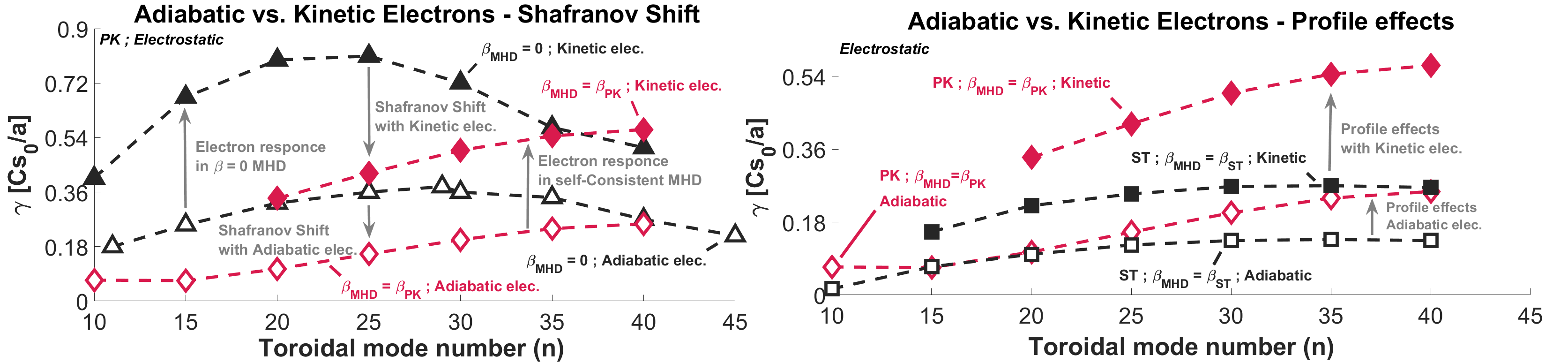}
\caption{\label{FIG:electron_models} \it
ITG growth rate $\gamma$ per toroidal mode number $n$ for adiabatic and kinetic electrons in the electrostatic limit.}
\end{center}
\end{figure}

%======================================================================
In this section we perform electrostatic simulations, and compare the plasma response with two electron models: adiabatic and drift kinetic \cite{lanti2020orb5}. Figure \ref{FIG:electron_models} (left panel) shows that for all cases, the kinetic electron response results in about double the growth rate found with adiabatic electrons. Thus, we recover using a global gyrokinetic approach the results of Rewolt and Tang \cite{rewoldt1989toroidal} and Dominski et al \cite{dominski2015non}. This result is attributed to the non-adiabatic response of the trapped electrons, at least in collisionless plasmas. %Interestingly, collisions tend to bring the growth rate back down by inducing adiabatic-like response away from mode rational flux surfaces. 

To study the effect of Shafranov shift on the ITG growth rate, we compare in the left panel of Figure \ref{FIG:electron_models} ITG cases driven by peaked gradients with either adiabatic or kinetic electrons in  either \textit{self-consistent} magnetic geometries with $\beta_{MHD}=\beta_{PK}$ or \textit{inconsistent}  $\beta_{MHD}=0$ equilibrium. For both kinetic and adiabatic electrons the additional Shafranov shift has a stronger stabilization at the longer wavelengths (lower toroidal mode numbers). As a result the most unstable mode, i.e. the mode with the highest growth rate, shifts to much shorter wavelengths (higher toroidal mode numbers). When Shafranov shift is consistently accounted for, the shape and the peak of the dispersion relation remain independent of the electron model.

Next, in the right panel of Figure \ref{FIG:electron_models} we compare between two cases where the Shafranov shift is \textit{self-consistent} with the plasma profiles. We find that for the electrostatic ITG, the gradient drive overcomes the Shafranov shift stabilization with a proportionally stronger response from the kinetic electrons as before. From this point forward, we will use drift kinetic electrons in all our cases. 

%\FloatBarrier
%======================================================================
\subsubsection{Electromagnetic effects}
\begin{figure} [h!]
\begin{center}
\includegraphics[width=0.495\textwidth]{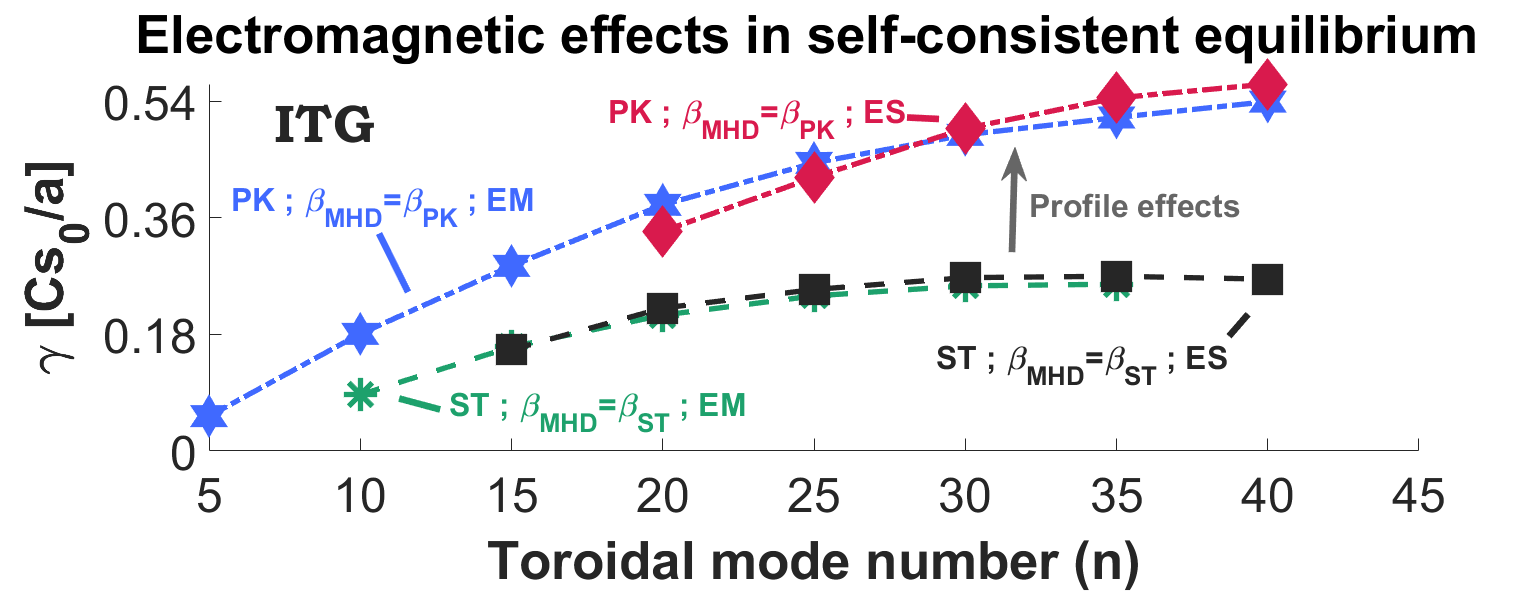}
\includegraphics[width=0.495\textwidth]{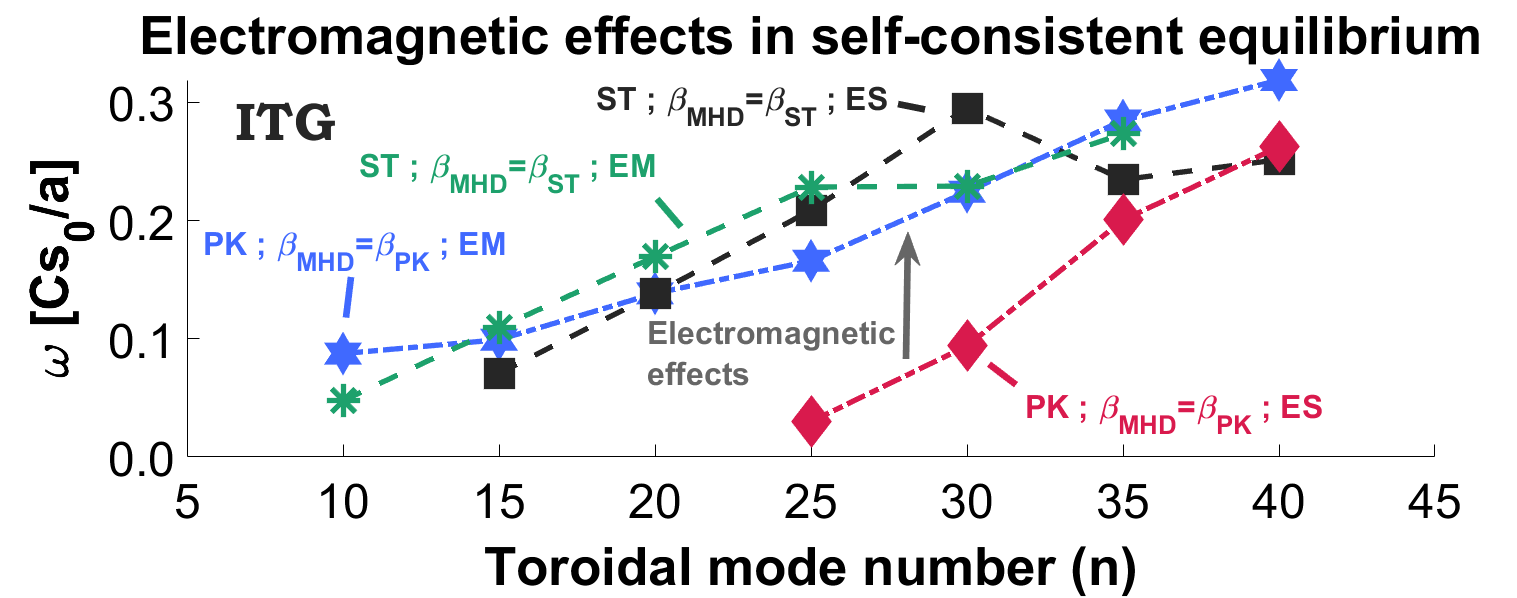}
\caption{\label{FIG:ESvsEM} \it
ITG dispersion relation for electrostatic and electromagnetic cases with kinetic electrons. Cases with \textit{self-consistent} MHD equilibria show very little electromagnetic effects on the growth rates and some effect on the frequency. }
\end{center}
\end{figure}

%======================================================================
In this section we study finite $\beta$ effects in cases where the species pressure is \textit{self-consistent} with the MHD equilibrium, i.e. $\beta_{MHD} = \beta_{ST}$ for standard profiles and $\beta_{MHD} = \beta_{PK}$ for the peaked profiles. Figure \ref{FIG:ESvsEM} shows that there are little to no electromagnetic effects on the ITG growth rate for the weaker standard profiles, while the frequency of the strongly driven ITG is pushed down by Shafranov shift and pushed up by electromagnetic effects.    

\FloatBarrier
\subsubsection{ITG destabilization by Shafranov shift (standard profiles)}
\begin{figure} [h!]
\begin{center}
\includegraphics[width=0.495\textwidth]{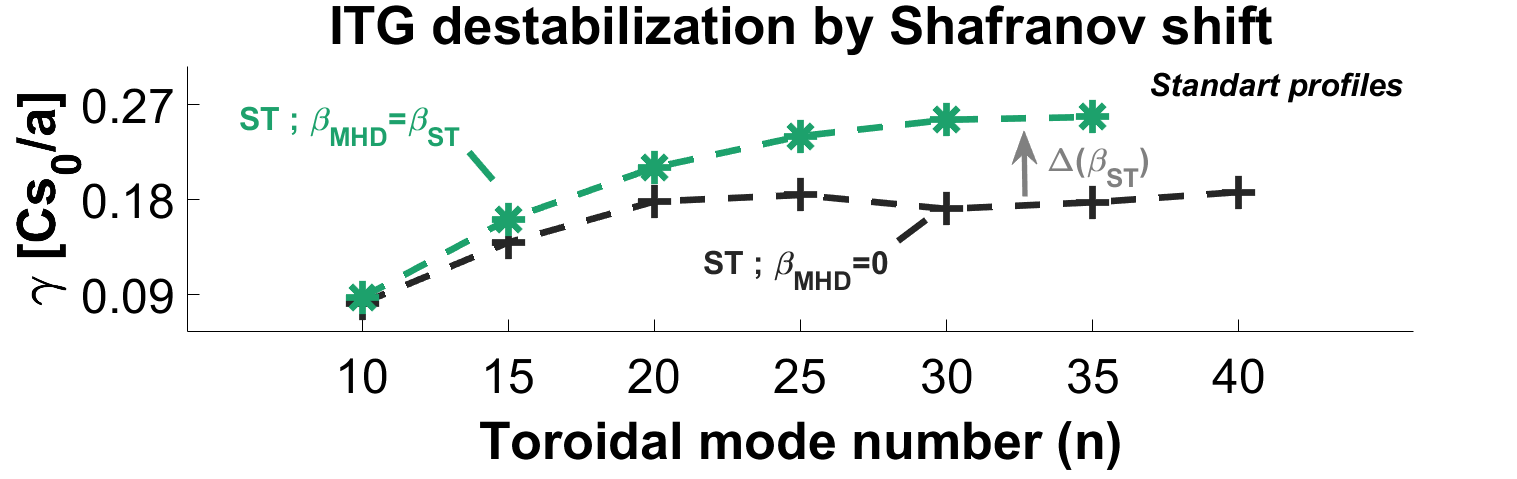}
\includegraphics[width=0.495\textwidth]{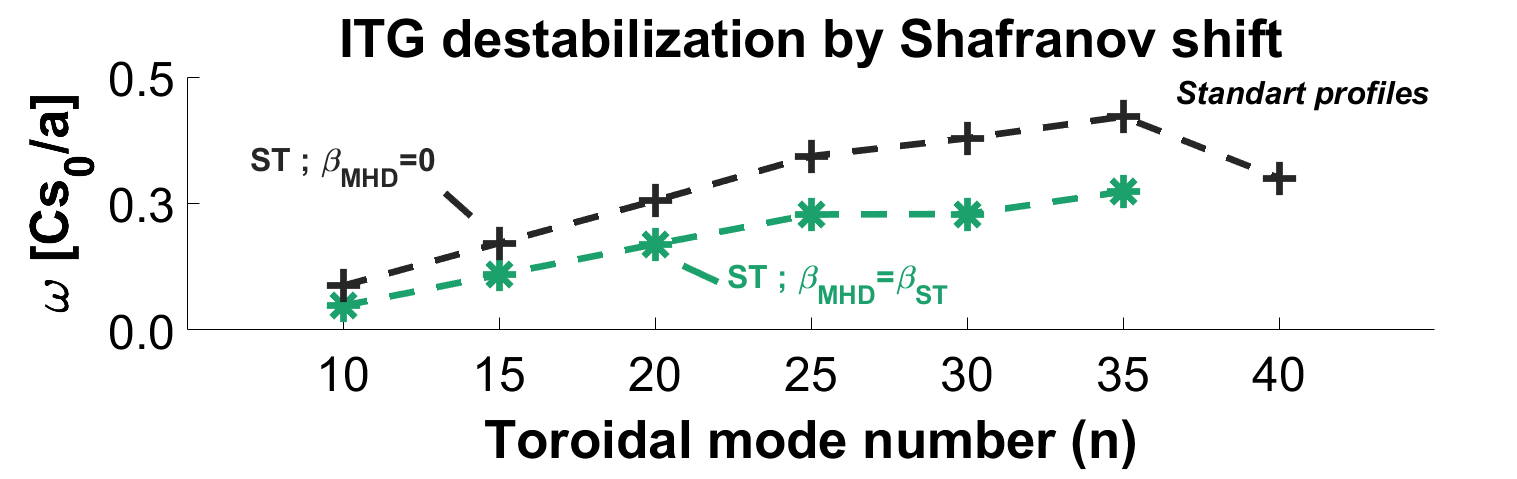}
\caption{\label{FIG:ITG_Shaf} \it
ITG dispersion relation driven by standard profiles in two magnetic geometries. \textit{Inconsistent} $\beta_{MHD} = 0$, and \textit{self-consistent} $\beta_{MHD} = \beta_{ST}$ ideal MHD equilibria.}
\end{center}
\end{figure}

\begin{figure} [h!]
\begin{center}
\includegraphics[width=0.235\textwidth]{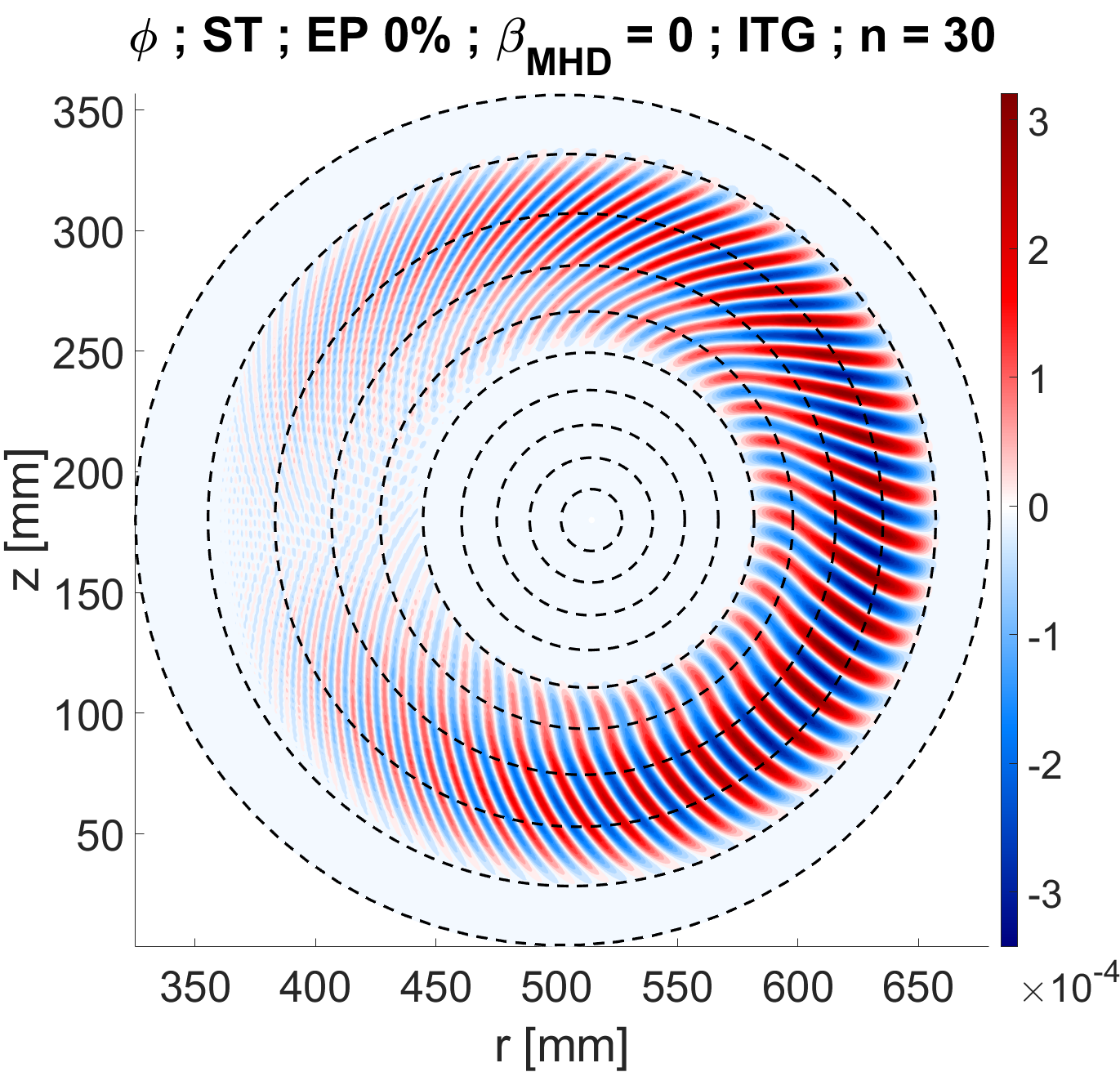}
\includegraphics[width=0.235\textwidth]{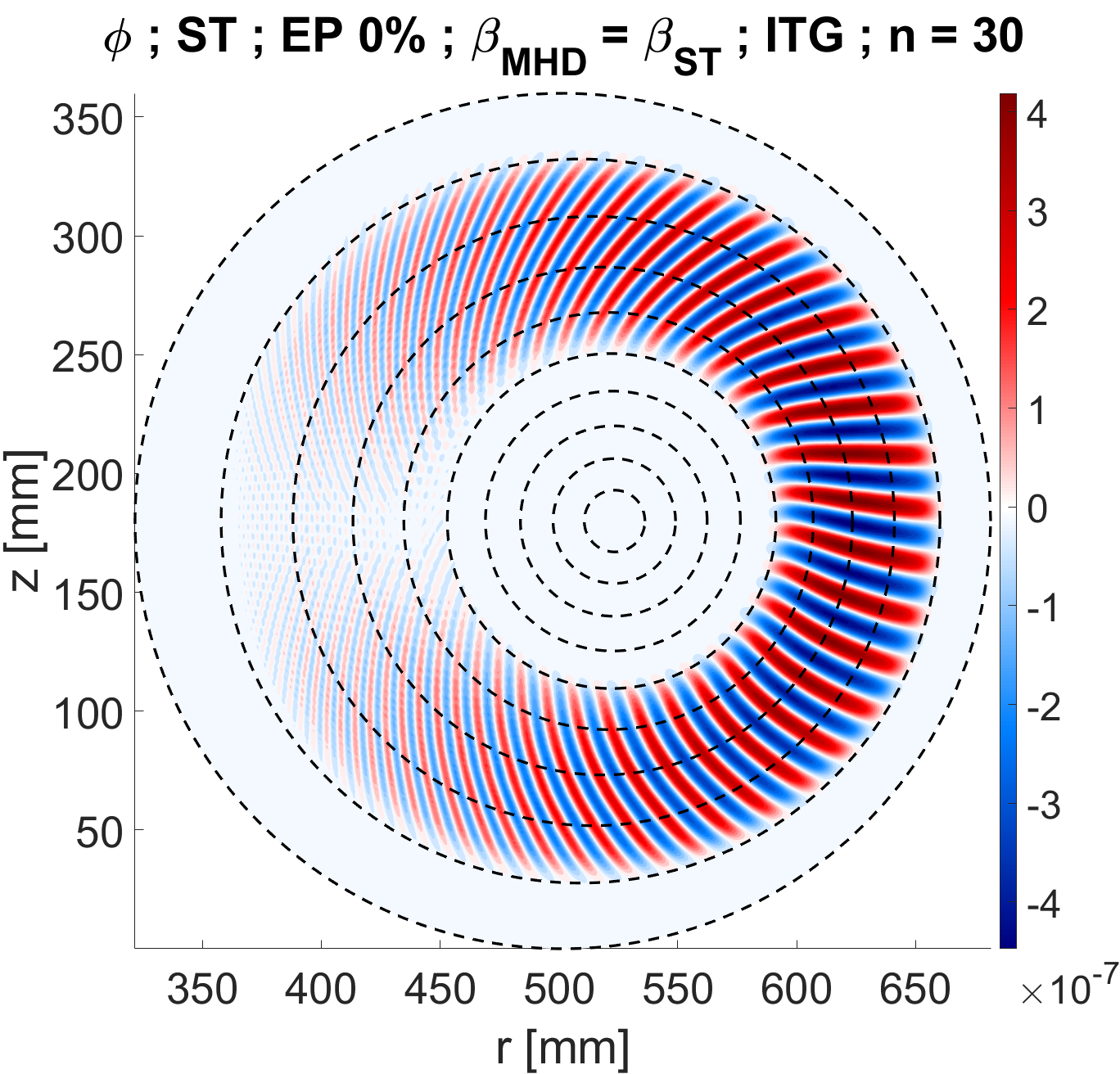}
\includegraphics[width=0.245\textwidth]{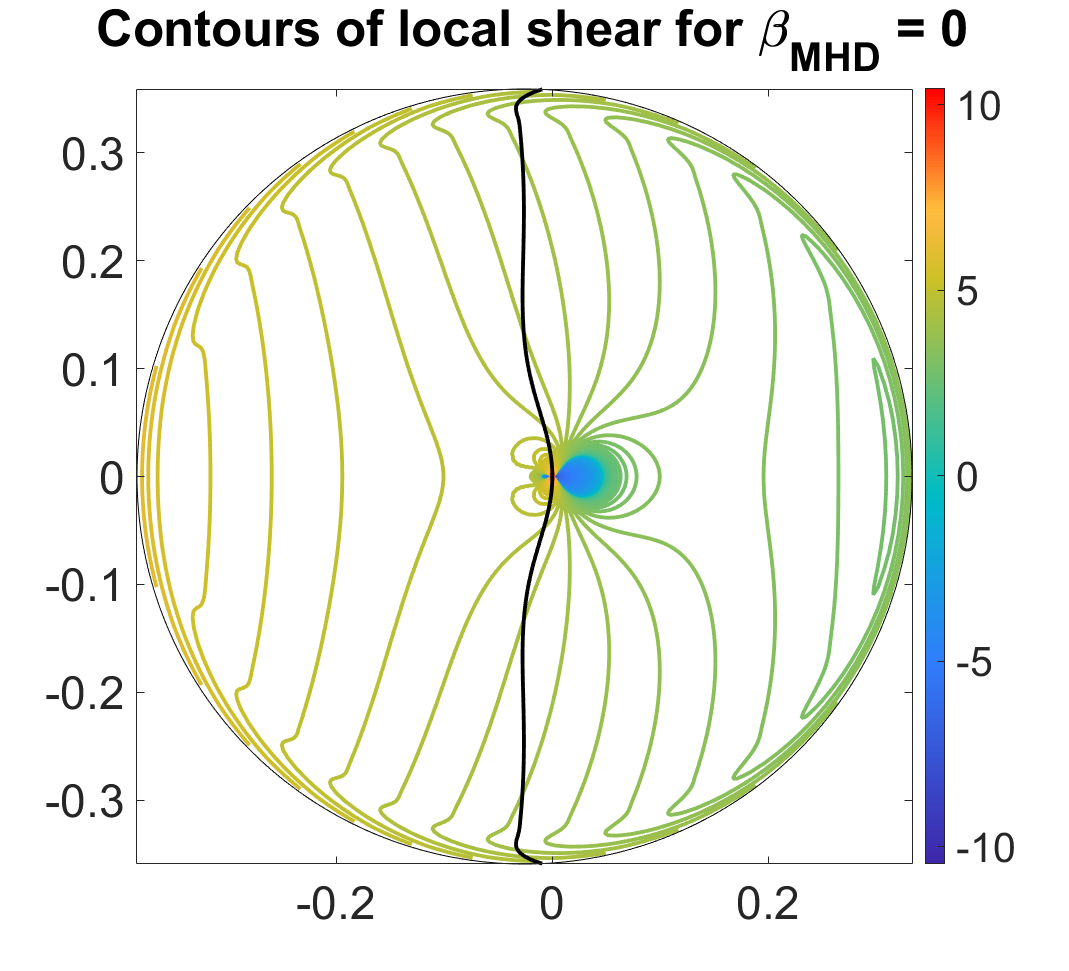}
\includegraphics[width=0.245\textwidth]{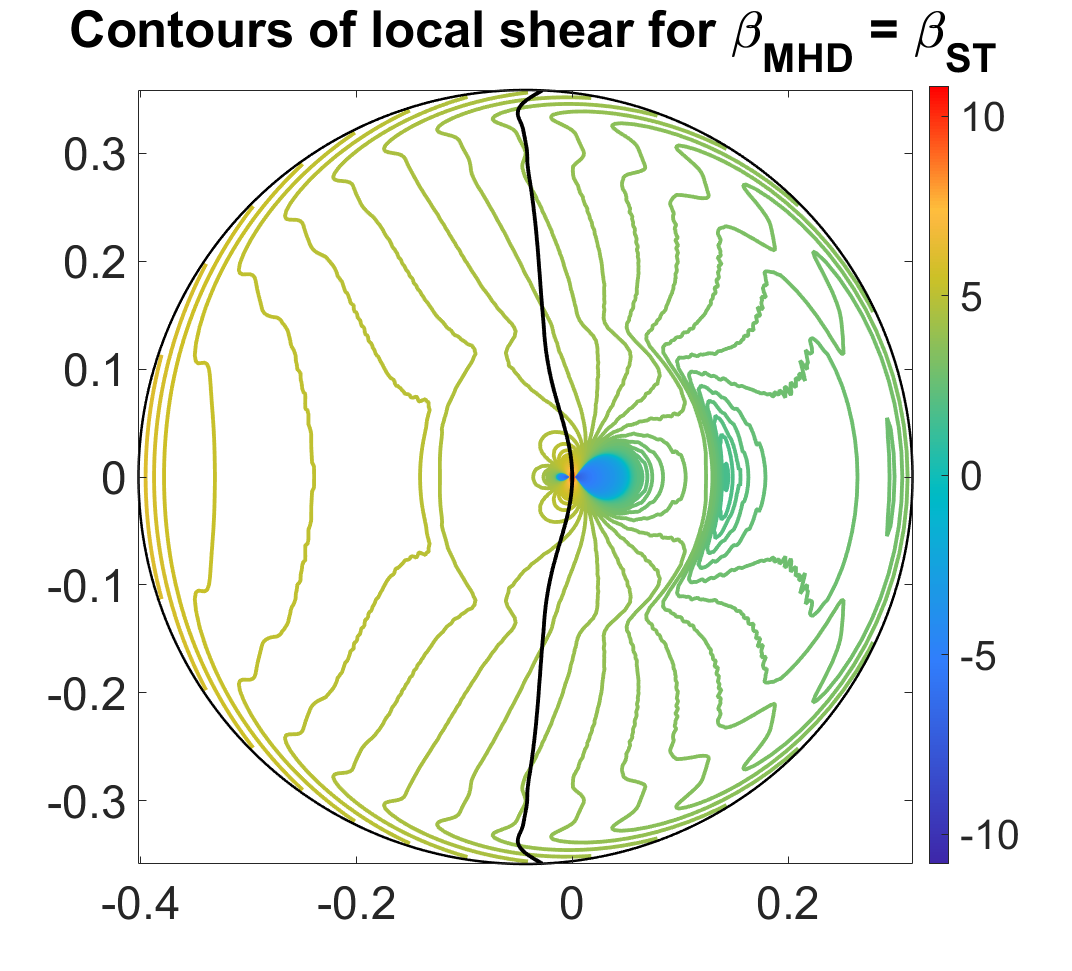}
\caption{\label{FIG:ITG_Shaf_mode_shaer} \it
Poloidal cross-sections of $\phi$ the perturbed electrostatic potential by the $n=30$ ITG mode in $\beta_{MHD}=0$ and $\beta_{MHD}=\beta_{ST}$ cases. On the right, contours of local shear and a black line to marks the border between "good" and "bad" curvature.}
\end{center}
\end{figure}

%======================================================================
In Figure \ref{FIG:ITG_Shaf} we plot the dispersion relation for an ITG driven by standard profiles in two MHD equilibria. Comparing the \textit{inconsistent} $\beta_{MHD} = 0$ with the \textit{self-consistent} $\beta_{MHD}=\beta_{ST}$ case, we show Shafranov shift destabilization which is stronger at the shorter wavelengths.  
Similar result were obtained by \cite{niiro2023plasma} in the local flux tube limit, showing that in consistent MHD equilibria, Shafranov shift effects can destabilize the ITG. 
Next, for brevity we focus on the $n=30$ mode for both cases. Figure \ref{FIG:ITG_Shaf_mode_shaer} shows a poloidal cross-section of the electrostatic potential perturbation due to the ITG mode. Both modes peak at a similar radial location. However, the ballooning angle of the mode in the \textit{self-consistent} $\beta_{MHD}=\beta_{ST}$ equilibrium is $3$-times smaller (less tilted structure at the outboard mid plane) than the ballooning angle of the mode in the $\beta_{MHD}=0$ equilibrium. This mode structure can explain the increase in growth rate with Shafranov shift. 

To better understand this, we plot in Figure \ref{FIG:ITG_Shaf_mode_shaer} the local magnetic shear \cite{lewandowski1995local,marinoni2021brief} and the $n=30$ mode structure for both cases. The former is a property of the magnetic geometry, while the latter is the system's response. The mode structure in the $\beta_{MHD}=\beta_{ST}$ case is less sheared (tilted) compared to the mode structure in the $\beta_{MHD}=0$ case. Looking at the local shear plots, we see that the increase Shafranov shift adds positive local shear which slightly destabilizes the mode. 

\FloatBarrier
%======================================================================
\subsubsection{Internal kink mode stabilization by Shafranov shift (peaked profiles)}
%======================================================================
In the following sections we show how the same peaked profiles can destabilize different modes depending on the Shafranov shift in the system. First, we use a circular concentric magnetic equilibrium (\textit{ad hoc} in ORB5 \cite{lanti2020orb5}) - which is not a consistent solution of the Grad-Shafranov equation and has zero Shafranov shift.

In this system, we find an internal kink mode \cite{bussac1975internal} driven by the peaked pressure gradients. In our case $\beta_{ORB5} = 0.002 > m_e/m_i = 0.001$, indicating that the ion Larmor radius is bigger than the electron skin depth $\rho_i>\delta_e$ which puts us in the in the collisionless regime \cite{antlitz2026comparison} where electron inertia helps destabilize the mode \cite{porcelli1991collisionless}. Furthermore, our poloidal $\beta_p \approx 0.68$ is bigger than the ideal-MHD stability limit of $\beta_{p,crit} = 0.3(1 - (5/3) \ (r_{peak}/a)) \approx 0.13$ indicating that the mode is MHD unstable \cite{antlitz2026comparison} (see equations 1 \& 48 in that work).  

\begin{figure} [h!]
\begin{center}
\includegraphics[width=0.495\textwidth]{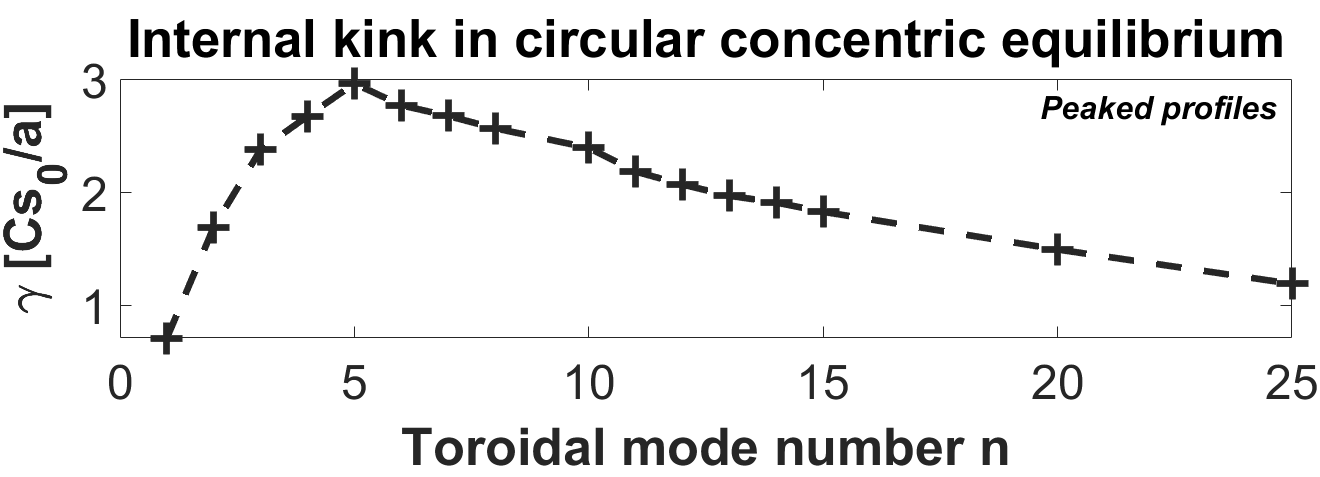}
\includegraphics[width=0.495\textwidth]{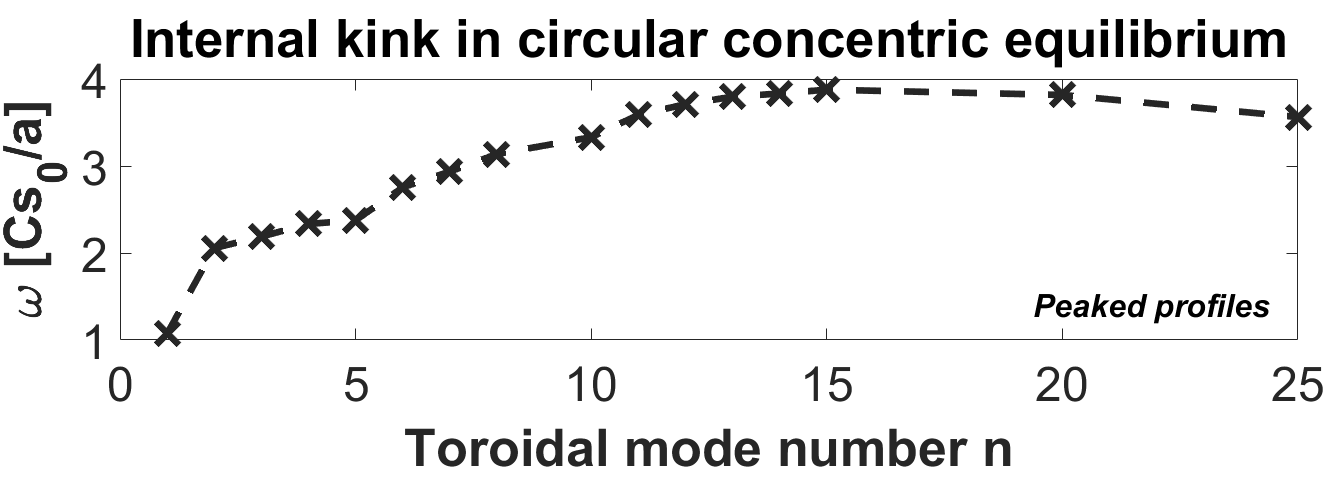}

\includegraphics[width=0.22\textwidth]{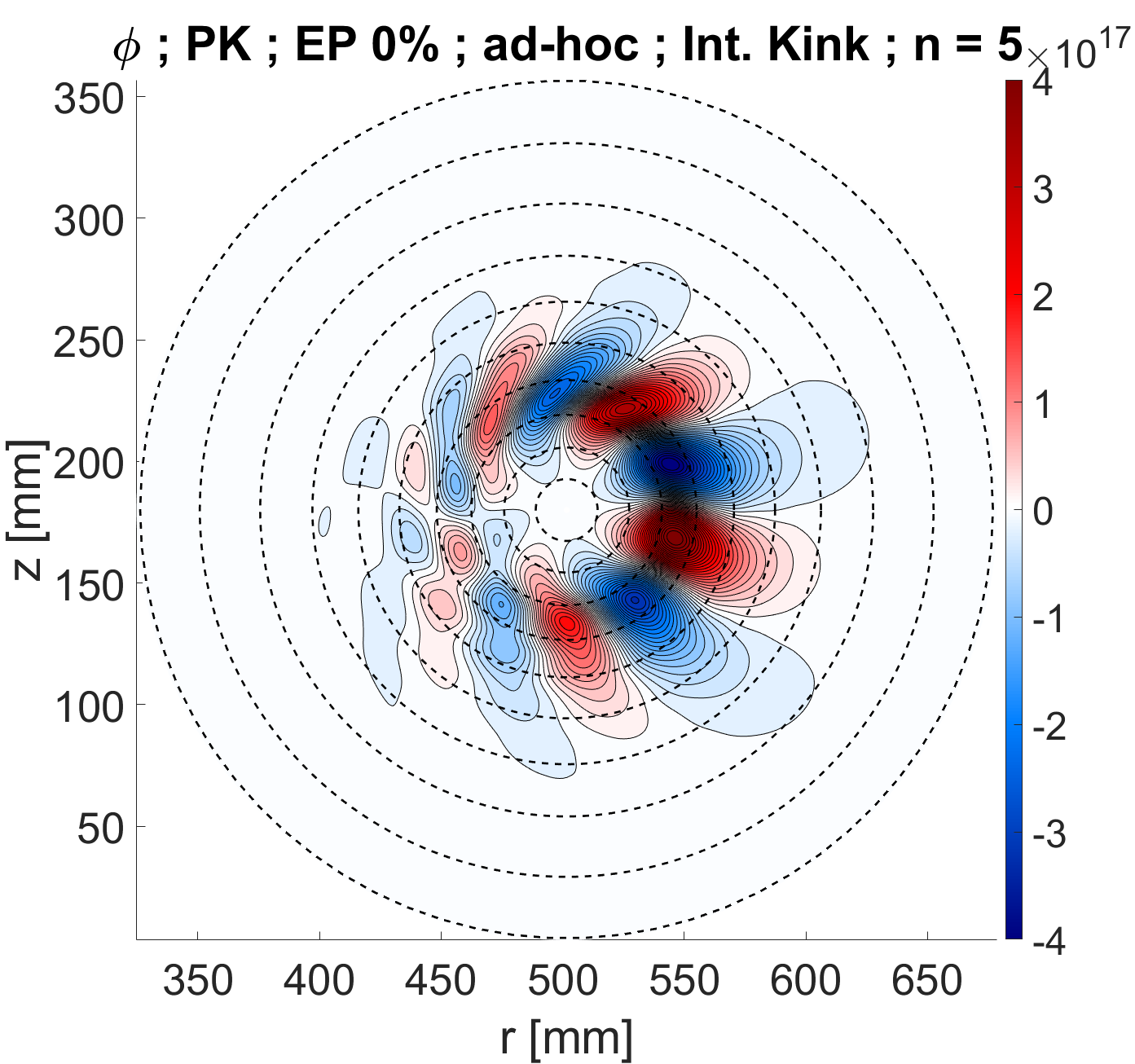}
\includegraphics[width=0.27\textwidth]{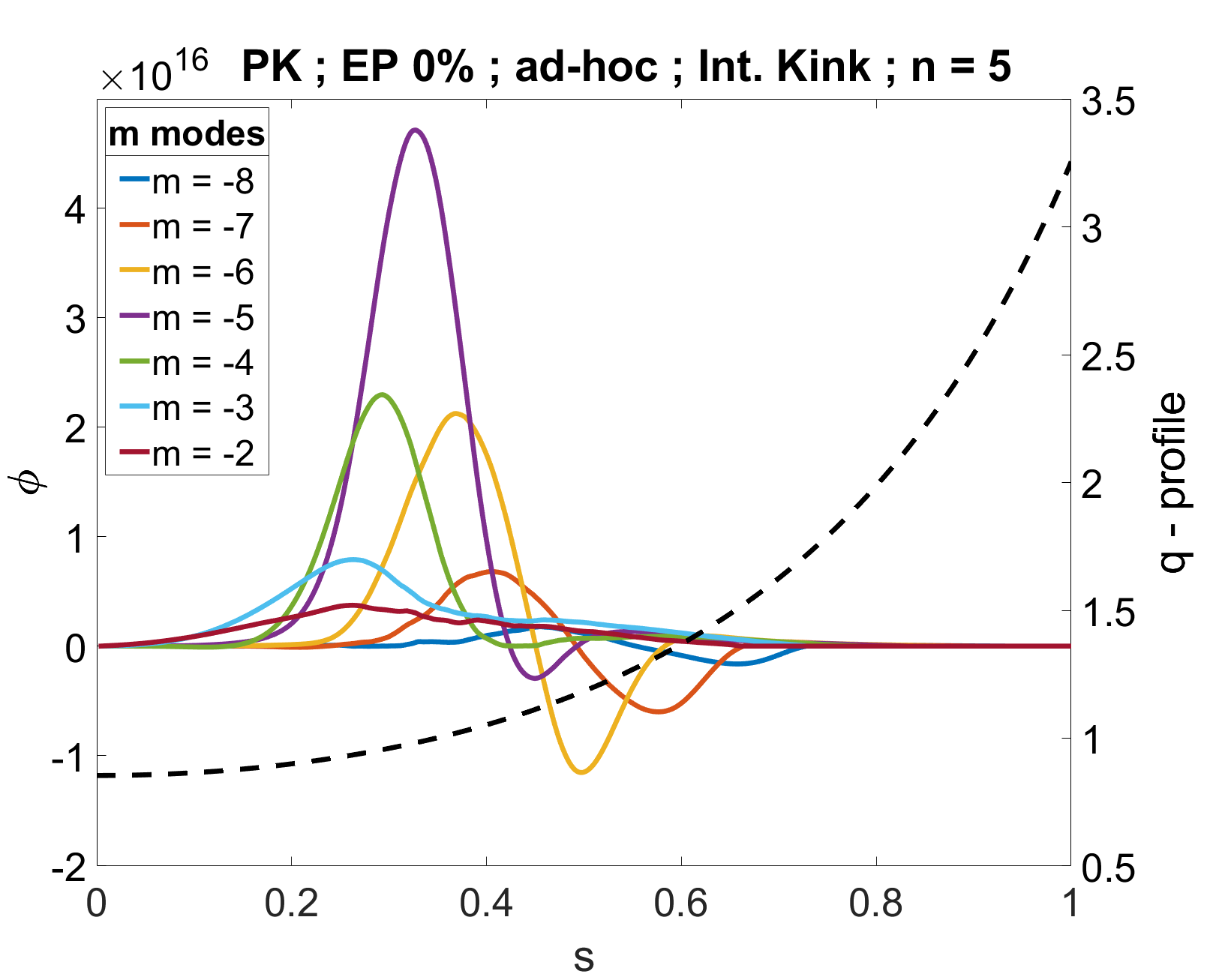}
\includegraphics[width=0.22\textwidth]{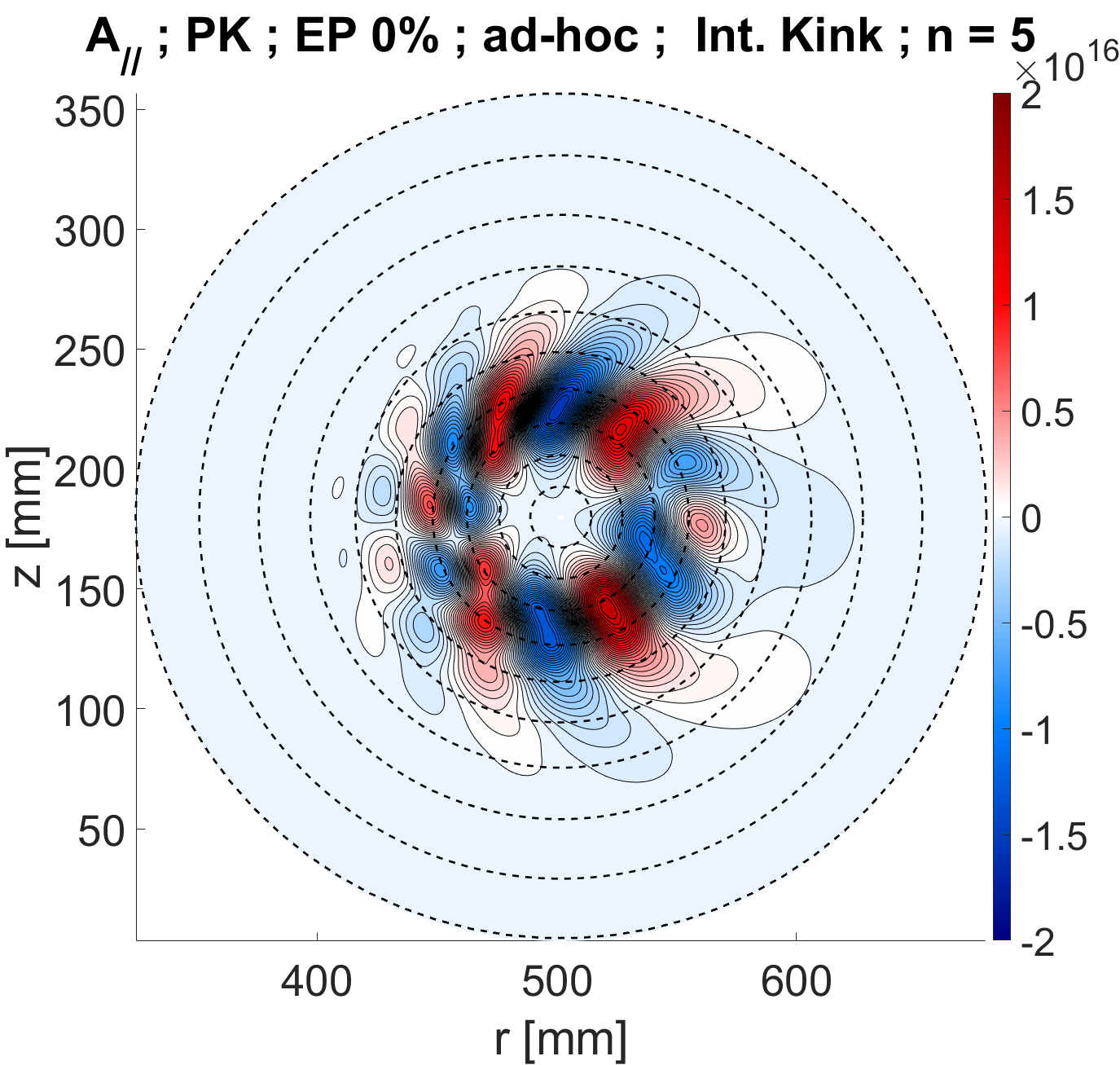}
\includegraphics[width=0.27\textwidth]{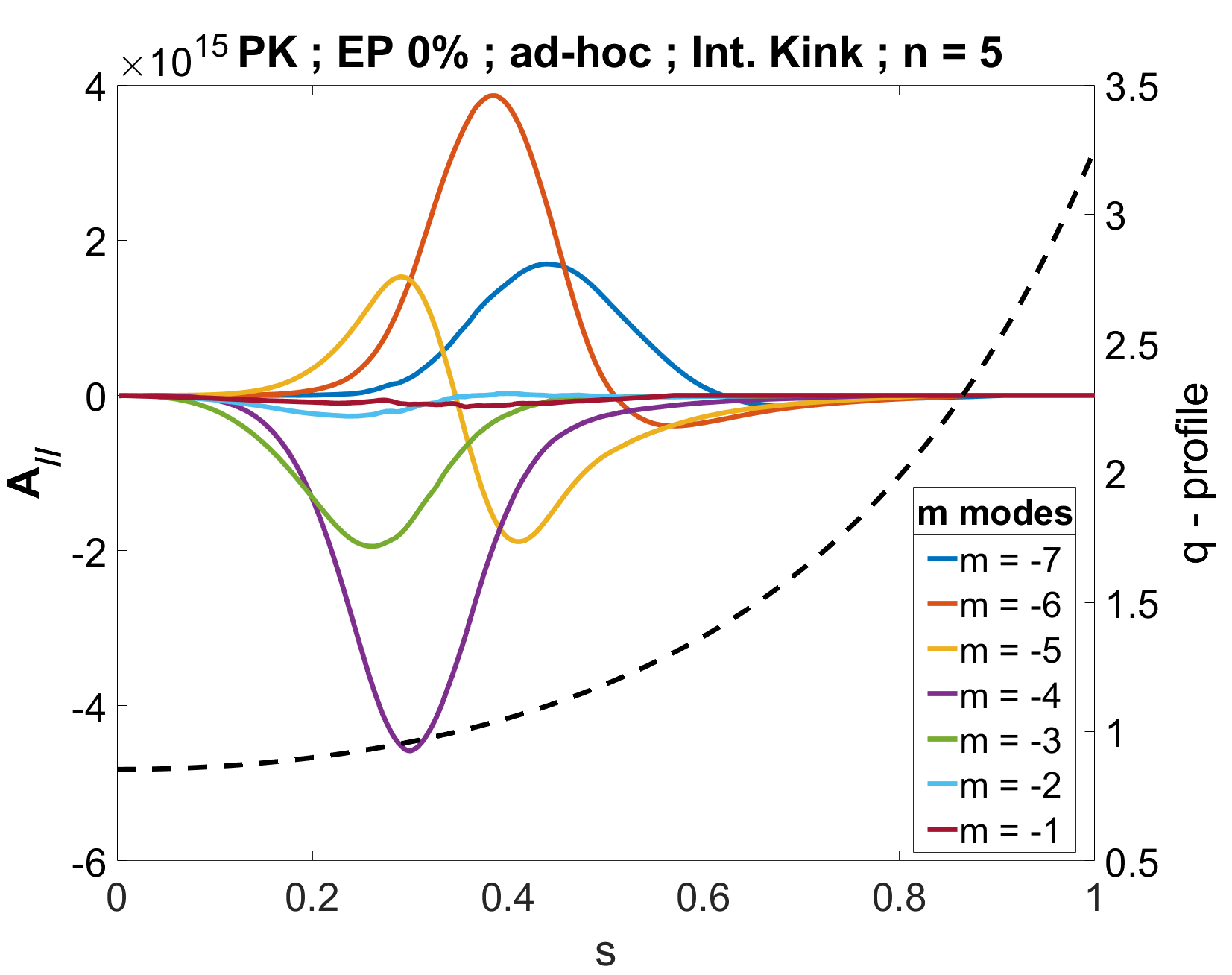}
\caption{\label{FIG:internal_kink} \it
Internal kink mode dispersion relation for peaked profiles in an \textit{ad hoc}, i.e. circular concentric equilibrium. The most unstable, $n=5$, mode structure in $\phi$ and $A_{\parallel}$.}
\end{center}
\end{figure}

In Figure \ref{FIG:internal_kink} we show the dispersion relation and the mode structure which peaks on the $q=1$ surface with a dominant poloidal harmonic equal to the toroidal mode number, $m_{max} = n$. In many cases the mode of interest is the $n=1,m=1$ mode. However all $n=m$ modes have $k_\parallel \rightarrow 0$ on the $q=1$ mode rational surface, and we find here that $n=5,m=5$ dominates the spectrum. We present the structure of this mode in $\phi$ and $A_{\parallel}$, noting that the toroidal harmonics have a similar structure.

Several key features like the wide shape of the main poloidal harmonic, and a sign reversal of the $A_{\parallel}$ side bands $[m-1,m+1]$ at the $q=1$ mode rational surface, where found by JOREK and ORB5 as well. In line with $\beta_p>\beta_{p,crit}$, they indicate a fluid like instability \cite{antlitz2026comparison} (see Figs. 9, 10 of this paper).
%The frequency of the mode is Alfv\'enic and appears to be diamagnetic (linearly scaling with $n$) in the range of toroidal modes $n$ where the kink peaks, which is expected from a pressure driven fluid-like mode. 

This highly unstable mode (with a maximal growth rate 3 times higher than the KBM), is very sensitive to Shafranov shift stabilization. In our case, the minimal geometric effect found in a $\beta_{MHD}=0$ magnetic equilibrium is sufficient to stabilize the mode. 

\FloatBarrier
%======================================================================
\subsubsection{KBM stabilization by Shafranov shift (peaked profiles)}
\begin{figure}
\begin{center}
\includegraphics[width=0.495\textwidth]{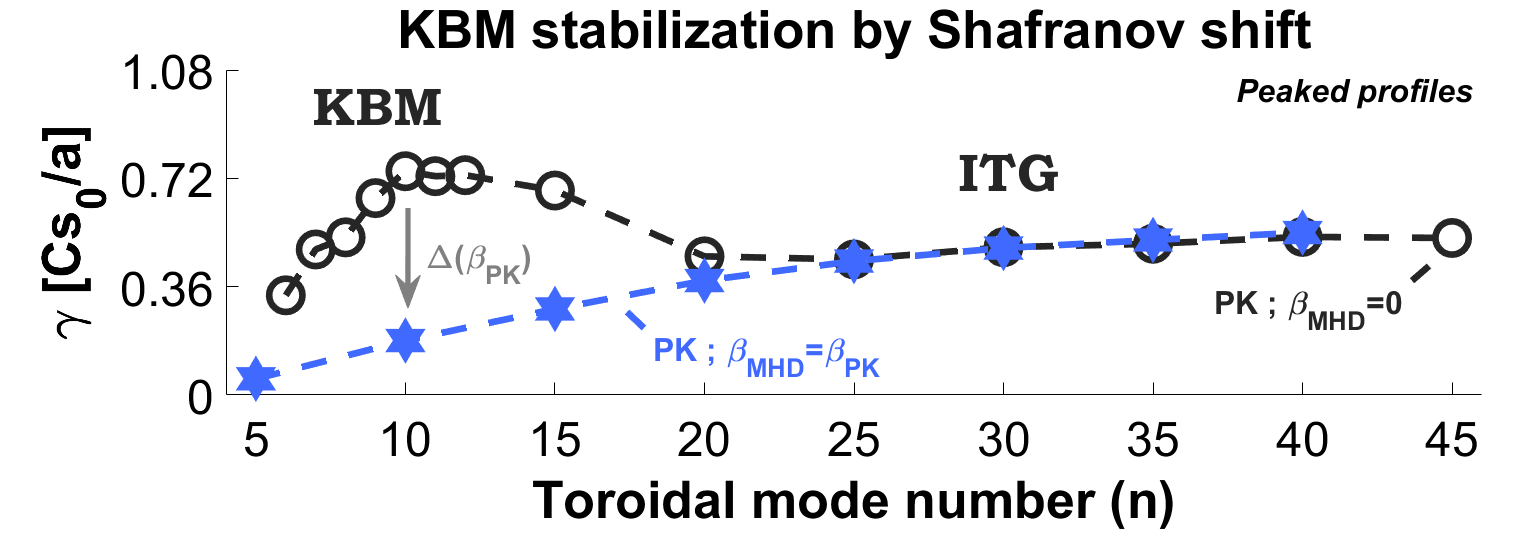}
\includegraphics[width=0.495\textwidth]{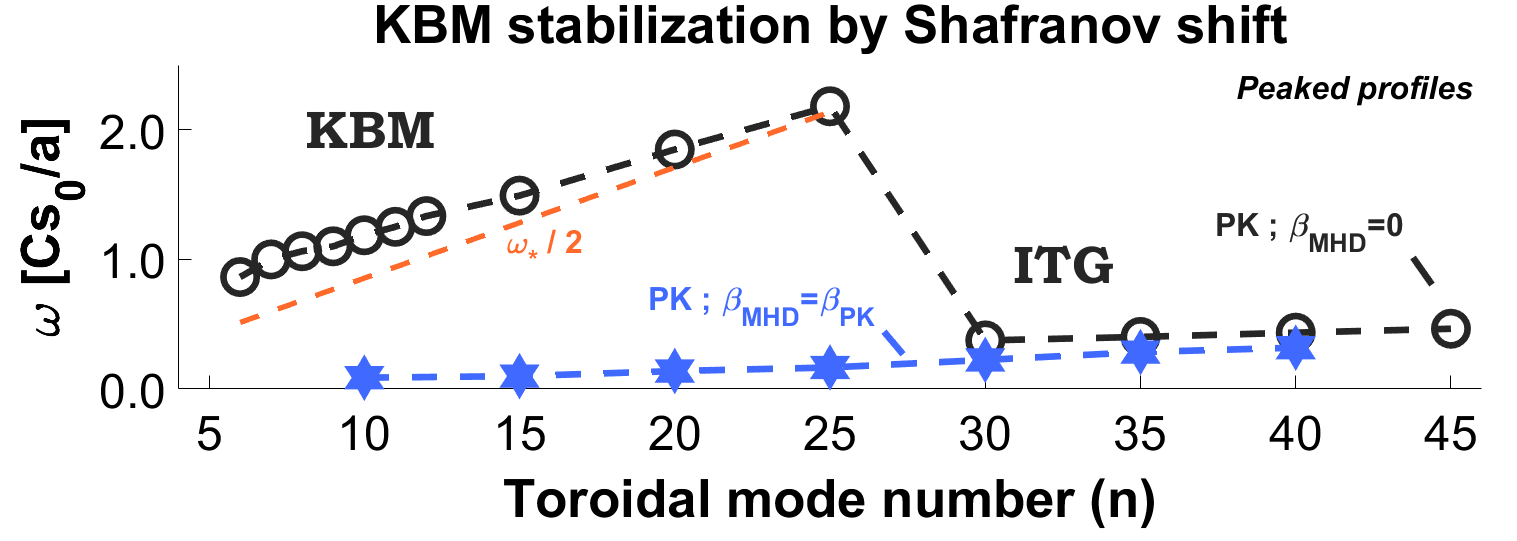}
\caption{\label{FIG:KBM_Shaf} \it
ITG and KBM dispersion relations for peaked profiles. KBMs are excited in the \textit{inconsistent} $\beta_{MHD}=0$ equilibrium, and are stabilized by Shafranov shift present in the \textit{self-consistent} $\beta_{MHD}=\beta_{PK}$ MHD equilibrium}
\end{center}
\end{figure}

%======================================================================
In a $\beta_{MHD} = 0$ MHD equilibrium, the small Shafranov shift stabilizes the internal kink mode, and the peaked profiles excite another mode instead. We identify the mode as KBM based on its frequency and ballooning mode structures presented in Figure \ref{FIG:KBM_Shaf}. KBMs are expected to rotate at about half the diamagnetic frequency $\omega_{*}$ which scale (for field aligned modes that peak at similar radial locations) with the toroidal mode number $n$. Where $\omega_{*} = \frac{nT_0R}{e}(R/L_T + R/L_n)$, as seen in the right panel.
%$\omega \sim \omega_{*,i}/2 \sim k_\perp \sim m_{a/2} \sim n q(a/2)$, as seen in the right panel.
In this case, the KBM is stabilized by the Shafranov shift in the \textit{self-consistent} equilibrium, which shows the underlying ITG instability instead. Qualitatively similar results were obtained by Aleynikova et. al. \cite{aleynikova2018kinetic}, (see Figs. 10, 11 of that paper), while Collar et. al. \cite{martin2020comparing} had to adjust the ideal-MHD equilibrium to reduce its stabilization and obtain unstable KBMs. 

\FloatBarrier
%======================================================================
\subsection{Part II - Energetic particles}
%======================================================================
Energetic particles are a small and hot minority in the plasma, which as a result have a substantial kinetic pressure that should be accounted for in the MHD equilibrium reconstruction. In this work we define a \textit{self-consistent} MHD equilibrium as one that includes in its force balance the pressure contributions from all the species i.e. $ \beta_{MHD} = \sum_{j}^{N} \beta_{j}$, where $N$ is total number of species. 
In our cases, we have EP fractions of $ 1\%-3 \%$ that are 120 hotter than the bulk ions, $ \tau_{EP}(s_0=0.5)=120$. As a result they have a comparable contribution to the Shafranov shift, i.e. $\beta_{EP} \approx \beta_{PK}$. Specific details about the species profiles are found in figure \ref{FIG:Combined_profiles_particles}.   

%======================================================================
\subsubsection{KBM stabilization by kinetic effects}
%======================================================================
\begin{figure}  [h!]
\begin{center}
\includegraphics[width=0.495\textwidth]{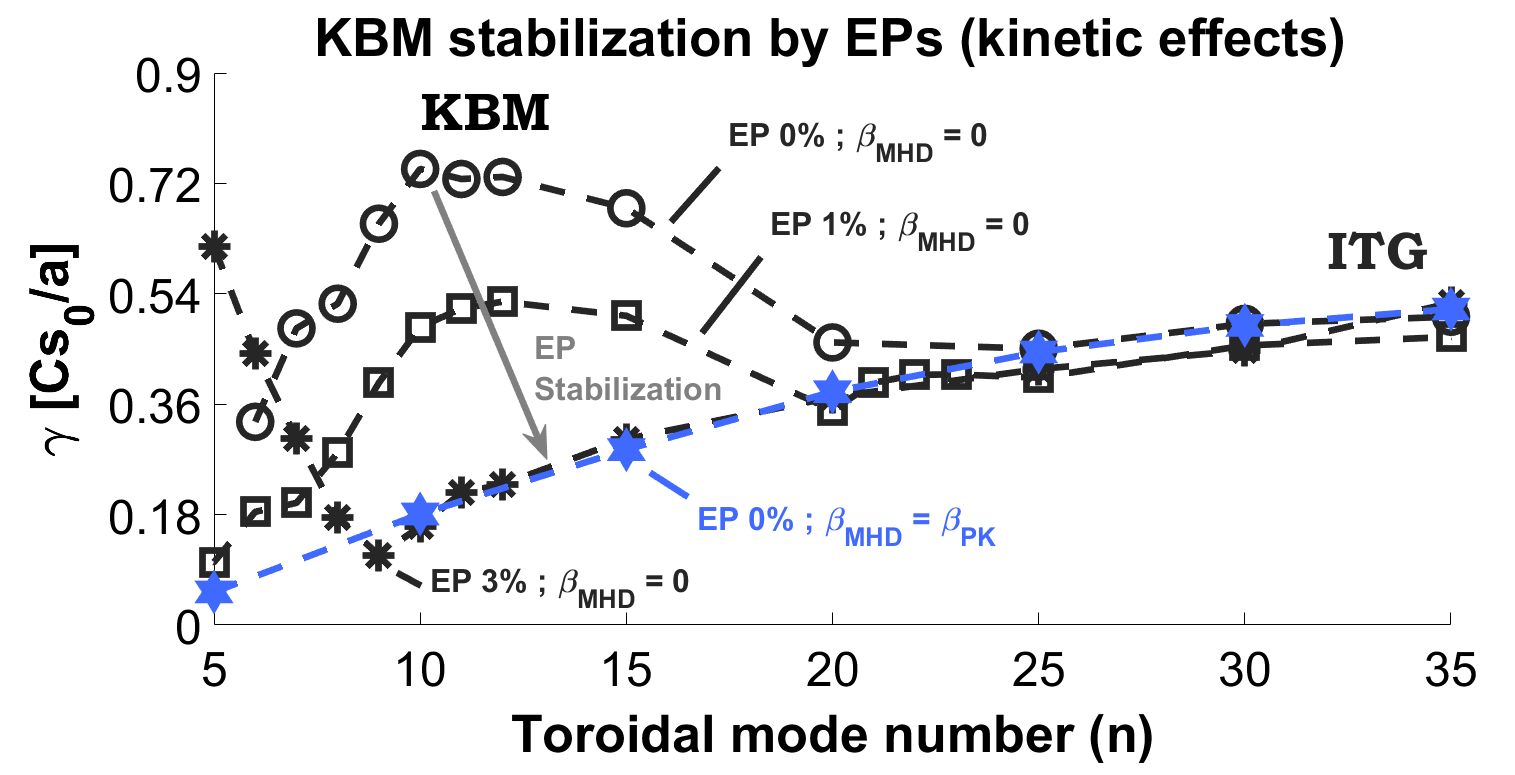}
\includegraphics[width=0.495\textwidth]{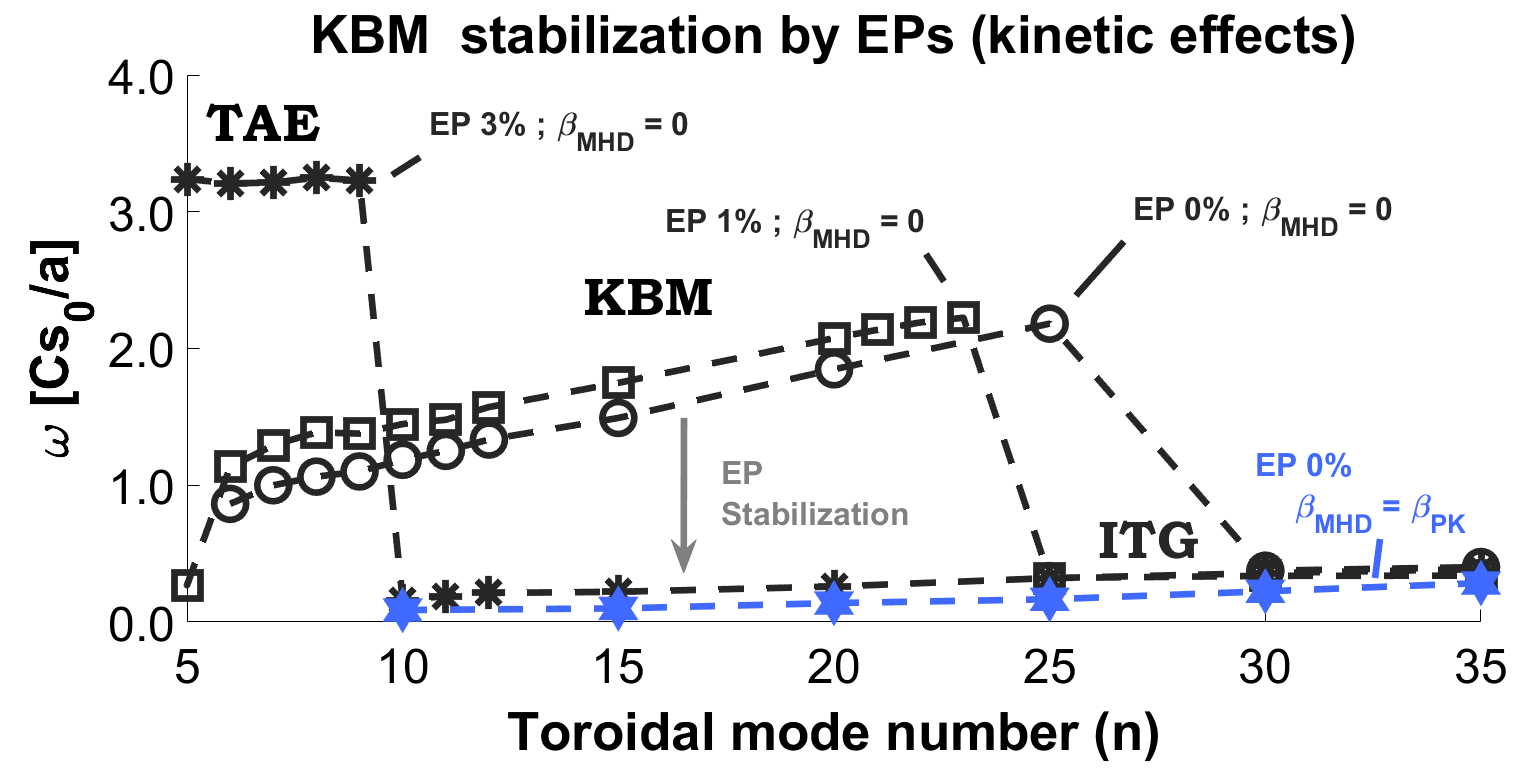}
\caption{\label{FIG:KBM_EP} \it
KBM is excited in a $\beta_{MHD}=0$ MHD equilibrium and is stabilized by the inclusion of $1\%-3\%$ of EPs as a third species.}
\end{center}
\end{figure}

We map the dispersion relation of the KBMs in $\beta_{MHD}=0$ MHD equilibrium in order to study the kinetic effect of EPs. We use an \textit{inconsistent} magnetic equilibrium with $\beta_{MHD}=0$ because the Shafranov shift in the \textit{self-consistent} case will stabilize the KBM by itself. In Figure \ref{FIG:KBM_EP} we show that the KBM is kinetically stabilized by the energetic particles, partially by $1\%$ EPs and completely with $3\%$ EPs in the system. This stabilization is in line with earlier works which used reduced modeling mainly based on the ballooning approximation \cite{salewski2025energetic}. However, this KBM is stable in \textit{self-consistent} MHD equilibria, thus we turn our attention to EP effects on the underling ITG microinstability.

\begin{figure}
\begin{center}
\includegraphics[width=0.22\textwidth]{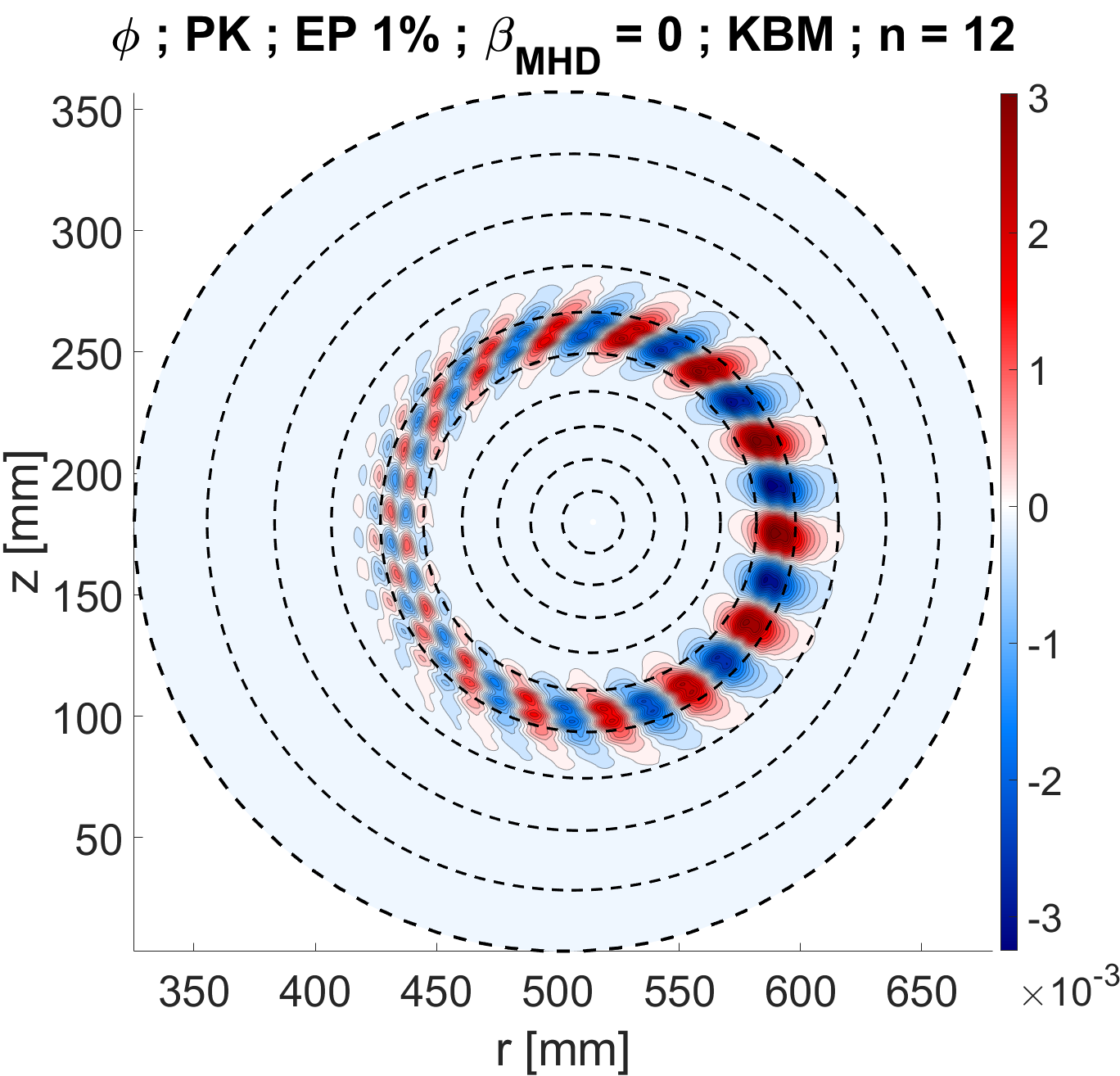}
\includegraphics[width=0.27\textwidth]{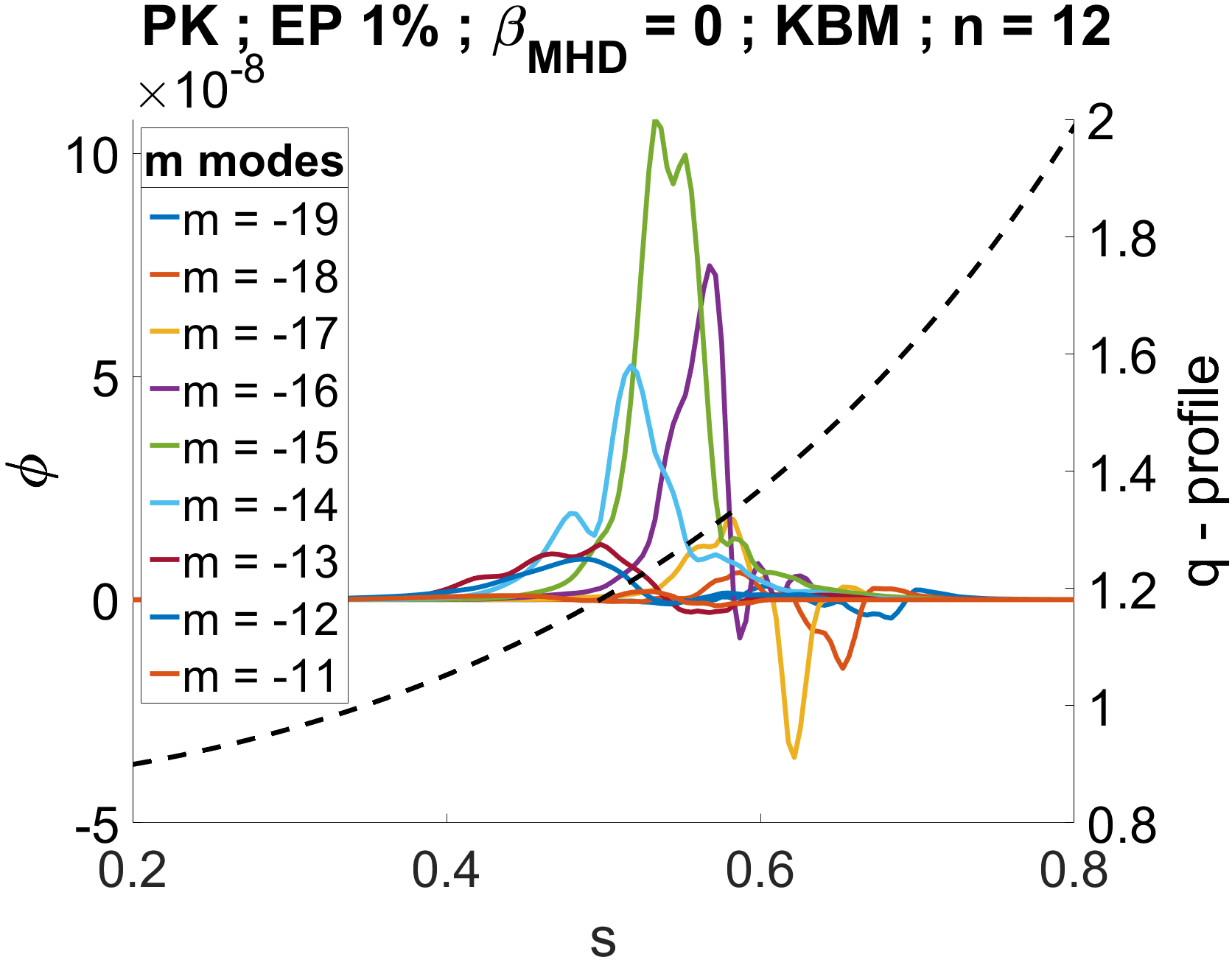}
\includegraphics[width=0.22\textwidth]{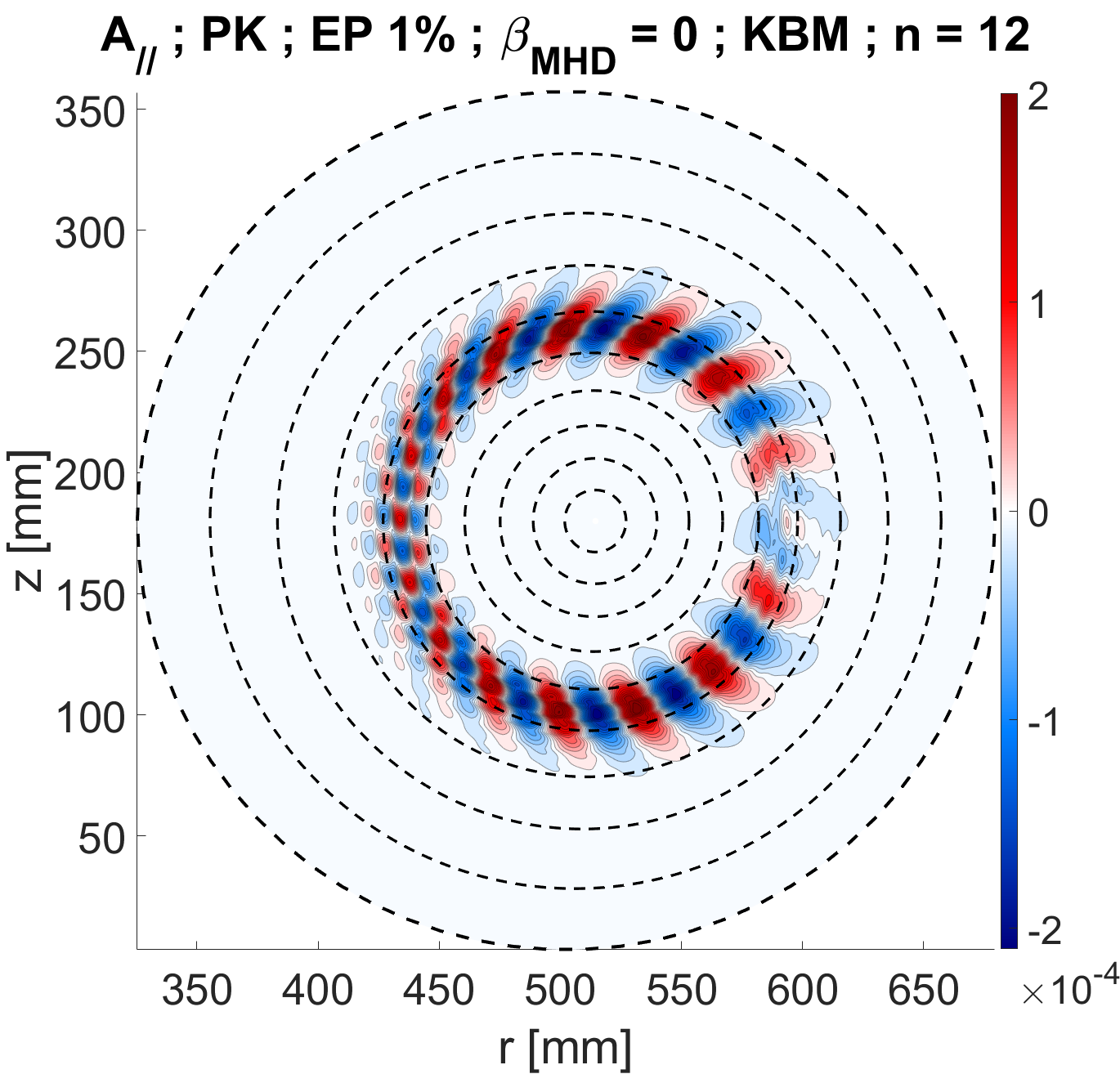}
\includegraphics[width=0.27\textwidth]{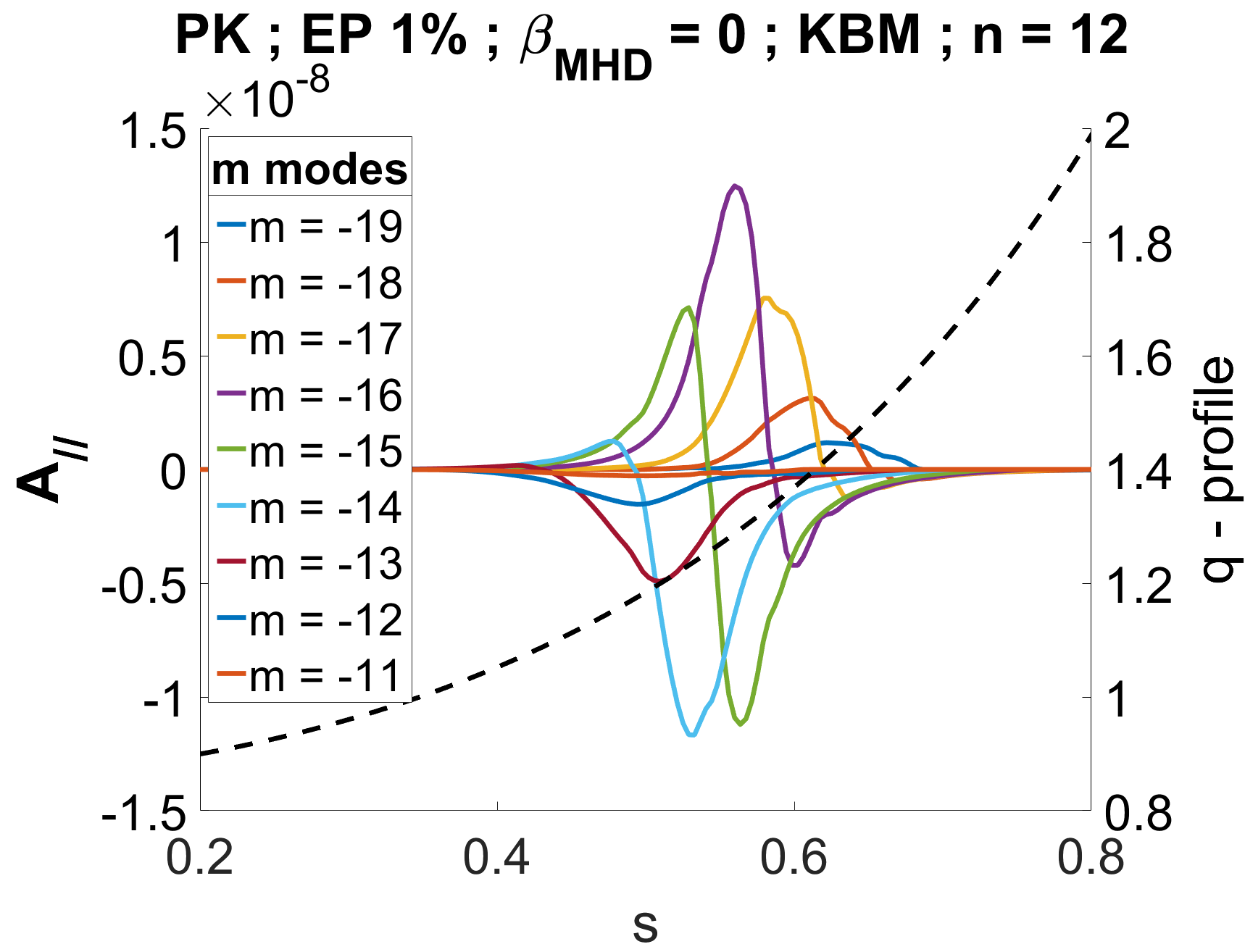}
\caption{\label{FIG:KBM_struct} \it
KBM, $n=12$, mode structure in $\phi$ and $A_{\parallel}$ for cases with $\beta_{MHD} = 0$ and $1\%$ EPs in the plasma. We might attribute the distorted poloidal mode structure $m$ in $\phi$ to kinetic effects.}
\end{center}
\end{figure} 

\FloatBarrier
%======================================================================
\subsubsection{ITG stabilization by energetic particles}
%======================================================================
\subsubsection{Direct stabilization by kinetic effects}
\begin{figure} [h!]
\begin{center}
\includegraphics[width=0.495\textwidth]{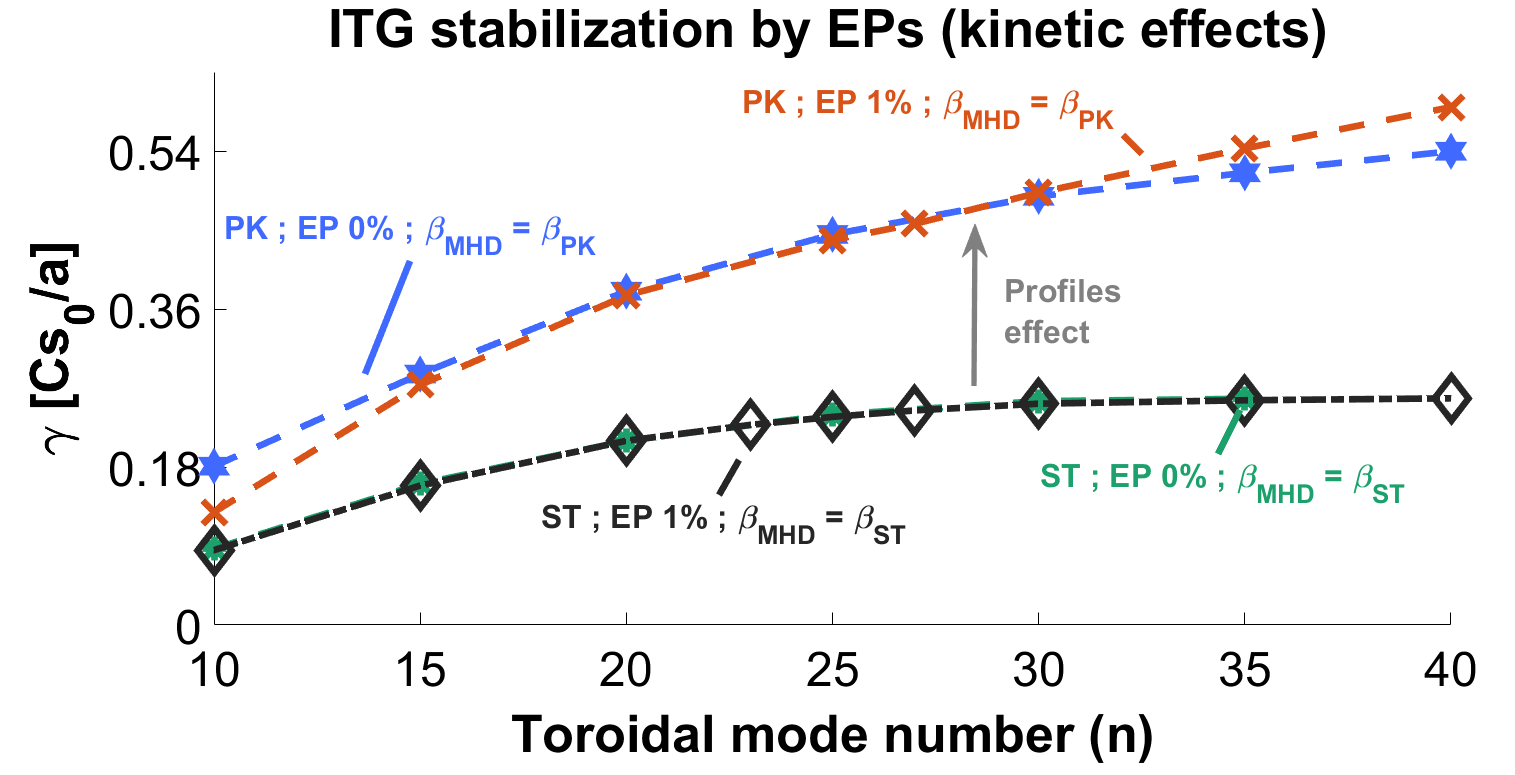}
\includegraphics[width=0.495\textwidth]{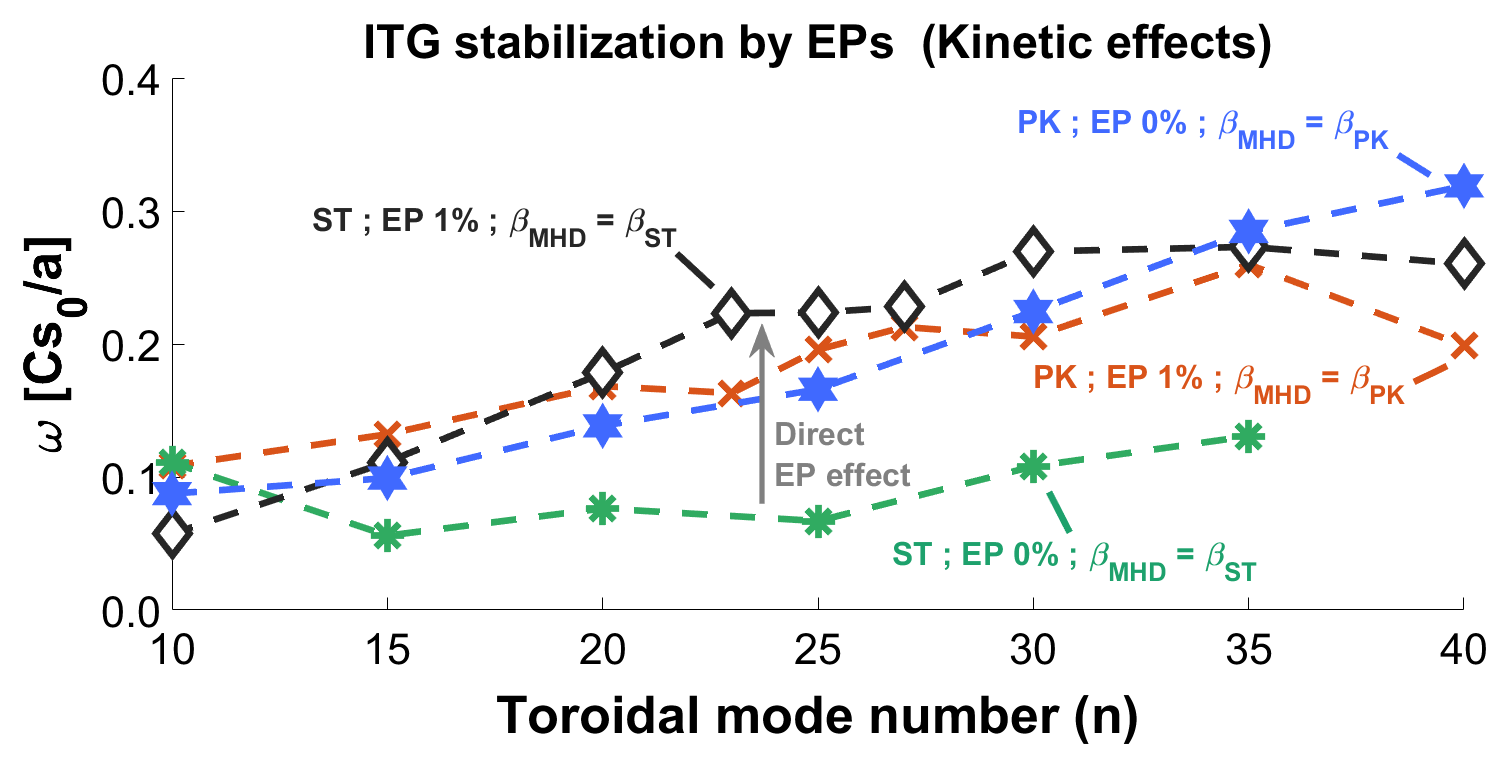}
\caption{\label{FIG:ITG_EP_Not_shaf} \it
ITG dispersion relations for cases with the same MHD equilibrium, with and without EPs. There is little to no direct effect of the EPs on the ITG growth rate, and an up-shift on the ITG frequency with standard profiles.}
\end{center}
\end{figure}

%======================================================================
First we study the kinetic effect of EPs on the ITG modes without accounting for their pressure in the magnetic equilibrium, i.e. $\beta_{MHD}$. Unlike the KBM (Fig. \ref{FIG:KBM_EP}), the ITG shows little to no direct effect of EPs on the growth rate (Fig. \ref{FIG:ITG_EP_Not_shaf}). In the case with the weaker ITG drive, i.e. standard profiles, the additional EP population acts to increase the mode frequency somewhat similar to the electromagnetic effects shown in \ref{FIG:ESvsEM}. These results are inline with other works, both local and global, which looked at hot EPs $\sim 100$ times the thermal ions \cite{di2018fast, ivanov2026dynamics}.

%======================================================================
\subsubsection{Indirect stabilization by Shafranov shift}
\begin{figure} [h!]
\begin{center}
\includegraphics[width=0.495\textwidth]{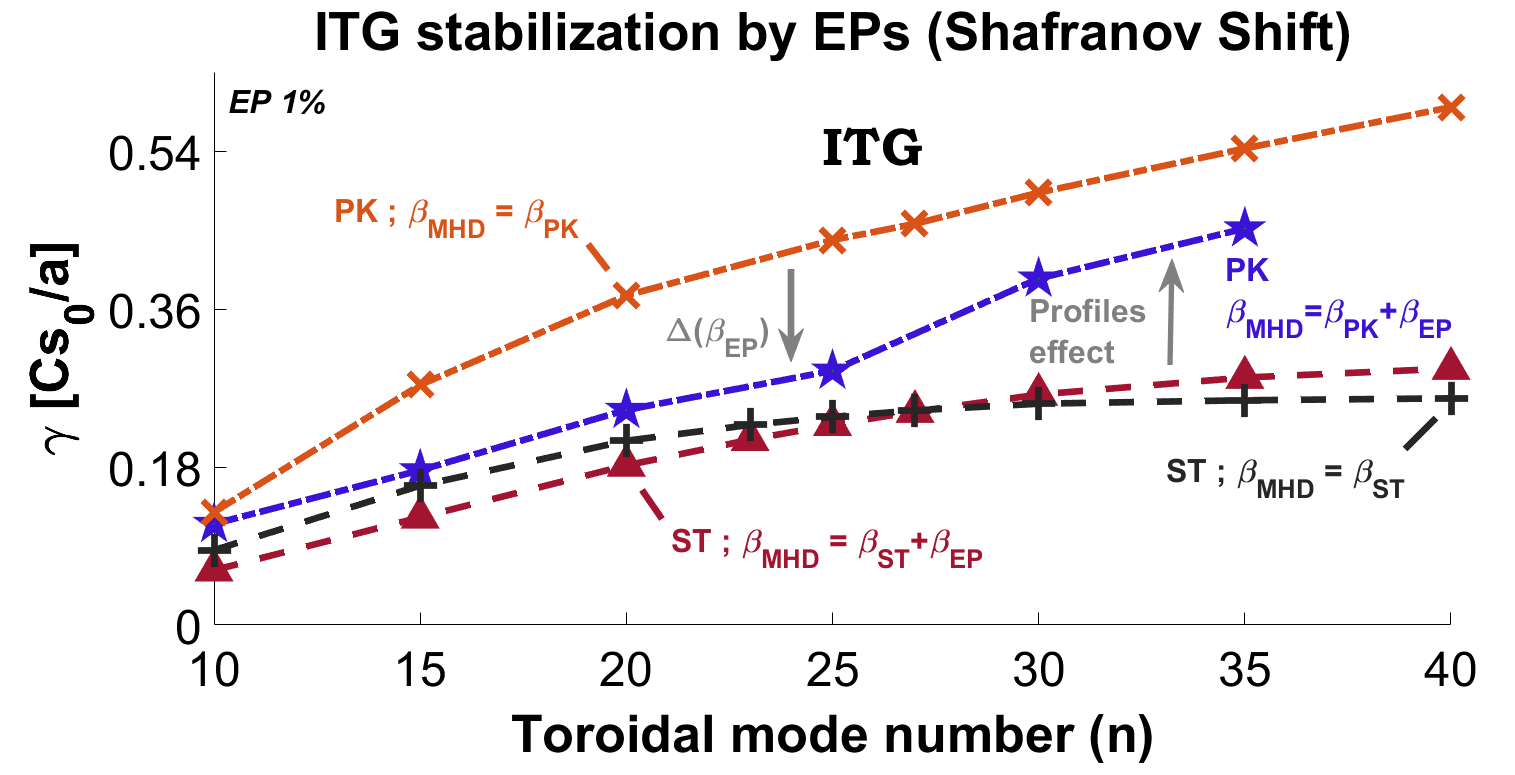}
\includegraphics[width=0.495\textwidth]{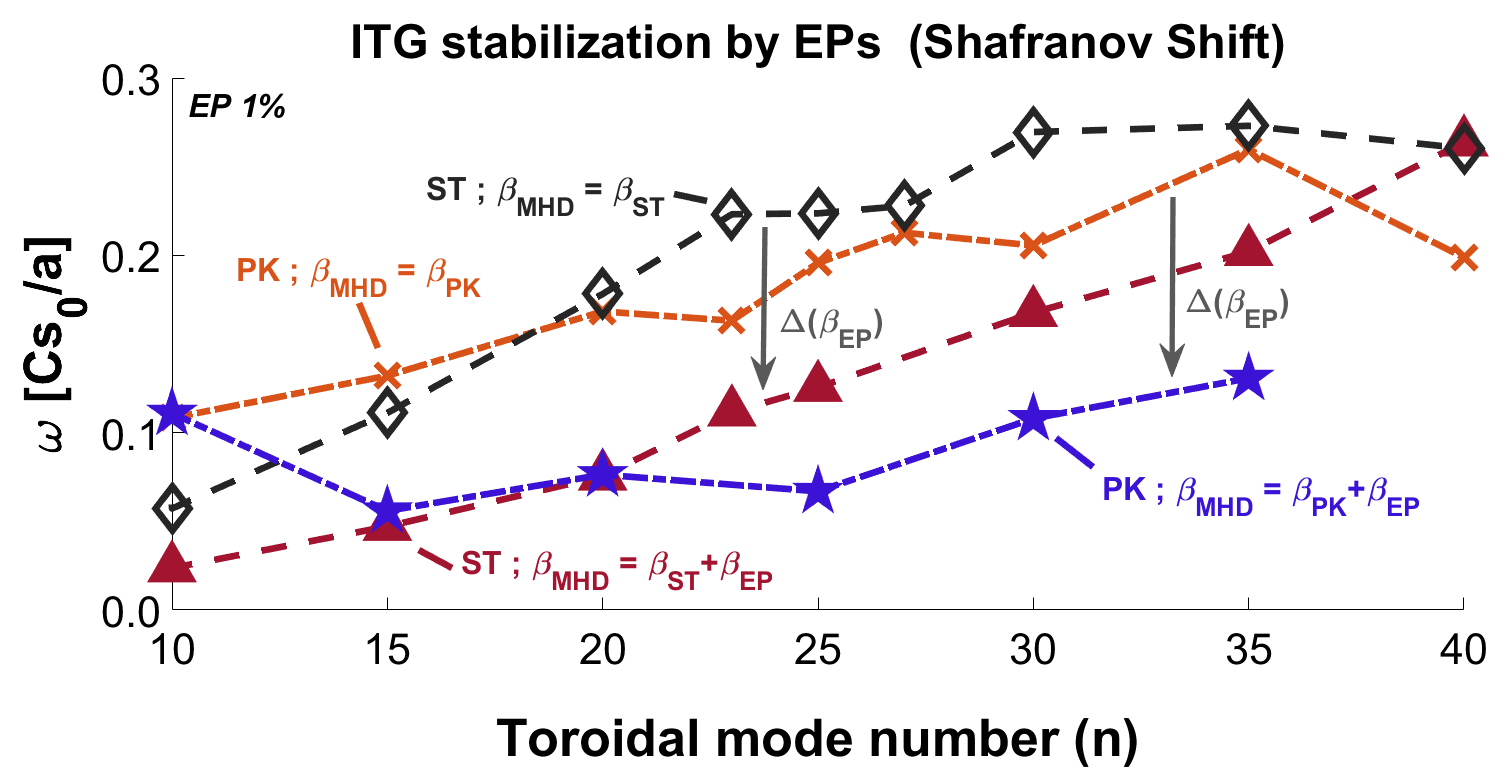}
\caption{\label{FIG:ITG_EP_Shaf} \it
ITG dispersion relations for cases with and without EP pressure $(\beta_{EP})$ as part of the MHD equilibrium. The Shafranov shift effect on the growth rate is more pronounced in the longer wavelengths (lower n). The frequency spectrum shows a consistent downshift due to the additional Shafranov shift. Here $\Delta(\beta_x)$ is the Shafranov shift effect due to $\beta_x$ component in $\beta_{MHD}$}
\end{center}
\end{figure}

%======================================================================
With $1\%$ EPs, that have $\beta_{EP}\approx\beta_{PK}$, the main effect of EPs on ITG modes is Shafranov shift stabilization through the equilibrium response as shown in Figure \ref{FIG:ITG_EP_Shaf}. The confined energetic particles add $\beta_{EP}$ to the plasma pressure $\beta_{MHD}$, independently of the profiles. Thus, increasing the Shafranov shift of the system without adding to the ITG drive - unlike the effect of increasing the profiles, which contribute to both. As a result, we see in Figure \ref{FIG:ITG_EP_Shaf} a strong decrease in ITG growth rate, especially at the longer wavelengths.

\begin{figure} [h!]
\begin{center}
\includegraphics[width=0.22\textwidth]{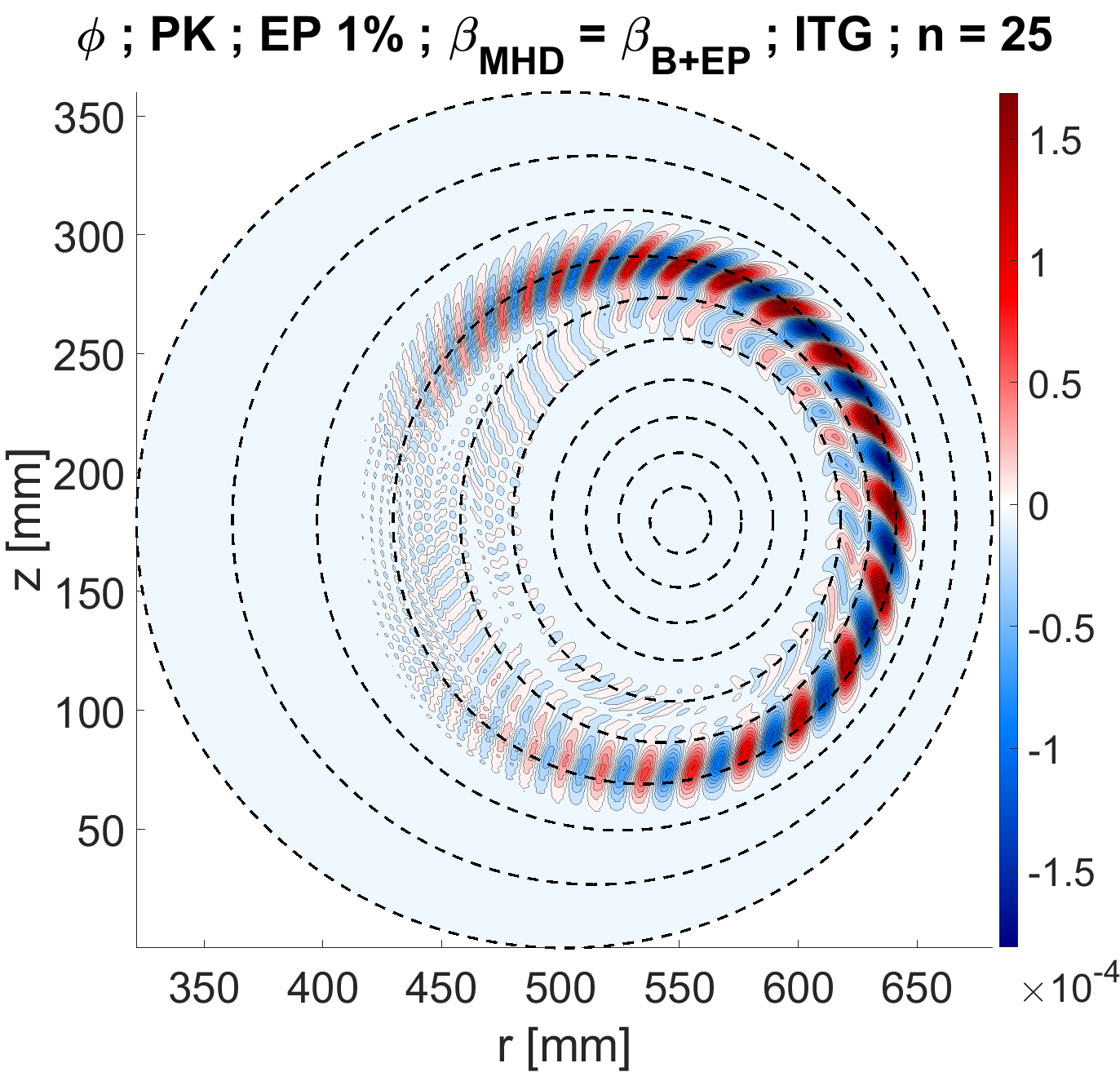}
\includegraphics[width=0.27\textwidth]{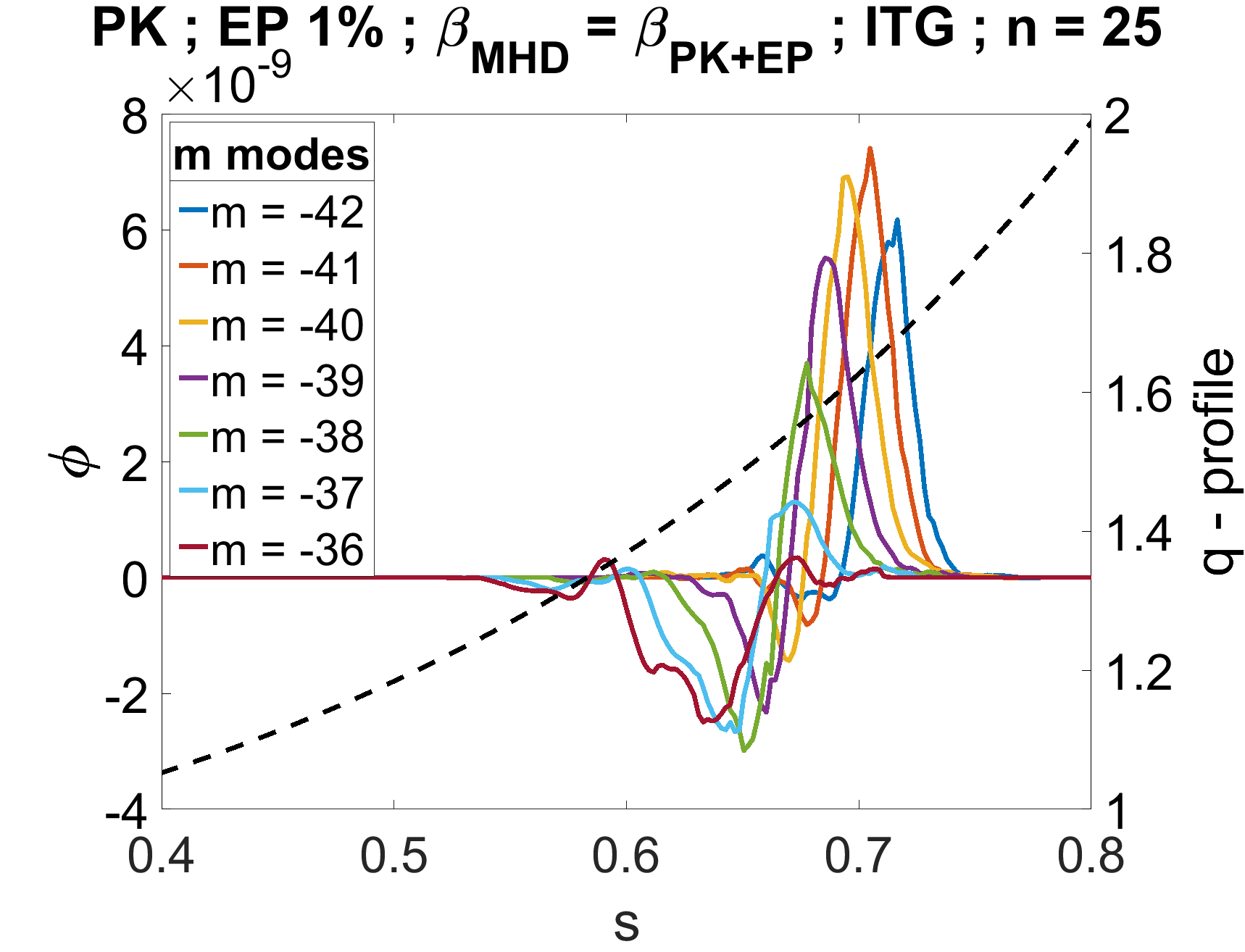}
\includegraphics[width=0.22\textwidth]{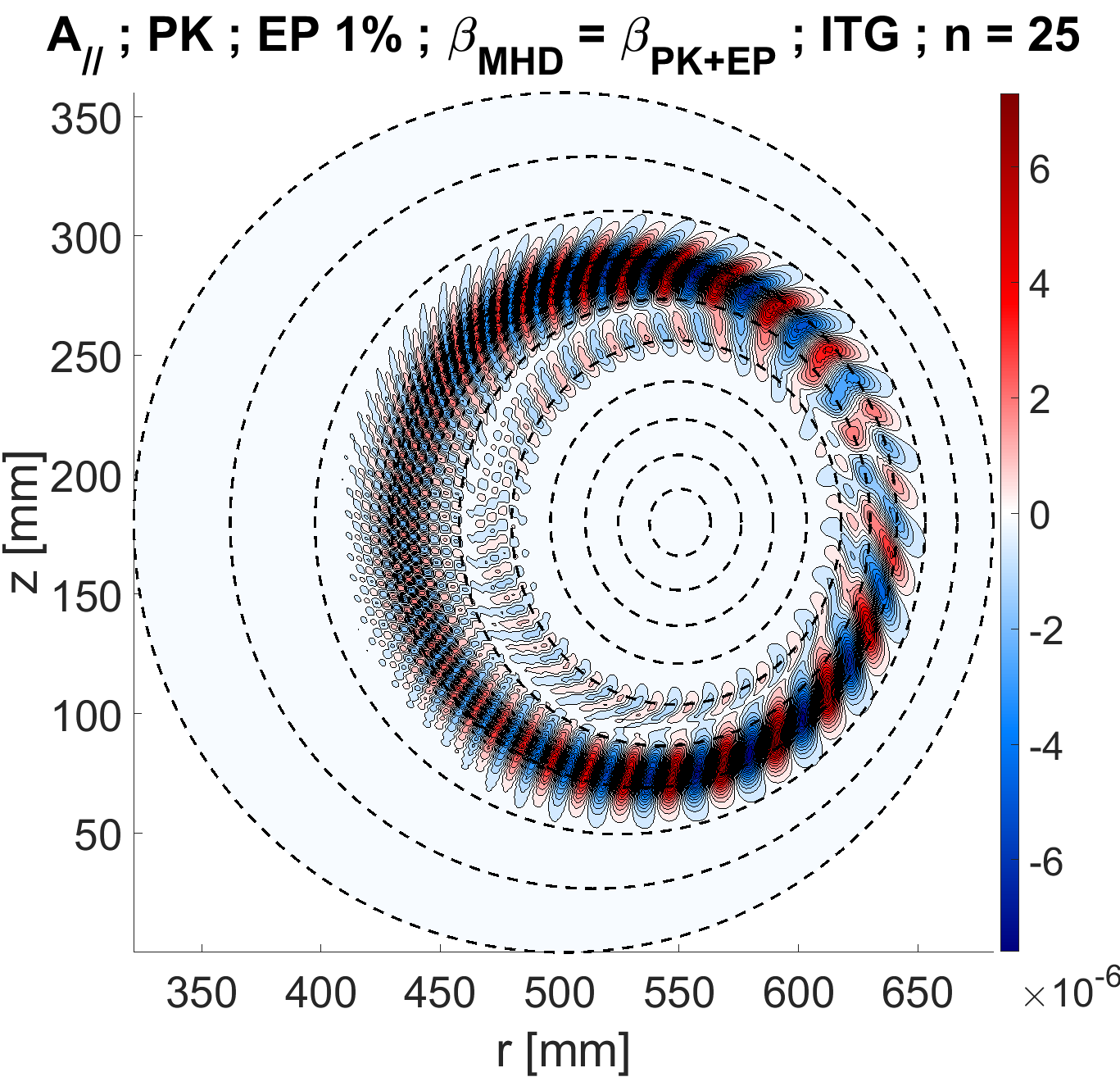}
\includegraphics[width=0.27\textwidth]{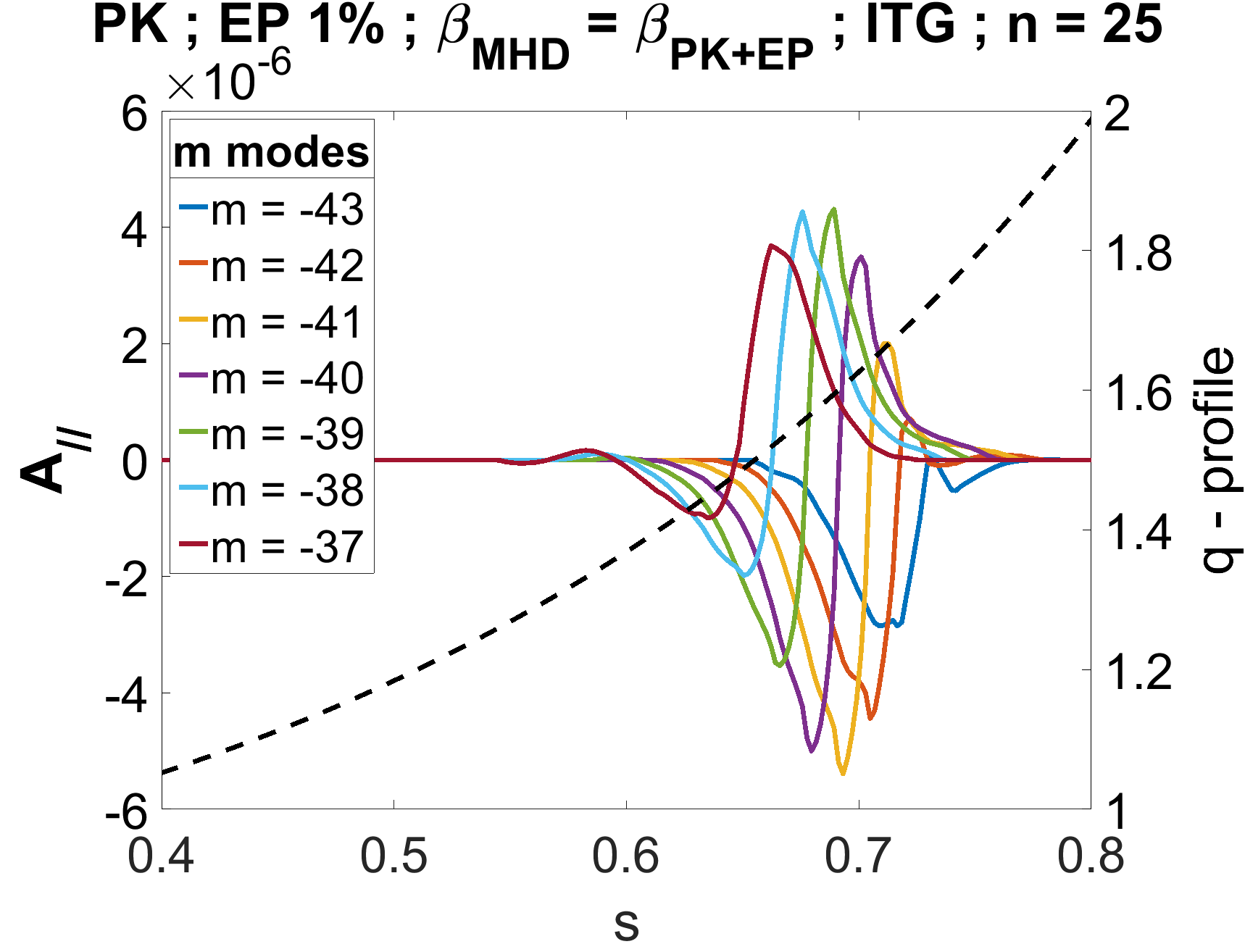}
\caption{\label{FIG:ITG_EP_struct} \it
Mode structure in $\phi$ and $A_{\parallel}$ of the $n=25$ mode in the ITG instability. The combined effects of bulk gradients and EPs fraction on ITG results in the PK cases being more unstable than ST cases, in contrast to the trend found for TAEs}
\end{center}
\end{figure}

The $\beta_{EP}$ is the same for both standard and peaked (profiles) cases, however, the effect on the ITG growth rate is quite different. The "weakly" driven case with the standard profiles experiences weak Shafranov shift effect, i.e. the growth rate slightly decreases in the longer wavelengths and increases for the higher mode numbers with a crossing point in the range $n=25-30$. While the "strongly" driven modes by the peaked profiles experiences a much stronger stabilization and do not regain the \textit{inconsistent} growth rate in the measured range. Figure \ref{FIG:ITG_EP_struct} shows on the $n=25$ mode structure in $\phi$ and $A_{\parallel}$, which due to Shafranov shift is more radially localized than the ITG in Figure \ref{FIG:ITG_Shaf_mode_shaer}. %Overall, the growth rate of the case with standard profiles acts as a "limit" for the system, with the Shafranov shift acting to restore it (Fig \ref{FIG:ITG_Shaf} \& Fig \ref{FIG:ITG_EP_Shaf}). and for all cases seen here (Figs. \ref{FIG:ITG_Shaf}, \ref{FIG:KBM_Shaf}, and \ref{FIG:ITG_EP_Shaf}), Shafranov shift decreases the ITG frequency. 

%The combined effects of bulk gradients on the drive, and Shafranov shift stabilization of the ITG instability, results in the PK cases being more unstable than ST cases, in contrast to the trend found for TAEs (Fig. \ref{FIG:lin_TAE_GR_EP_scan}) which are mainly stabilized by increased bulk gradients. 
\FloatBarrier
%======================================================================
\subsection{Part III - Alfv\'en Eigenmodes}
%======================================================================
In a fluid description, one finds that toroidal magnetic geometry couples the $m$ and $m+1$ poloidal harmonics, leading to the formation of so called gaps in the shear Alfv\'en wave (SAW) continuum. In these gaps, Alfv\'en Eigenmodes can resonate undamped, at radial locations where $nq = (m + 1/2)$ with a typical frequency of $\omega_0 \cong \omega_A/2q$. Where $\omega_A = v_A/R$ is the Alfv\'en frequency, $v_A = B_0/\sqrt{\mu_0\rho_m}$ is the Alfv\'en velocity, and $\rho_m$ is the mass density. These are global (low-$n$), electromagnetic modes called Toroidal Alfv\'en Eigenmodes (TAEs). In this section we will first present the Alfv\'en continuum with a representative $n=2$ TAE followed by an analysis the TAE dispersion relation.  

%======================================================================
\subsubsection{Alfv\'en continuum}
\begin{figure} [h!]
\begin{center}
\includegraphics[width=0.495\textwidth]{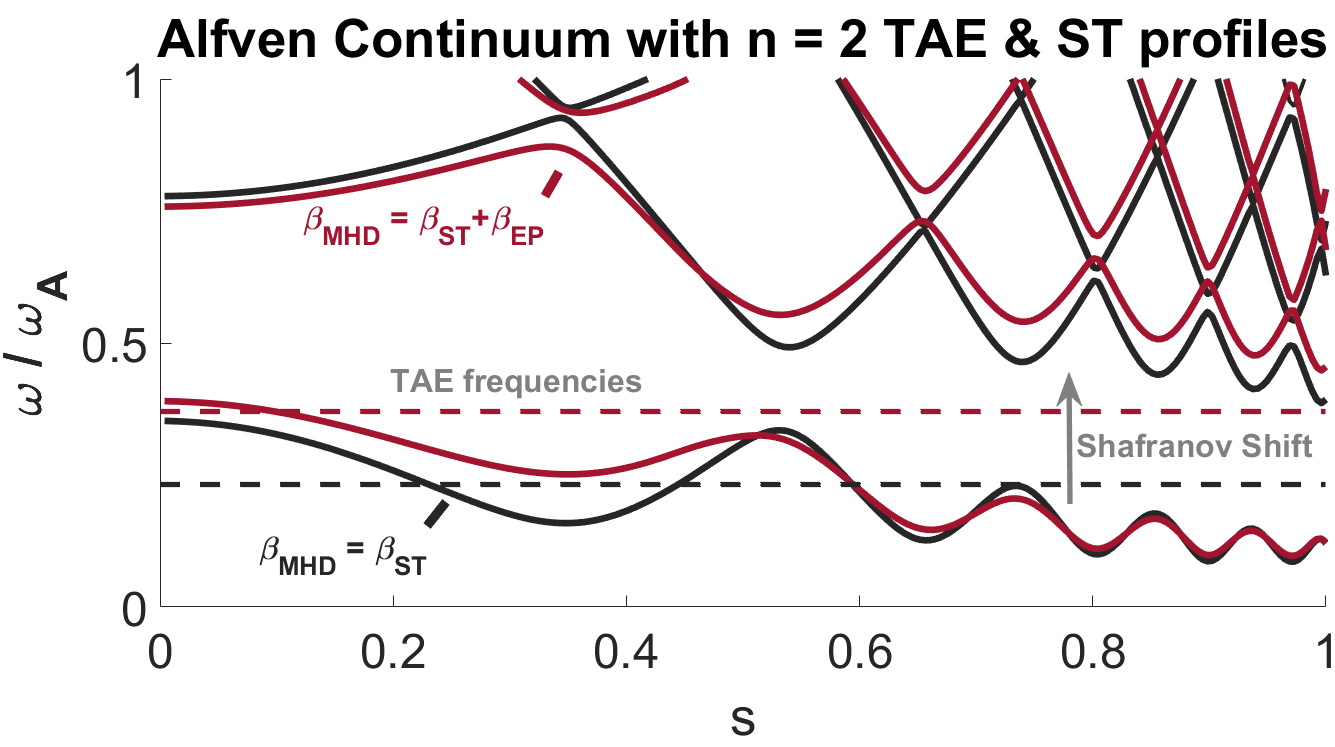}
\includegraphics[width=0.495\textwidth]{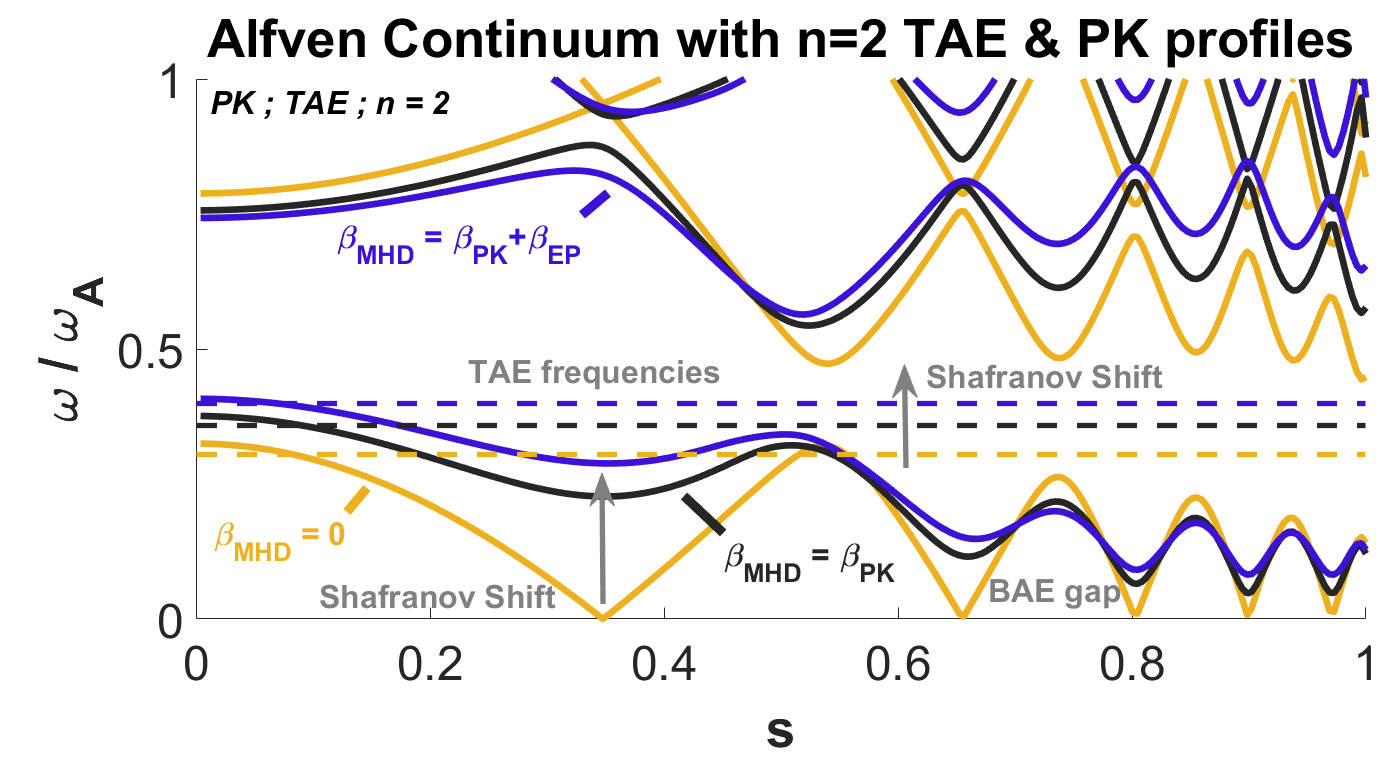}

\caption{\label{FIG:Alf_cont} \it 
Alv\'en continuum (under the slow-sound approximation \cite{chu1992numerical}) for the $n = 2$ mode with standard (left) and peaked (middle) thermal profiles. The measured frequencies of the $=2$ modes are plotted in doted lines, which match the TAE gaps well.}
\end{center}
\end{figure}

%======================================================================
In Figure \ref{FIG:Alf_cont} we plot the Alfv\'en continuum under the slow-sound approximation \cite{chu1992numerical} for the most unstable $n=2$ TAE. Each curve represents a different MHD equilibrium. On top of the continuum plots we add the measured frequencies of the unstable $n=2$ modes. Thus, together with the mode's structure and EP destabilization we identify them as TAEs. Moreover, we see how the increase in TAE frequency (plasmas response) is linked to the modification of the Alfv\'en continuum by Shafranov shift and finite $\beta$ effects.

The slow sound approximation is $v_A^2 \gg c_s^2$, where $v_A = B_0/\sqrt{\mu_0\rho_m}$ is the Alfv\'en velocity and $c_s = \sqrt{T_e/m_i}$ is the ion sound speed. Which can also be rewritten as $1 \gg (c_s/v_A)^2 = \beta$. Applying this approximation to the shear Alfv\'en continuum equations results in a simplified coupling between the sound and the Alfv\'en continua through the geodesic curvature and plasma pressure. As can be seen in Figure \ref{FIG:Alf_cont}, this assumption is sufficient to capture several feature of the full finite $\beta$ Alfv\'en continuum: $(1)$ opening of a lower BAE gap \cite{heidbrink1999beta}, $(2)$ up-shift of the TAE gap, and $(3)$ widening (opening) of the TAE gap closer to the edge. All in line with the results of Chu et al \cite{chu1992numerical} (see eq. 16 of that paper). A more general treatment of the coupling and the use of Alfv\'eninicty as a diagnostic of the branch polarization, can be found in the works of Falessi et al \cite{falessi2019shear} and \cite{falessi2020polarization}.  

\begin{figure} [h!]
\begin{center}

\includegraphics[width=0.215\textwidth]{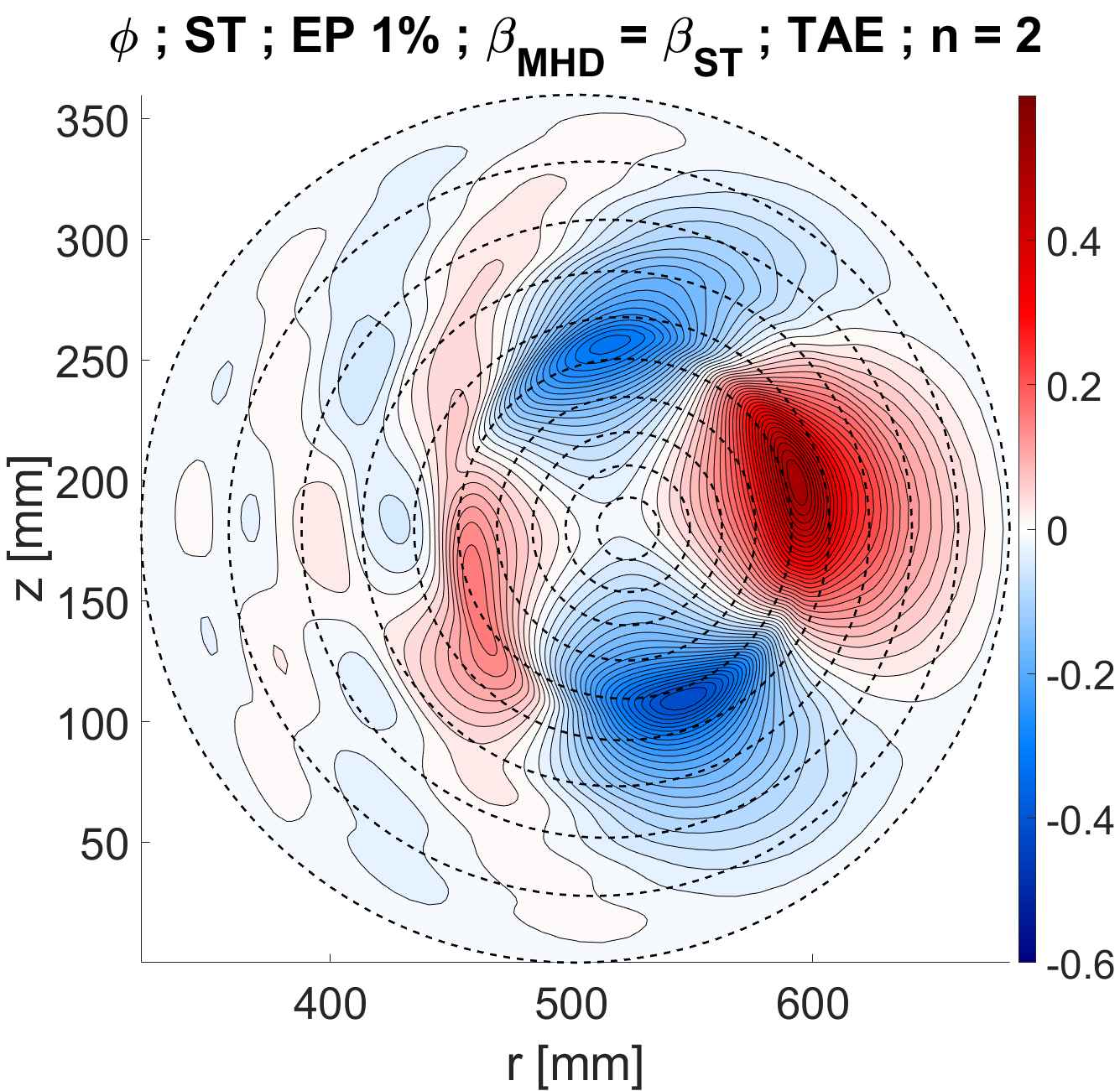}
\includegraphics[width=0.275\textwidth]{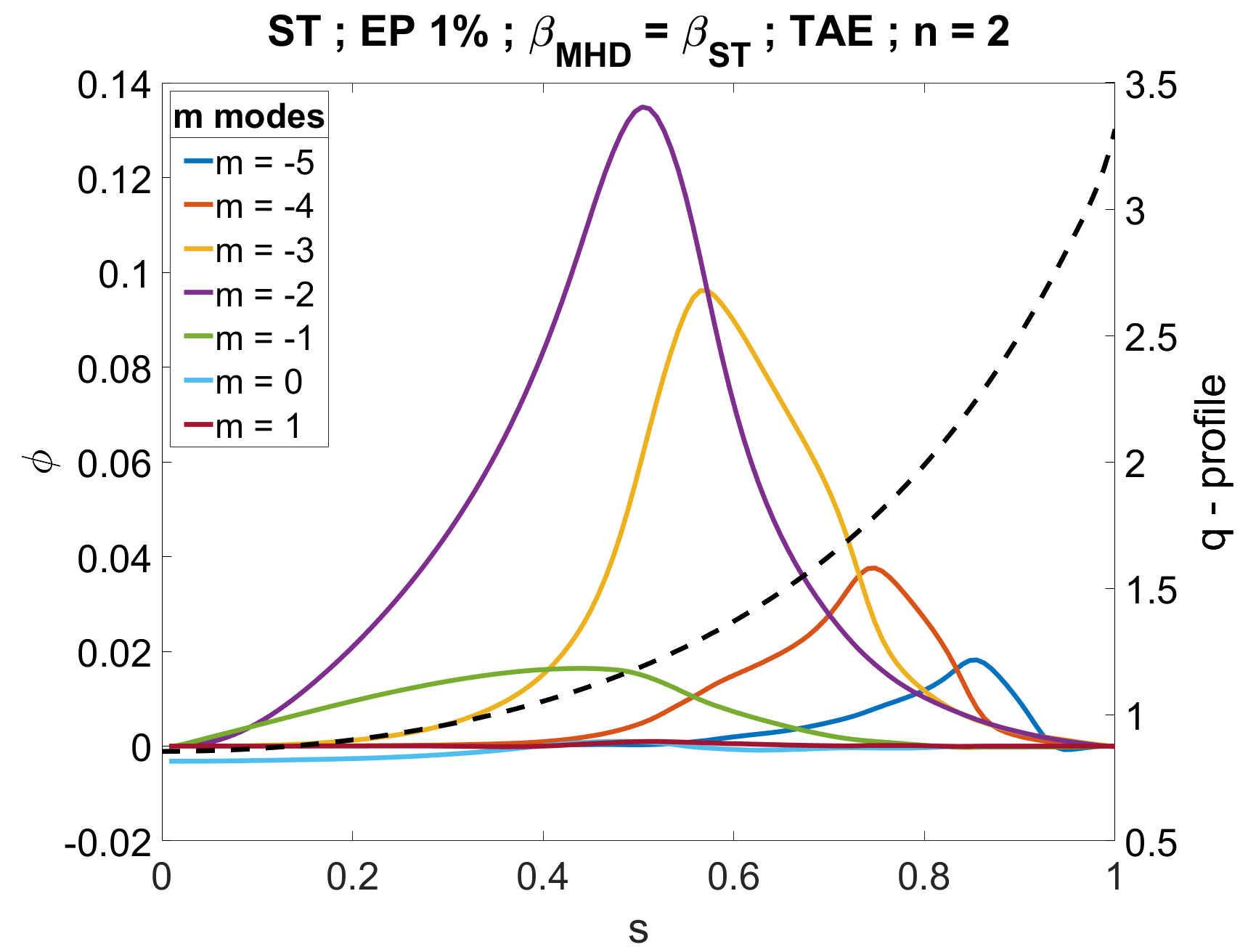}
\includegraphics[width=0.22\textwidth]{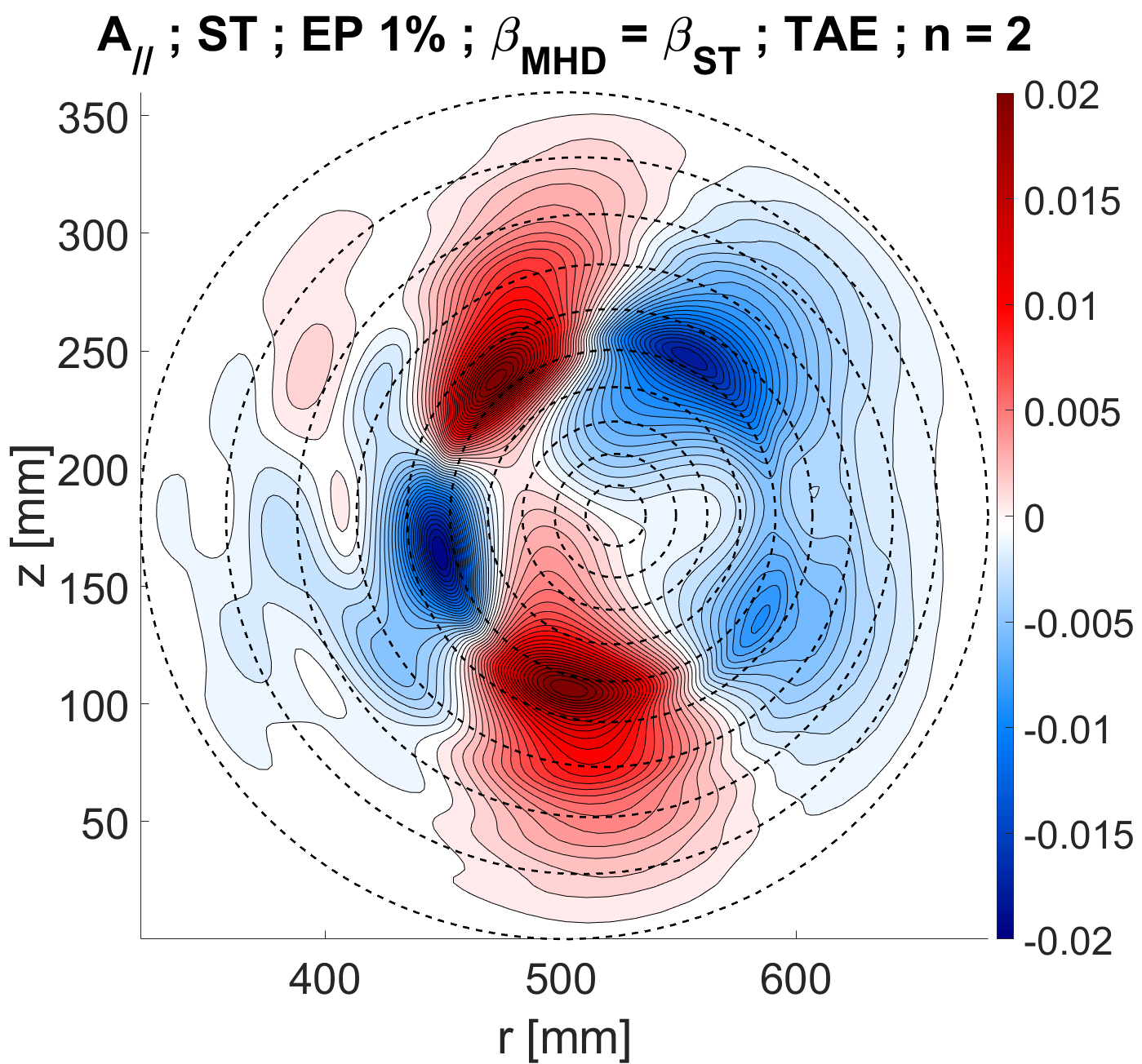}
\includegraphics[width=0.27\textwidth]{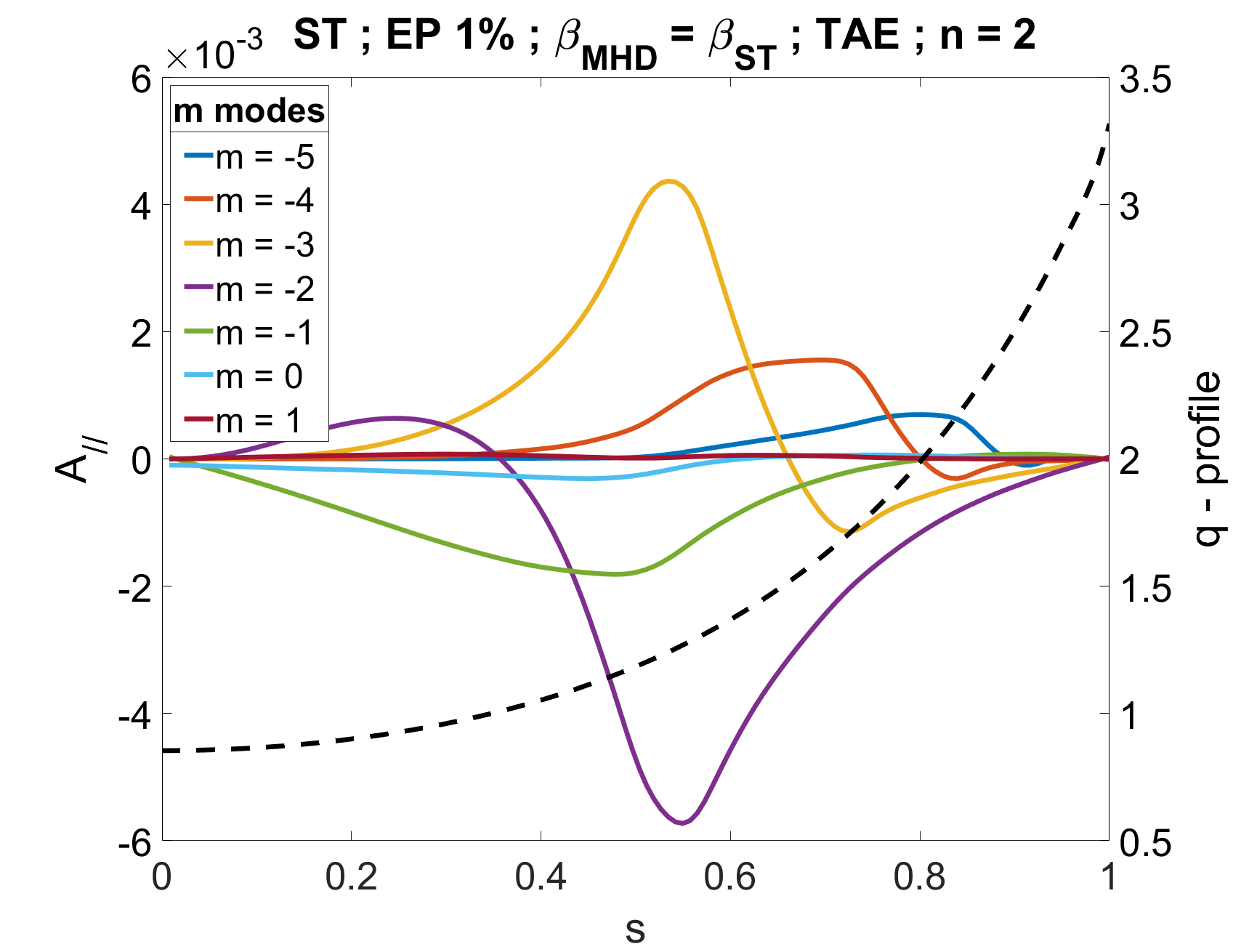}

\includegraphics[width=0.22\textwidth]{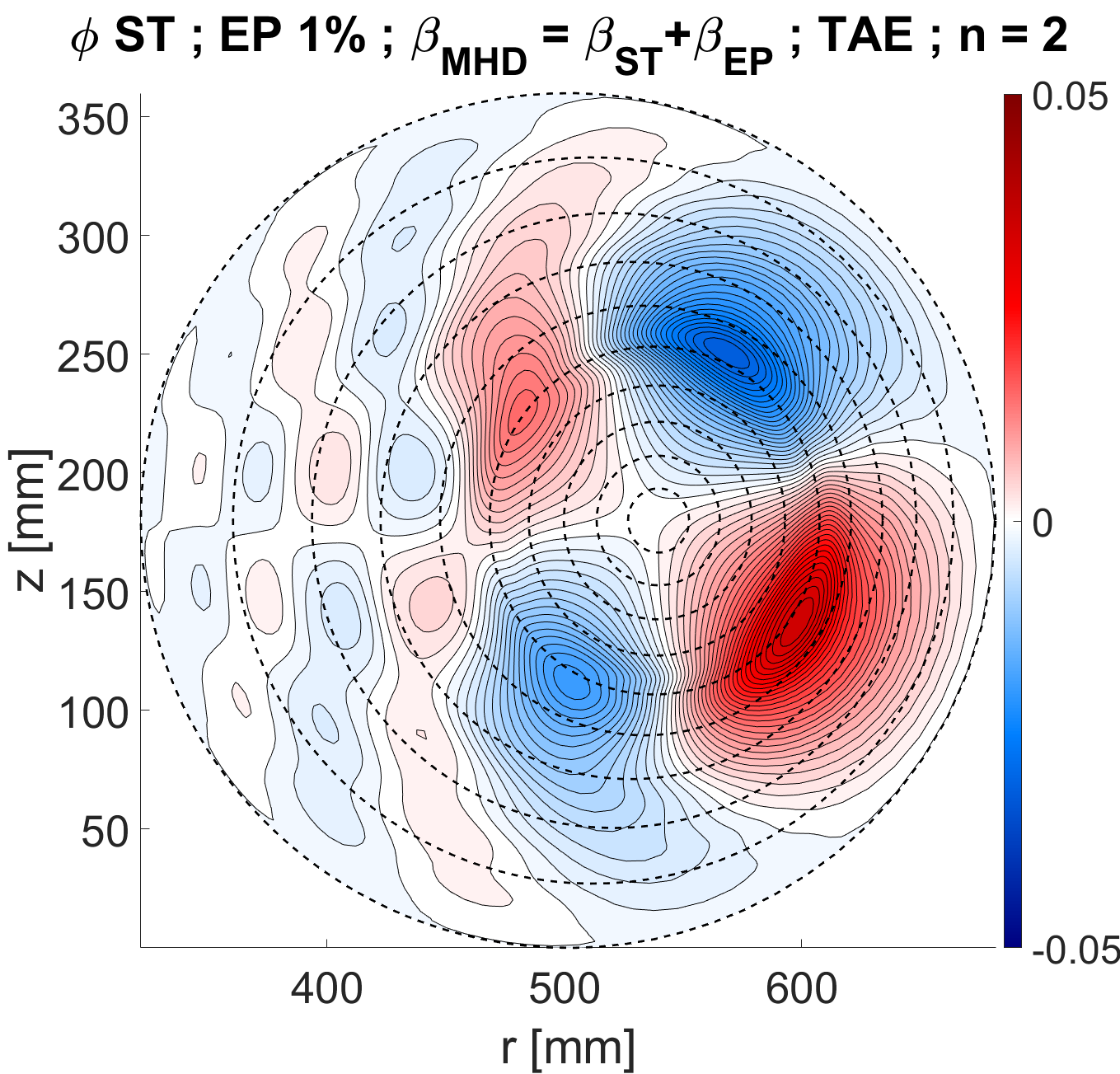}
\includegraphics[width=0.27\textwidth]{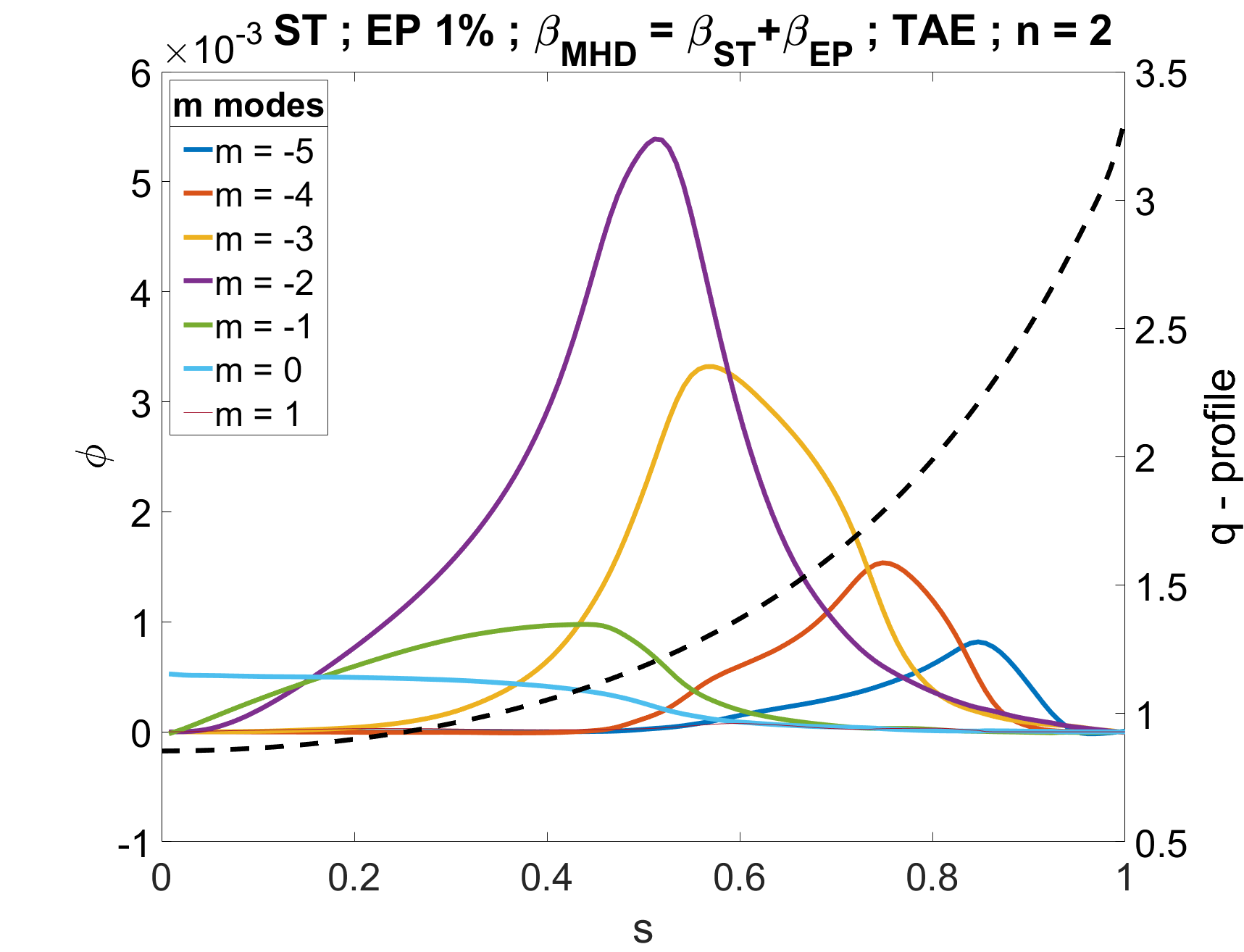}
\includegraphics[width=0.22\textwidth]{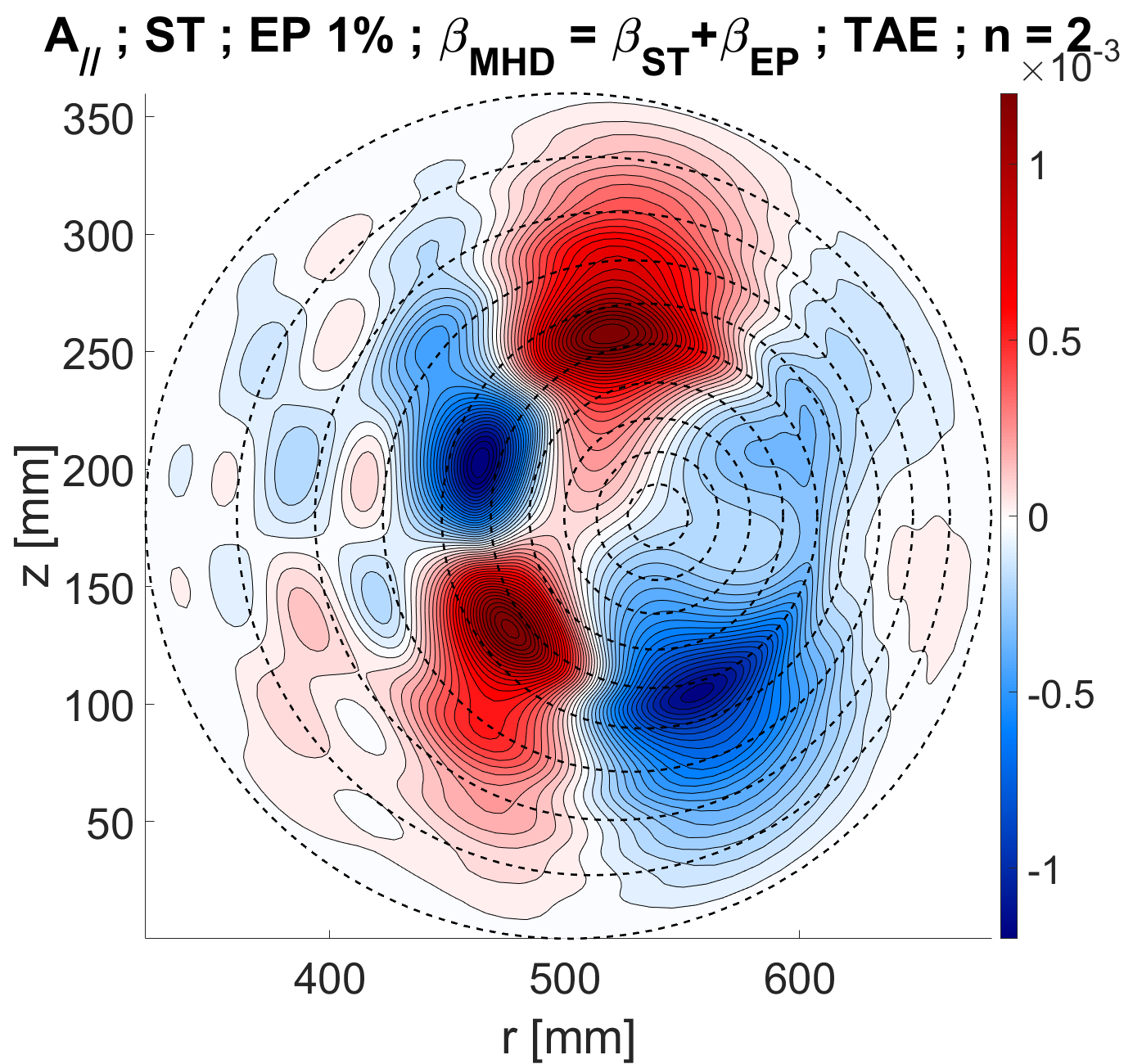}
\includegraphics[width=0.27\textwidth]{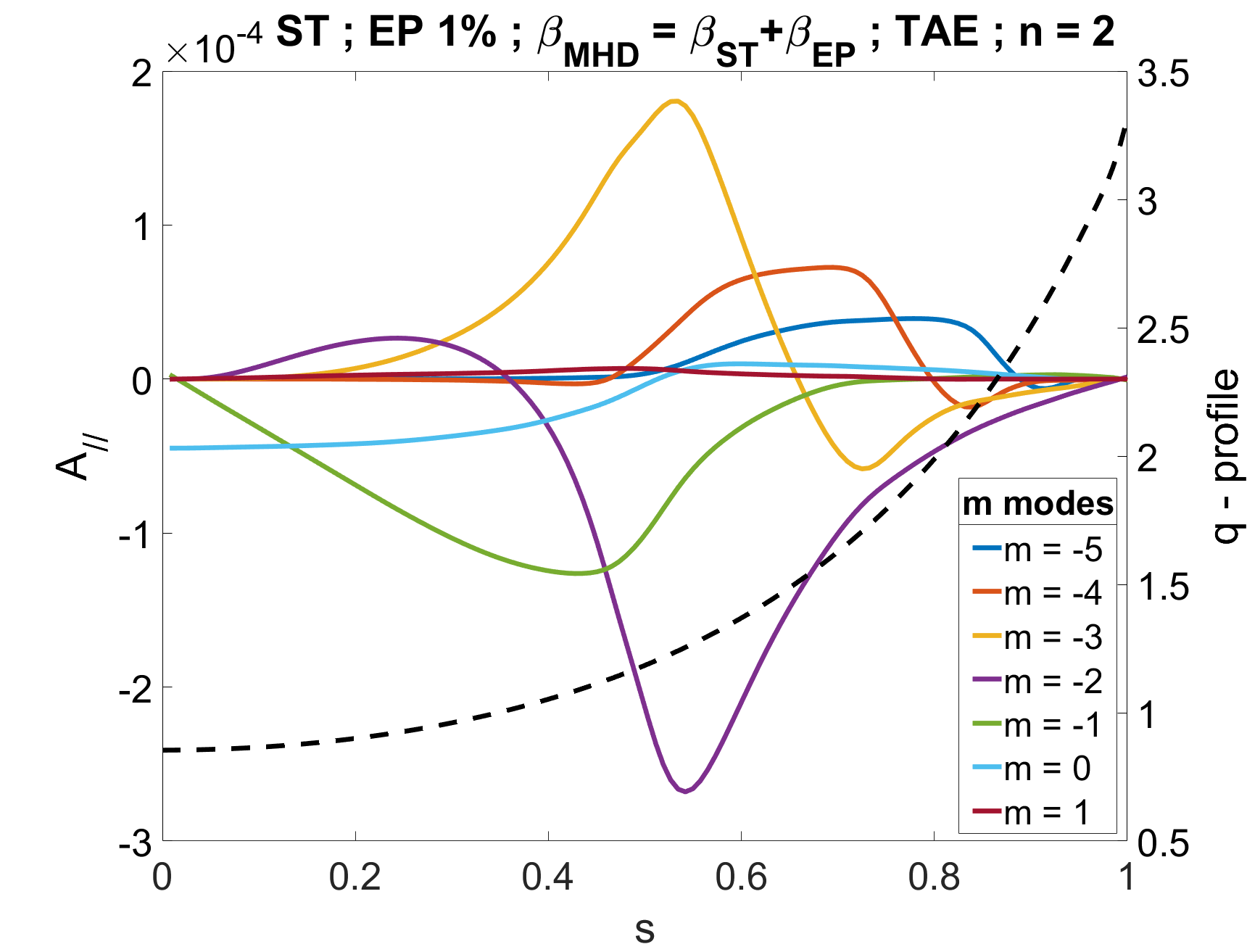}

\caption{\label{FIG:ST_STEP_TAE} \it 
TAE mode structure for \textit{inconsistent} case with $\beta_{MHD}=\beta_{ST}$ on the top row, and \textit{self-consistent} case with $\beta_{MHD}=\beta_{ST}+\beta_{EP}$ on the bottom row.}
\end{center}
\end{figure}

In Figure \ref{FIG:Alf_cont} (left) we see a large increase in TAE frequency with Shafranov shift for the cases with standard profiles. Based on the continuum plot we attribute the change in TAE frequency to Shafranov shift opening the gap near the edge which allows the mode to exist closer to the middel of the gap with less damping. Despite the change in frequency, there is no significant impact of the additional $\beta_{EP}$ on the poloidal mode structure as seen in Figure \ref{FIG:ST_STEP_TAE}. 

\FloatBarrier
%======================================================================
\subsubsection{Energetic particles excite TAEs}
\begin{figure} [h!]
\begin{center}
\includegraphics[width=0.495\textwidth]{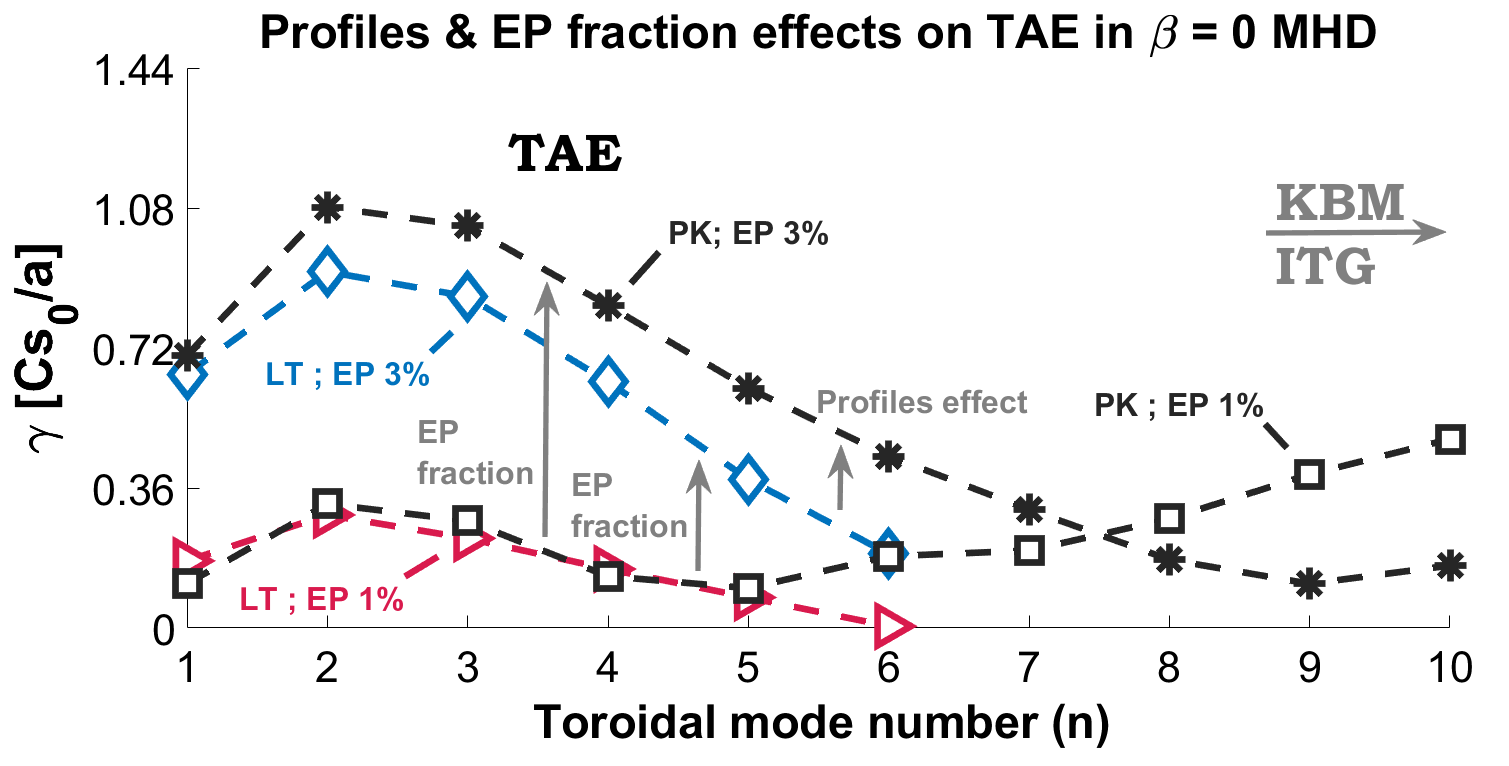}
\includegraphics[width=0.495\textwidth]{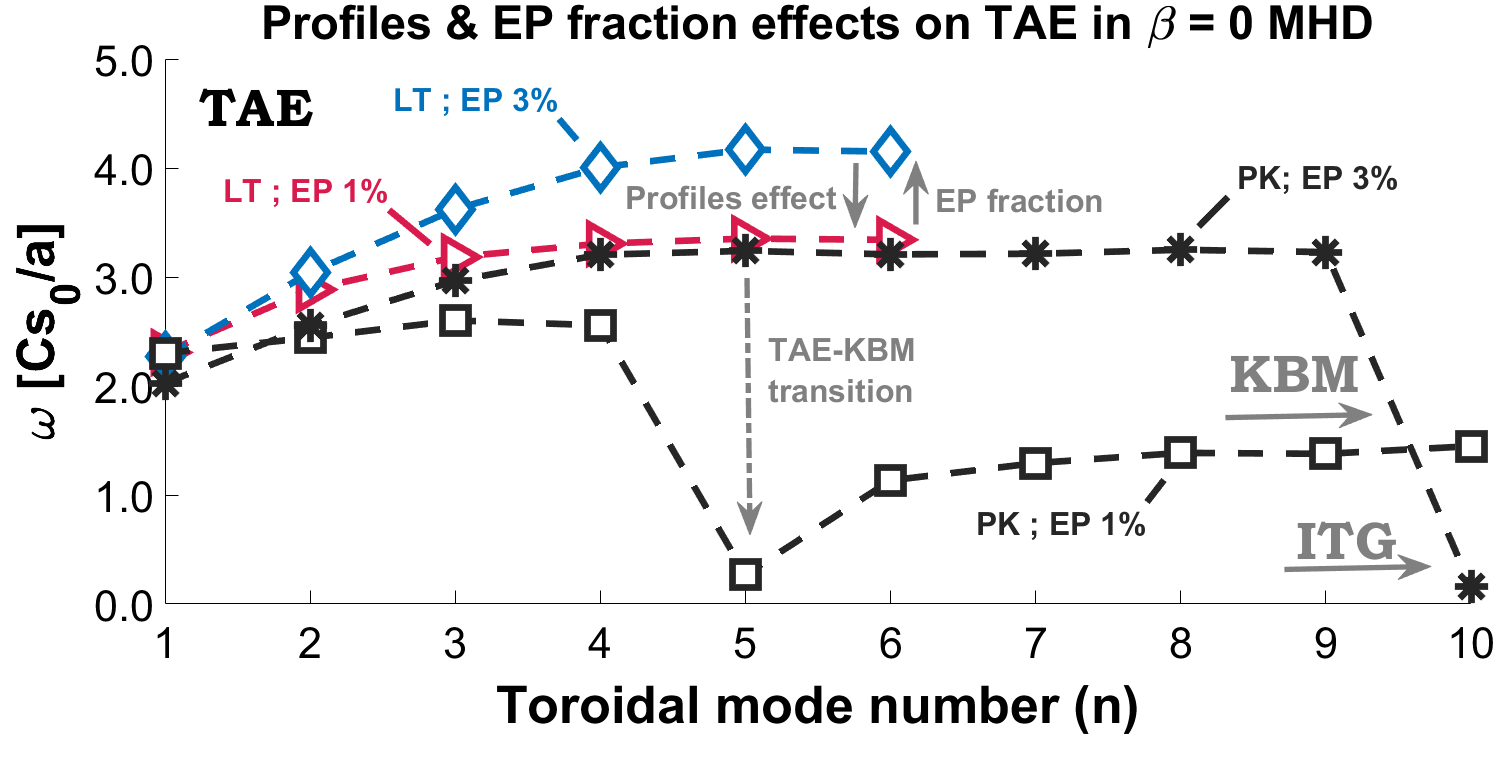}
\caption{\label{FIG:TAE_dispersion} \it
TAE dispersion relation in $\beta_{MHD}=0$ magnetic equilibrium, with peaked (PK) and low temperature (LT) profiles. For LT profiles $R/L_T=R/L_n = -6.23.$}
\end{center}
\end{figure}

%======================================================================
The energetic particles in our system have Alfv\'enic parallel velocities which match the resonance conditions to excite TAEs across the entire plasma width. For example, Figure \ref{FIG:Combined_profiles_particles} shows the radial profiles of $v_{th}/v_A$ for the EPs, ions and electrons the a case with peaked profiles. In this hot plasma we mainly excite the $v_A/3$ side band resonance.  
In Figure \ref{FIG:TAE_dispersion} we plot the TAE spectrum in a $\beta_{MHD}=0$ magnetic equilibrium. With an increase in EP fraction we see a proportional increase in TAE growth rate and therefore a broadening of the TAE spectrum, where for $3\%$ EPs the TAE spectrum extends to $n = 9$. The TAEs driven by $1\%$ EPs show a more complex behavior with a TAE to KBM transition occurring at $n=5$. For all our cases, the TAE frequency increased with EP fraction. This affect is stronger for shorter toroidal wavelengths, an opposite trend to the one found by Biancalani et al \cite{biancalani2016linear} (in Figure 6 of that paper). In order to test the effects of kinetic pressure on TAEs we chose to change the temperature while leaving the density profile unchanged. Because it is the density which affects the Alfv\'en continuum and we wish to separate variables in our analysis. For this reason we have simulated cases with a low temperature profile (LT) identical to the density profile in a peaked case. We also find a small affect of the plasma temperature on the TAE growth rate, that is proportional to the EP fraction. Where for all examined cases the TAE frequency decreased with the plasma temperature. 

\FloatBarrier
%======================================================================
\subsubsection{Shafranov shift effects on TAE stability}
\begin{figure} [h!]
\begin{center}
\includegraphics[width=0.495\textwidth]{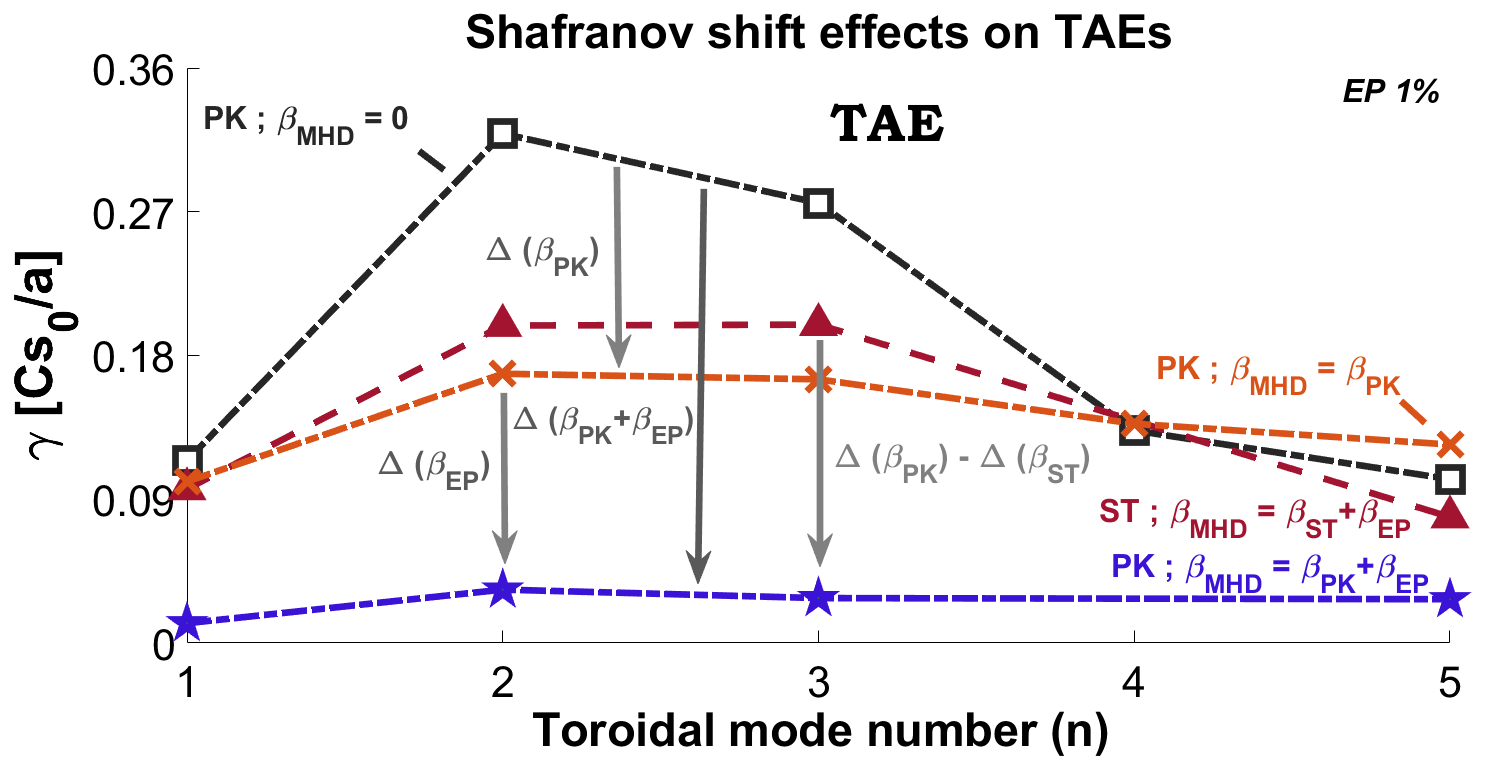}
\includegraphics[width=0.495\textwidth]{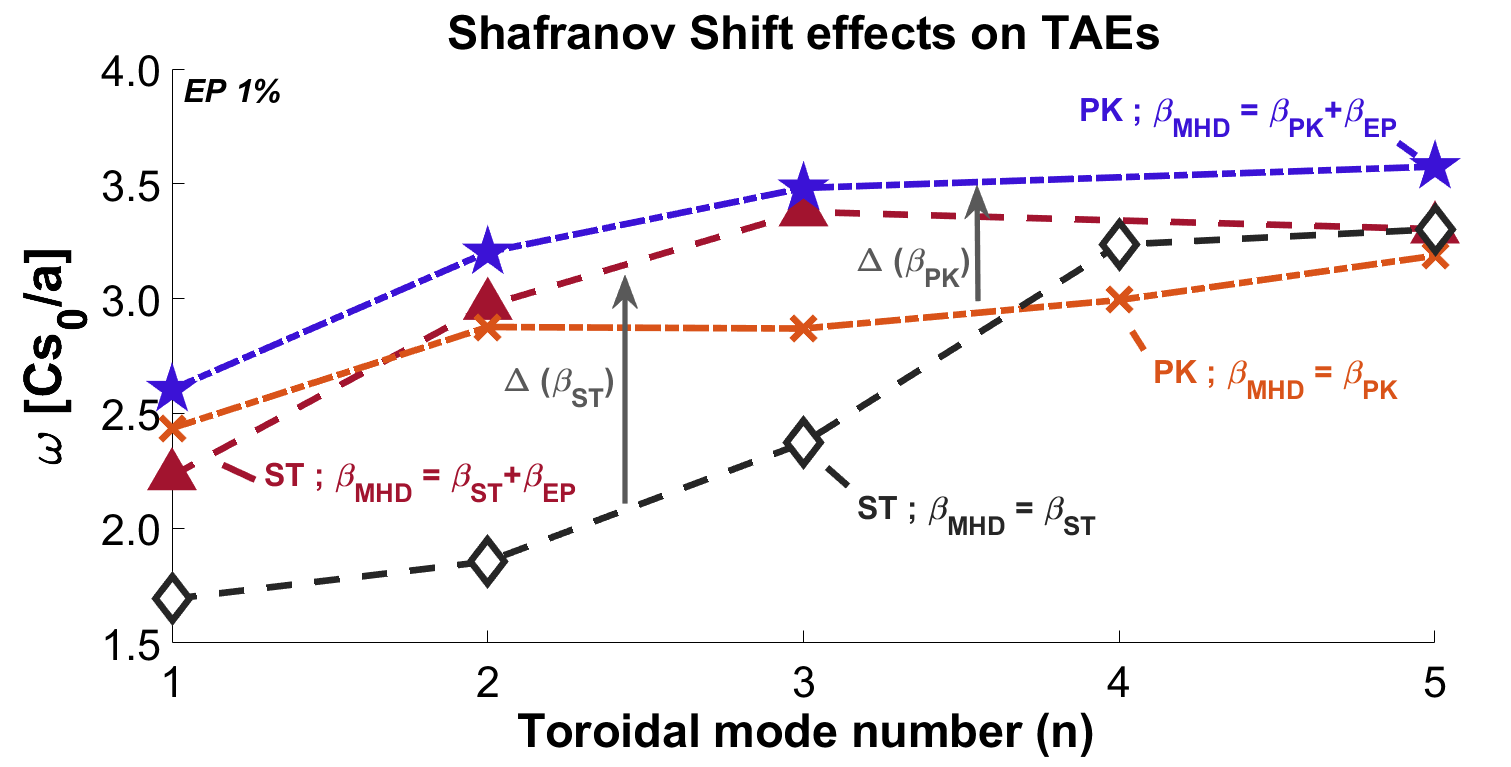}
\caption{\label{FIG:TAE_Shaf}\it TAE dispersion relations driven by $1\%$ EPs, in magnetic equilibria with $\beta_{MHD}=[0;\beta_{th};\beta_{th}+\beta_{EP}]$. Here $\Delta(\beta_x)$ is the Shafranov shift due to $\beta_x$ component in $\beta_{MHD}$. 
}
\end{center}
\end{figure}

%======================================================================
Next we add Shafranov shift effects to the magnetic equilibrium, where in all cases we include $1\%$ EPs. Thus we have two main sources of kinetic pressure - the thermal bulk plasma and the EPs. In this case, unlike for the ITG, the thermal profiles mainly stabilize through Shafranov shift, while the EPs contribute to both the drive and the damping of the mode. In figure \ref{FIG:TAE_Shaf} we see how sensitive global TAEs are to Shafranov shift stabilization with a $90\%$ reduction in growth rate for the main modes. Unlike in the ITG case, for the TAE Shafranov shift increases the mode's frequency with an asymptotic plasma response towards an upper limit. 

\FloatBarrier
%======================================================================
\subsubsection{EP fraction effects on TAE stability}
\begin{figure} [h!]
\begin{center}
    \includegraphics[width=0.495\textwidth]{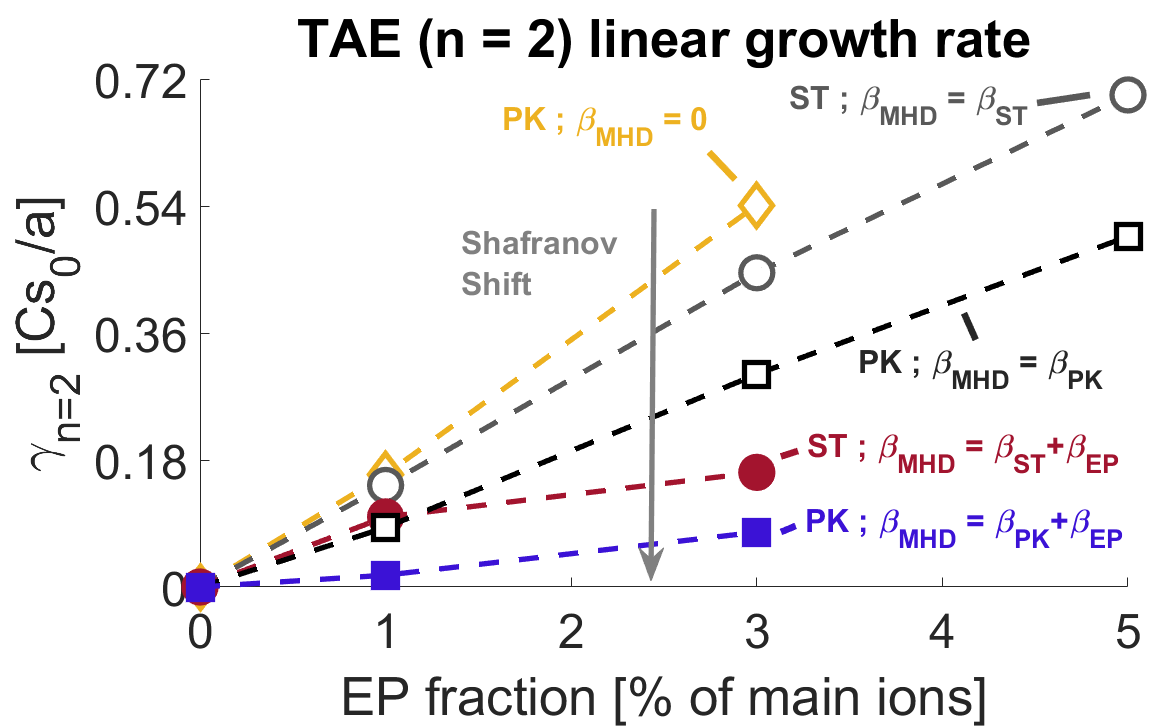}
    \caption{\label{FIG:EP_scan}\it The linear growth rate of the $n=2$ TAE as a function of: EP fraction, Shafranov shift and bulk plasma profiles. }
\end{center}
\end{figure}

%======================================================================
The EP fraction is an important operational parameter for fusion reactors. In practice it will be a controllable parameter resulting from a balance between the fusion reactions and the EP losses. Due to their highlly energetic nature, small changes to EP fraction translate to large changes to the system. We select the TAE with the highest growth rate, $n=2$, to represent the TAEs in the system and study the effects of different EP fractions. We perform a scan in the TAE growth rate as function of EP fraction, bulk plasma profiles and Shafranov shift. In Figure \ref{FIG:EP_scan} we plot the combined results. In \textit{inconsistent} magnetic equilibria With $\beta_{MHD}=0$ and with $\beta_{MHD}=\beta_{th}$, $\gamma$ increases linearly with EP fraction similar to the results obtained by Biancalani et al \cite{biancalani2016linear} (Fig. 9 of that work). However, in \textit{consistent} cases with $\beta_{MHD}=\beta_{th}+\beta_{EP}$, where we account for all the species pressure contributions, we see a saturation of the growth rate as a function of EP fraction. 

\FloatBarrier
%======================================================================
\subsection{Shafranov shift toroidal mode dependence}
%======================================================================

\begin{figure} [h!]
\begin{center}
\includegraphics[width=0.495\textwidth]{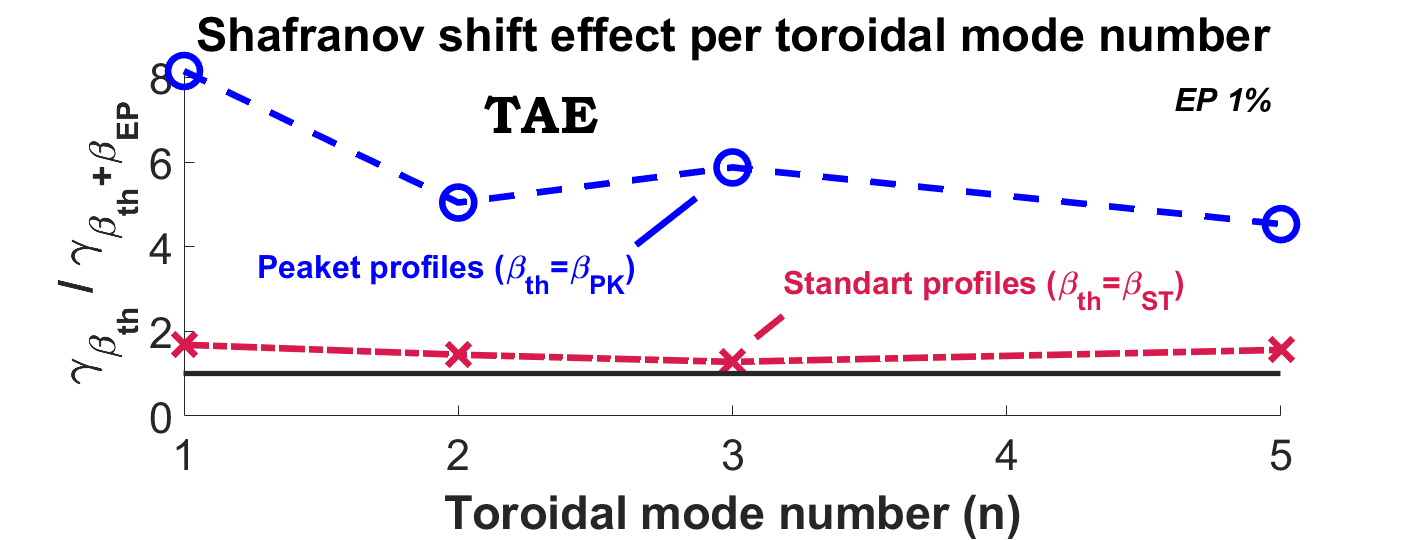}
\includegraphics[width=0.495\textwidth]{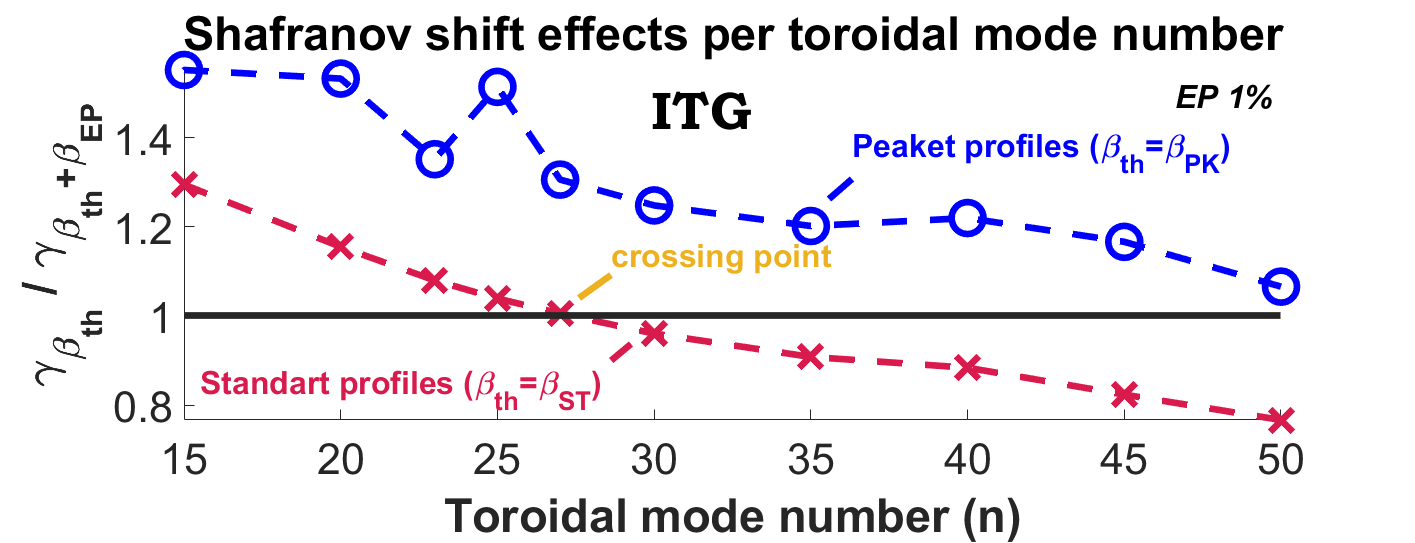}
\includegraphics[width=0.495\textwidth]{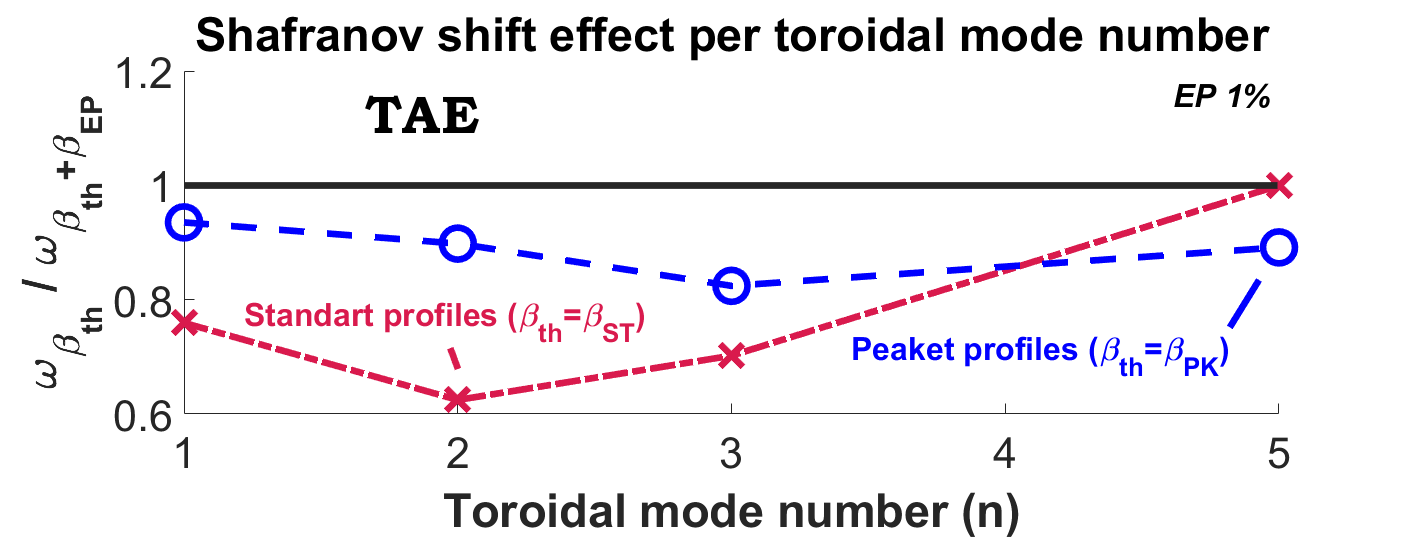}
\includegraphics[width=0.495\textwidth]{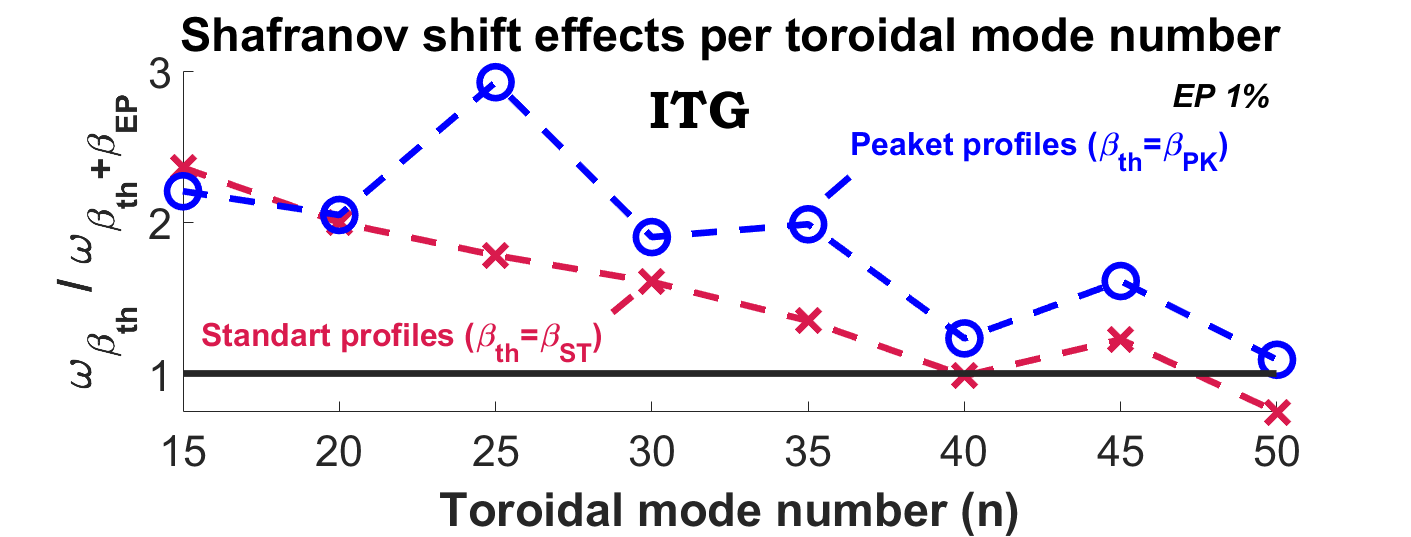}
\caption{\label{FIG:full_Shaf} \it
Toroidal mode number $n$ dependence of Shafranov shift effects on the growth rates $\gamma$ and frequencies $\omega$ of the unstable modes. Due to the separation in scales we separate between the TAE [1-5] and ITG [15-50] ranges.}
\end{center}
\end{figure}

In Figure \ref{FIG:full_Shaf} we plot the full dispersion relation of our cases with $1\%$ EPs. We show the ratio of growth rates (or frequencies) found in a $\beta_{MHD} = \beta_{th}$ MHD equilibrium, to the ones arising in a $\beta_{MHD} = \beta_{th} + \beta_{EP}$ equilibrium. We find that 
Shafranov shift effect on ITG are toroidal mode $n$ dependent, resulting in stabilization of the longer wavelengths and destabilization of the shorter wavelengths. The ITG is mainly an electrostatic mode driven by the bulk gradients, with the peaked bulk profiles adding more drive than stabilization. In the TAE cases, the wavelength dependence is much weaker, perhaps due to the limited spectrum of excited modes.

\FloatBarrier
%======================================================================
\section{Nonlinear physics}
%======================================================================
In this section we examine the nonlinear saturation level, profile relaxation and resulting fluxes of several representative cases of TAEs and ITGs. Focusing on Shafranov shift effects in the nonlinear phase, we exclude the axisymmetric plasma response, i.e. the $n=0$ modes. Thus we prevent the self-consistent formation of zonal flows and structures. Deeper investigation of the zonal flows and their impact on the nonlinear system can be found in \cite{rofman2026roleshafranovshiftzonal}. For our analysis we are interested in comparing, over a time averaged window, the nonlinear state of the system established by different instabilities. However, variations in linear growth rate and nonlinear saturation level between the instabilities prevent us from using the same time window for all our cases. For this reason we have used a comparable sized time window centered a while after the saturation phase of each instability. The specific time windows are found in Table \ref{tab:NL_t_Win}   

\begin{table}[!h]
\begin{center}
\caption{Nonlinear time averaging window}
\label{tab:NL_t_Win}
\begin{tabular}{ | c | c | c | c | c | } 
 \hline
Profiles & Ideal-MHD & Instability & t [$a/c_{s_0}$] & t [$\Omega^{-1}_{c_i}]\cdot10^4$s \\ [0.5ex]  
 \hline
 \hline
 ST & $\beta_{MHD} = \beta_{ST} \ , \ \beta_{ST}+\beta_{EP}$ & TAE [2] & 111 - 172 & 2 - 3.1 \\ 
 \hline
 ST & $\beta_{MHD} = \beta_{ST} \ , \ \beta_{ST}+\beta_{EP}$ & ITG [25] & 89 - 144 & 1.6 - 2.6 \\
 \hline
 \hline
 PK & $\beta_{MHD} = 0$ & TAE [2] & 66 - 110 & 1.2 - 2.0 \\
 \hline
 PK & $\beta_{MHD} = 0$ & KBM [12] & 66 - 128 & 1.2 - 2.3 \\
 \hline
 \hline
 PK & $\beta_{MHD} =  \beta_{PK} \ , \  \beta_{PK}+\beta_{EP}$ & TAE [2] & 194 - 278 & 3.5 - 5 \\
 \hline
 PK & $\beta_{MHD} =  \beta_{PK} \ , \  \beta_{PK}+\beta_{EP}$ & ITG [25] & 194 - 278 & 3.5 - 5 \\
 [1ex] 
 \hline
 %\hline

\end{tabular}

\end{center}
\end{table}

\subsubsection{TAE saturation}
\begin{figure} [h!]
\begin{center}
\includegraphics[width=0.495\textwidth]{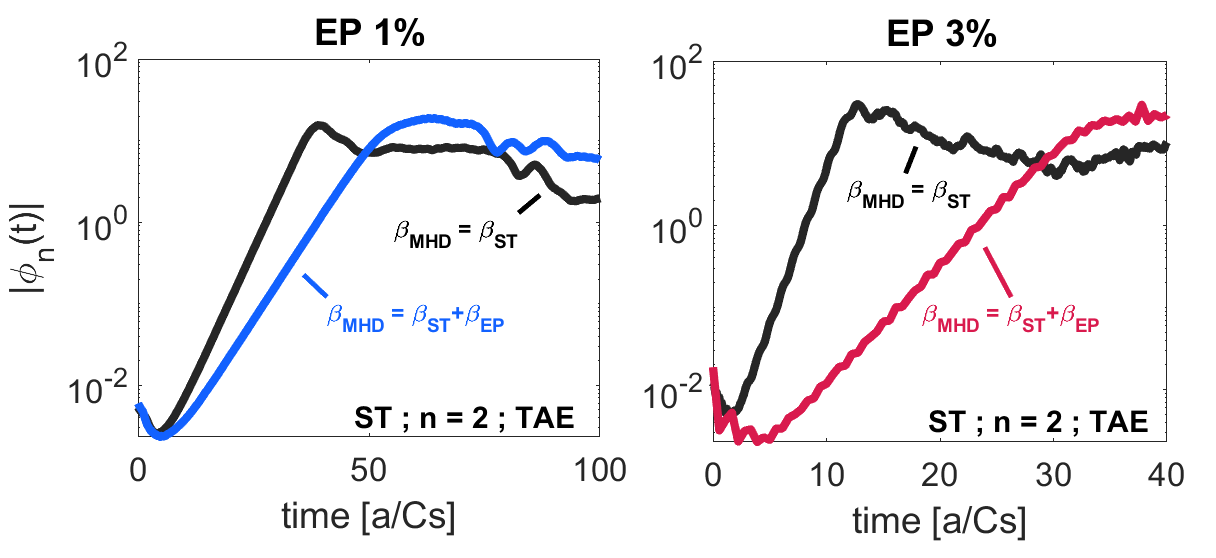}
\includegraphics[width=0.495\textwidth]{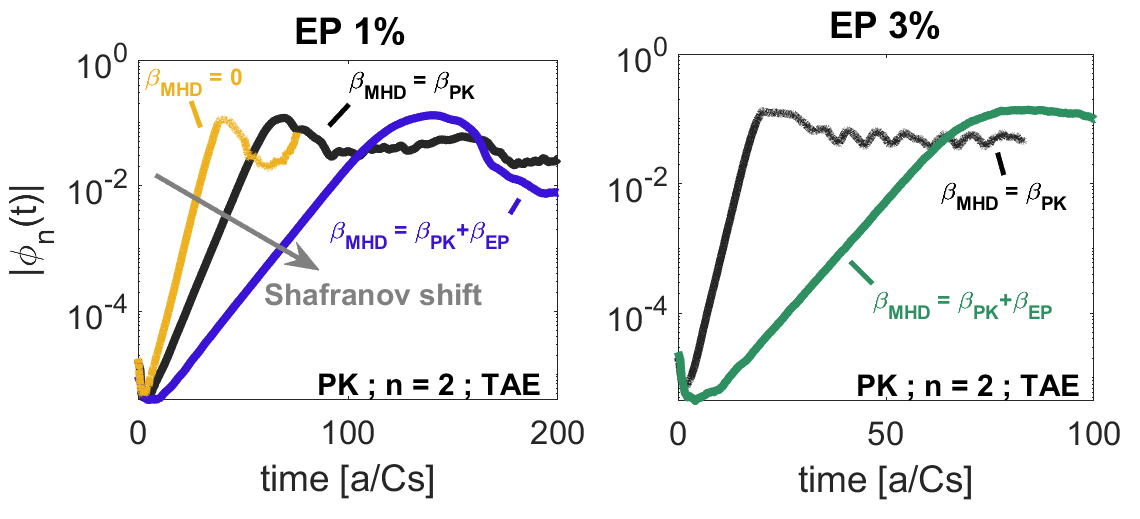}

\caption{\label{FIG:nonlinear_TAE_Sat} \it 
Time traces of the total electrostatic potential $|\phi_n|=\iint{|\phi_n(\theta^*,s,t)| J(\theta^*,s) d\theta^* ds }$, for standard (left) and peaked (right) thermal profiles and $1\%$ and $3\%$ EPs in consistent and not consistent ideal-MHD equilibria. The $\beta_{EP}$ changes according to the EP fraction.}
\end{center}
\end{figure}

In Figure \ref{FIG:nonlinear_TAE_Sat} we plot the total absolute electrostatic potential in a specific toroidal mode, $\phi_n=\iint{\phi_n(\theta^*,s,t) J(\theta^*,s) d\theta^* ds }$. We choose this approve over plotting the potential at a specific radial location to better capture the global nature of low $n$ modes and the ballooning nature of higher $n$ ITG modes.
Noticeably, even though the linear growth rates vary by an order of magnitude between the cases (Fig. \ref{FIG:EP_scan}), the nonlinear saturation levels remain nearly constant. Overall we find the TAE nonlinear saturation levels independent of: EP fraction, Shafranov shift, and thermal profiles.

%To expand our understanding of the nonlinear TAE dynamics we scan in EP fraction $[0\%,1\%,3\%]$, for 5 different MHD equilibria and two sets of profiles. Plotted in Figure \ref{FIG:nonlinear_TAE_Sat} on the right are the saturation levels vs. growth are for the $n = 2$ TAE. We show that for case with $1\%$ EPs, the saturation level increase with Shafranov shift. With a similar trend, a more complex behavior arises in cases with $3\%$ EPs, most likely due to the very strong Shafranov shift and local shear.  

\begin{figure} [h!]
\begin{center}
\includegraphics[width=0.9\textwidth]{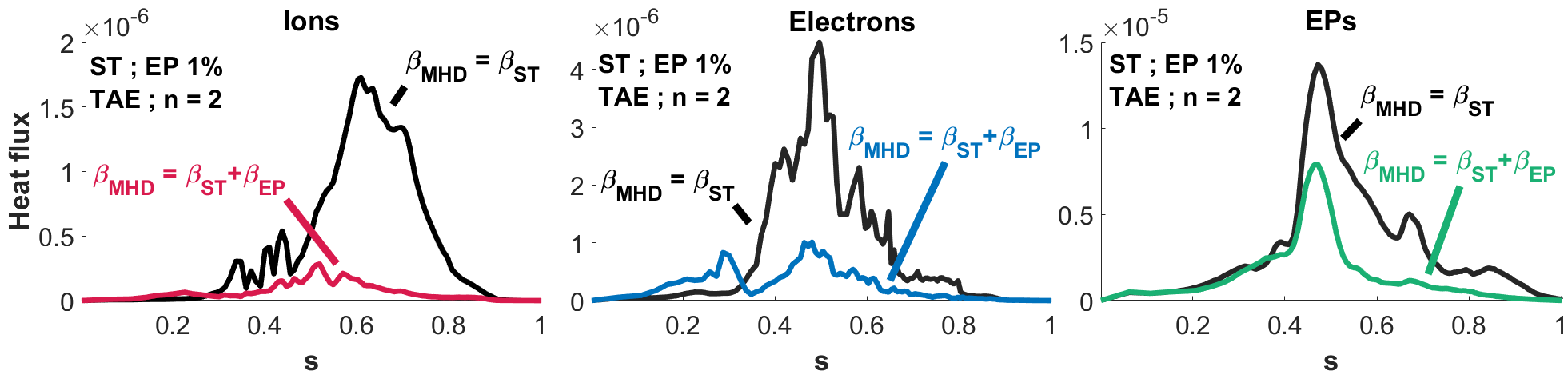}
\includegraphics[width=0.9\textwidth]{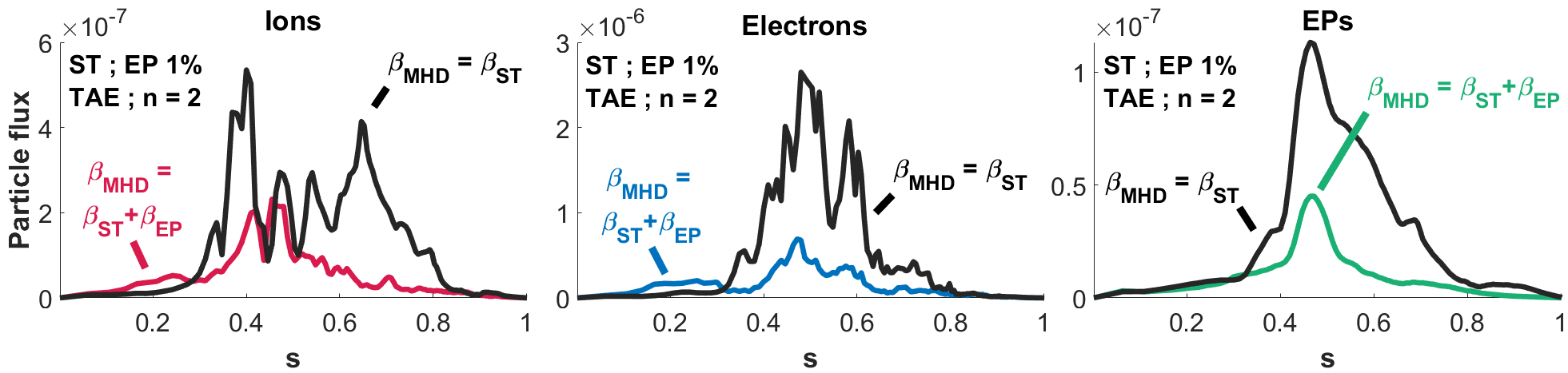}
\caption{\label{FIG:TAE_heat_part_fluxs} \it
Heat and particles fluxes driven by the $n=2$ TAE in a system with standard profiles.The fluxes are averaged over a time window $\Delta$t = $(111 - 172) \ [a/c_{s_0}]$ situated after the initial saturation phase. Note, the y-scale is different for each species and flux. The peaked cases which exhibit a similar behavior, are omitted for brevity.}
\end{center}
\end{figure}

These results raise the question whether or not linear stabilization translates to the heat and particle fluxes. In Figure \ref{FIG:TAE_heat_part_fluxs} we find strong Shafranov shift stabilization for all cases and fluxes. Indicating that for the TAEs, Shafranov shift stabilization effectively reduces the nonlinear transport.

The EP heat and particle fluxes show a relatively sharp feature at around $s=0.45$ which corresponds well to EP density gradient and the $\Delta^\prime$ profile presented in Figure \ref{FIG:Combined_profiles_particles}. Due to the resonant nature of the TAE instability we expect it to drive heat, and especially particle fluxes mainly in the EP channel \cite{salewski2025energetic, chen2016physics}. However we find that while the heat flux is dominated by the EP channel, a fact only extenuated by Shafranov shift, the particle flux is dominated by the electrons.

\FloatBarrier
%%%%%%%%%%%%%%%%%%%%%%%%%%%%%%%%%%%%%%%%%%%%%%%%%%%%%%%%%%%%%%%%%%%%%5
\subsubsection{ITG saturation }

\begin{figure} [!h]
\begin{center}
\includegraphics[width=0.65\textwidth]{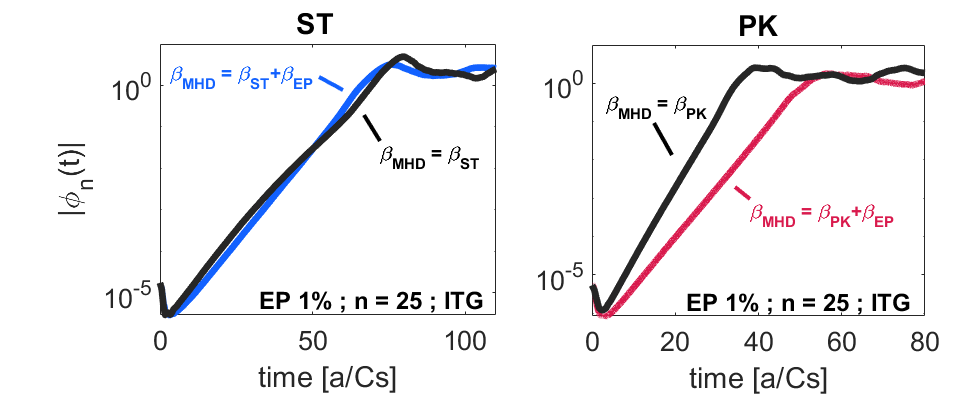}
\caption{\label{FIG:nonlinear_ITG_Sat} \it 
Time evolution of the total electrostatic potential $\phi_n=\iint{|\phi_n(\theta^*,s,t)| J(\theta^*,s) d\theta^* ds }$.}
\end{center}
\end{figure}

The time evolution of the ITG microinstability shown in Figure \ref{FIG:nonlinear_ITG_Sat} shows very similar results to the TAE, where the linear Shafranov shift stabilization has only a minor impact on the nonlinear saturation level.

\begin{figure} [!h]
\begin{center}
\begin{minipage}[c]{0.65\textwidth}
    \includegraphics[width=0.49\textwidth]{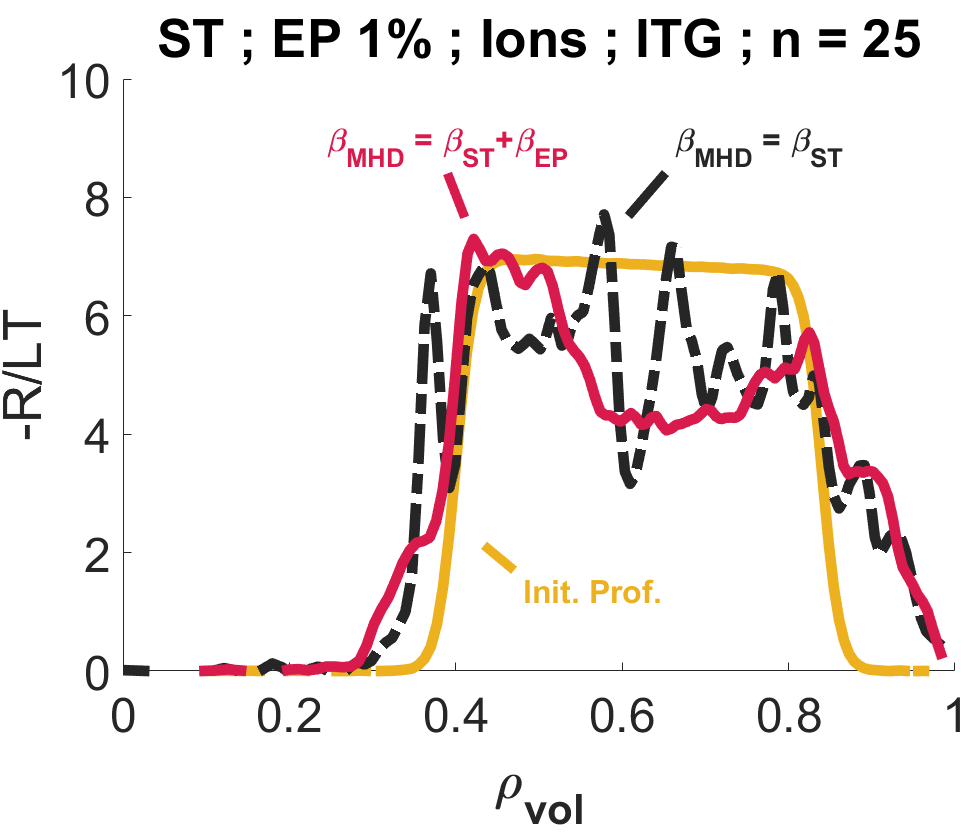}
    \includegraphics[width=0.49\textwidth]{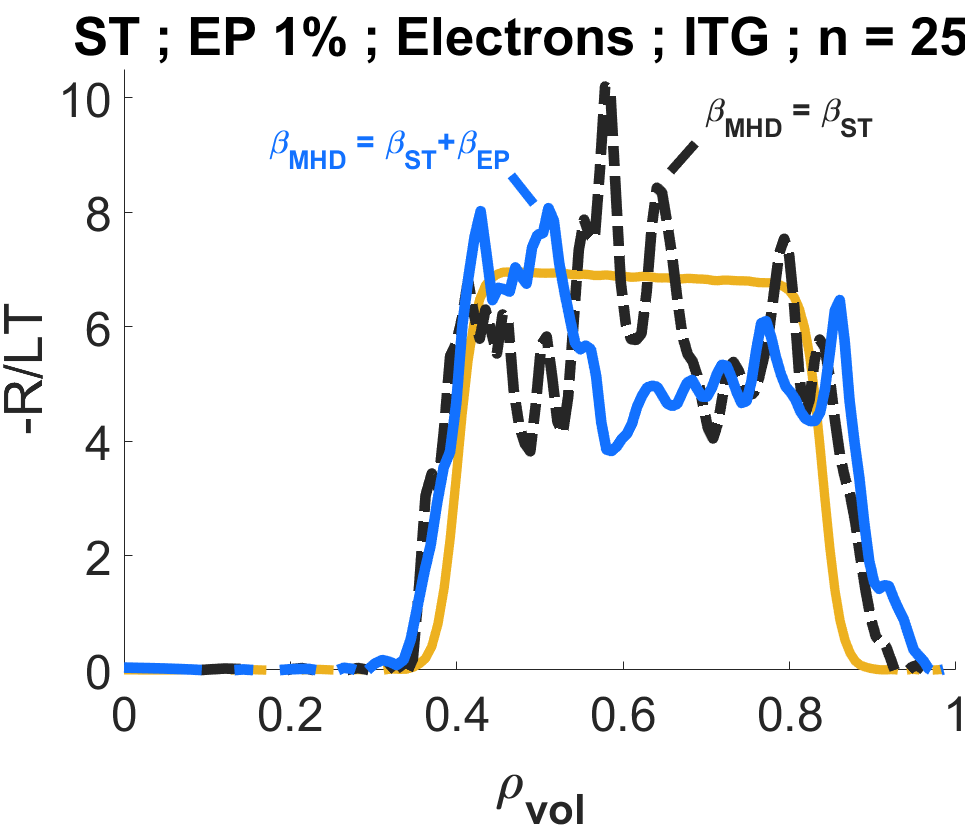}
\end{minipage}
\caption{\label{FIG:nonlinear_ITG_RLT} \it 
Time-averaged relaxation of the logarithmic temperature gradients after the initial saturation phase, separated between the ion (left) and electron channels (right). Presented for the ST case.}
\end{center}
\end{figure}

In Figure \ref{FIG:nonlinear_ITG_RLT} we plot the initial temperature profile for comparison to the time averaged profile taken after the nonlinear saturation phase. The Shafranov shift appears to help maintain the initial gradients which are closer to the core, up to $\rho_{vol} \approx 0.5 ; s \approx 0.6$. We find a correspondence between this region and the strong gradient region in $\Delta^\prime$.  

%Although Shafranov shift modify the profile relaxation quite dramatically, there is no it is difficult to extrapolate the influence on the turbulent fluxes. 

\begin{figure} [h!]
\begin{center}
\includegraphics[width=0.9\textwidth]{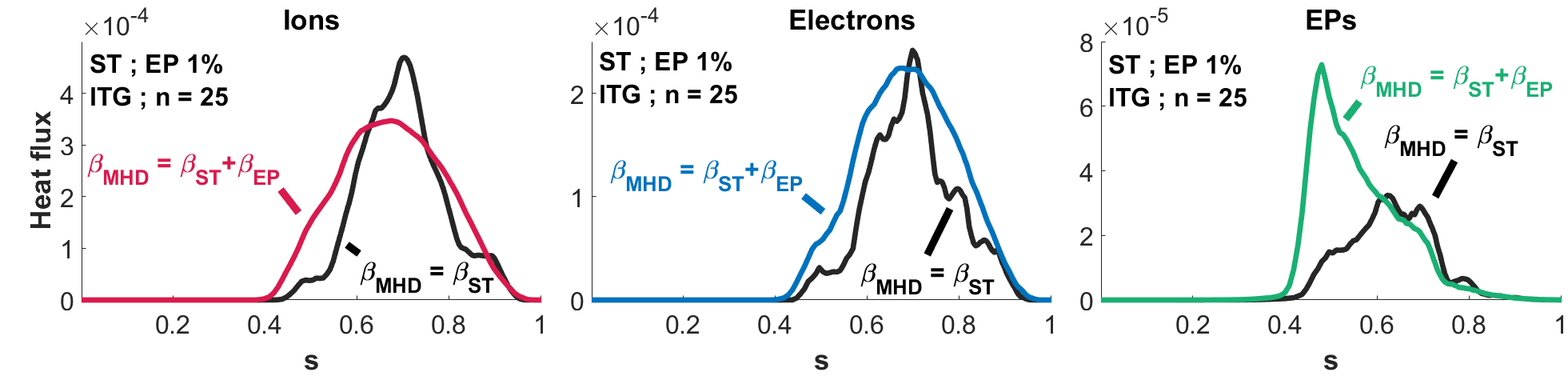}
\includegraphics[width=0.9\textwidth]{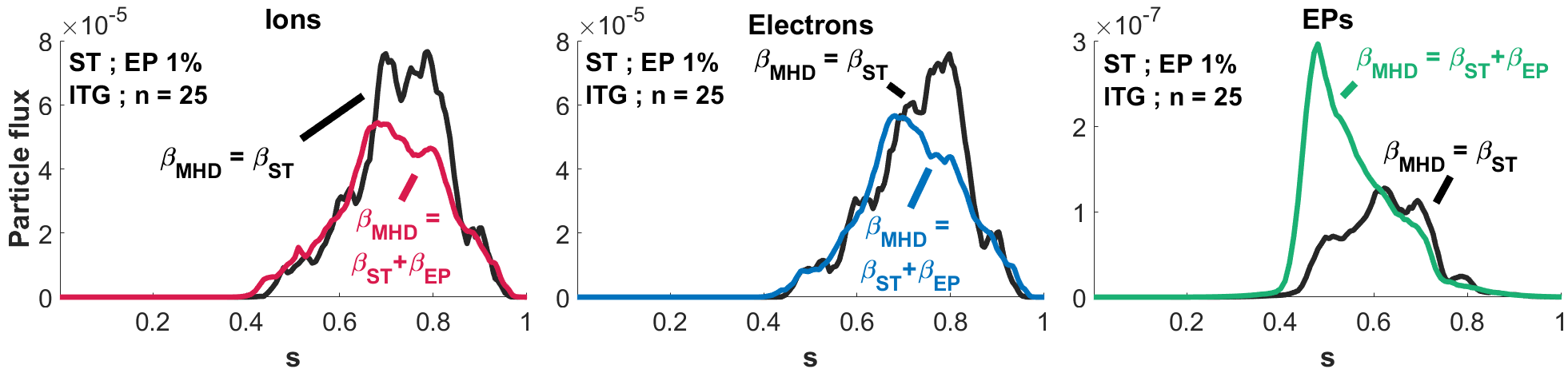}
\caption{\label{FIG:ITG_heat_part_fluxs} \it
Heat and particles fluxes driven by the $n=25$ ITG in a system with standard profiles. The fluxes are averaged over a time window $\Delta$t = $(89 - 144) \ [a/c_{s_0}]$ situated after the initial saturation phase. Note, the y-scale is different for each species and flux.}
\end{center}
\end{figure}

Indeed, the resulting fluxes plotted in Figure \ref{FIG:ITG_heat_part_fluxs}, show no clear Shafranov shift stabilization on the ITG induced fluxes. With the small EP flux, carried by the turbulence, even increasing as a result of including $\beta_{EP}$ in the equilibrium. Similar result for electromagnetic ITG in \textit{self-consistent} MHD equilibria where found by Ishizawa et al \cite{ishizawa2019persistence}.

\FloatBarrier
%======================================================================
\section{Conclusions}
%=====================================================================

In this work we found that to model a burning plasma system where AEs and microturbulence coexist in the presence of EPs, one must solve the electromagnetic Maxwell-Vlasov system, using kinetic electrons in a self-consistent magnetic equilibrium. Otherwise, as we have shown for the internal kink and KBM instabilities, simulations result in modes that are in fact stable in a self-consistent system. Thus, while individual instabilities can be investigated separately, the chosen set of assumptions should maintain the entire global instability spectrum.
Using the global gyrokinetic code ORB5, our analysis offers several additional key insights:

\textbf{Shafranov shift}
\begin{itemize}
    \item The Shafranov shift, even due to a small fraction of hot EPs, is sufficient to modify the global instability spectrum and should always be included in predictive simulations.
    \item In general, we find Shafranov shift effects are stronger at longer wavelengths.
\end{itemize}

\textbf{KBMs are stabilized by two key mechanisms:}
\begin{itemize}
    \item Direct wave-particle interaction with the EPs in $\beta_{MHD}=0$ ideal-MHD equilibrium.
    \item The Shafranov shift due to the thermal plasma in a self-consistent magnetic equilibrium without EPs, is sufficient to stabilize the mode, in contrast to the ITG case.
\end{itemize}

\textbf{TAEs}
\begin{itemize}
    \item In self-consistent equilibria both the EP drive and the Shafranov shift stabilization increase with EP fraction. These competing effects cause the TAE linear growth rate to saturate with EP fraction.
    \item The linear TAE growth rate is very sensitive to Shafranov shift stabilization. Resulting in up to $90\%$ reduction in growth rate in cases with self-consistent equilibria when compared to cases with $\beta_{MHD}=0$ ideal-MHD equilibrium.
    \item Nonlinear saturation levels were found to be practically independent of
    \begin{itemize}
        \item[] EP fraction
        \item[] Thermal profiles
        \item[] Shafranov shift
    \end{itemize}
    \item This strong reduction in linear growth rates does not impact the nonlinear saturation levels, however it does translate to lower EP transport and losses.
\end{itemize}

\textbf{ITGs}
\begin{itemize}
    \item The Shafranov shift due to the thermal plasma in a self-consistent magnetic equilibrium without EPs, has only a weak effect on the dispersion relation, in contrast to the KBM case.
    \item Direct wave-particle interactions with the EPs, have very little effect on the dispersion relation.
    \item Shafranov shift due to EPs significantly modifies the dispersion relation.
    \item The combined Shafranov shift effect due to the thermal plasma and EPs on the nonlinear turbulent transport is negligible.
    \item A model must account for kinetic electrons to capture the correct growth rates, and include electromagnetic effects to capture the correct frequency.
\end{itemize}

In this work, we chose to neglect the axisymmetric $n=0$ plasma response which otherwise dominates the system behavior during the non-linear phase. This topic is the focus of a separate work by Rofman et. al. \cite{rofman2026roleshafranovshiftzonal}.
While in this work we tested many of the frequently used assumptions, we have far from exhausted the list. Future works can build upon this systematic study to find the relative impact of additional physics, such as geometric shaping factors, distribution functions, and system size - $\rho^*$ effects.
The understanding of individual modes as part of the combined global system can be expanded in future works to systems where different instabilities coexist, free to interact with each other. Although some works combine many physical elements simultaneously to study more realistic systems. It is often challenging to interpret the results and identify the individual contributions of different physical elements. Therefore, the insight and subsequently the ability to apply such results beyond the case in point is limited. To gain better understanding of the nonlinear interactions in a burning plasma system, one first needs to understand the nonlinear evolution of individual modes under a globally consistent set of approximations. This work lays a well-explored foundation for more comprehensive nonlinear studies of mode-mode interactions between AE, drift waves like ITG, and EPs in a self-consistent system.

\section{Acknowledgment}
This work has been carried out within the framework of the EUROfusion Consortium, partially funded by the European Union via the Euratom Research and Training Programme (Grant Agreement No 101052200 — EUROfusion). The Swiss contribution to this work has been funded in part by the Swiss State Secretariat for Education, Research and Innovation (SERI). Views and opinions expressed are however those of the author(s) only and do not necessarily reflect those of the European Union, the European Commission or SERI. Neither the European Union nor the European Commission nor SERI can be held responsible for them. This work was supported in part by the Swiss National Science Foundation (Grant number 228309). This work is also supported by a grant from the Swiss National Supercomputing Center (CSCS) under project IDs s1232 $\&$ lp73. This work was supported in part by the Swiss National Science Foundation.

%======================================================================
%\section*{References}
%======================================================================
%\ldots
%\bibliographystyle{abbrv}
%\bibliography{References}
\printbibliography %Prints bibliography
\end{document}